\documentclass[prc,aps]{revtex4-2}
\usepackage{graphicx,hyperref}
\usepackage{color}
\begin{document}
\title{Phase transitions in {\it{\bf{N}}} = 40, 60 and 90 nuclei}

\author{A. Pr\'{a}\v{s}ek}\thanks{deceased} and \author{P. Alexa}
\address{Department of Physics, 
V\v{S}B -- Technical University Ostrava,
17. listopadu 2172/15, CZ-708 00 Ostrava, Czech Republic}

\author{D. Bonatsos}
\address{Institute of Nuclear and Particle Physics, N.C.S.R. Demokritos, GR-15310 Aghia Paraskevi, Attiki, Greece}

\author{G. Thiamov\'{a}}
\address{Universite Grenoble 1, CNRS, LPSC, Institut Polytechnique de  Grenoble, IN2P3, F-38026 Grenoble, France}

\author{D. Petrellis and P. Vesel\'{y}}
\address{Nuclear Physics Institute, Czech Academy of Sciences, CZ-250 68 \v{R}e\v{z} near Prague, Czech Republic}

\date{\today}
\begin{abstract}
In this paper we focus on three mass regions where first-order phase transitions occur, namely for $N=40$, 60 and 90. We investigate four isotopic chains (Se, Zr, Mo and Nd) 
in the framework of microscopic Skyrme-Hartree-Fock +  Bardeen-Cooper-Schrieffer calculations for 15 different parametrizations. The microscopic calculations show the typical behavior expected for first-order phase transitions. To find the best candidate for the critical point phase transition we propose new microscopic position and occupation indices calculated for positive-parity and negative-parity proton and neutron single-quasiparticle states around the Fermi level. The microscopic calculations are completed by macroscopic calculations within the Algebraic Collective Model (ACM), and compared to the experimental data for $^{74}$Se, $^{102}$Mo and $^{150}$Nd,  considered to be the best candidates for the critical point nuclei.
\end{abstract} 

\pacs{21.10.Re, 21.60.Ev, 23.20.Lv}

\maketitle

\section{Introduction}
\label{Intro}

Quantum phase transitions (QPT) \cite{Casten2009,Cejnar2009,Cejnar2010} and critical point symmetries (CPS) \cite{Iachello2000,Iachello2001,Casten2006,Casten2007} in even-even atomic nuclei belong to one of most studied topics in nuclear structure physics for several decades. They are observed in chains of isotopes in which the addition of two neutrons causes a radical change of the nuclear structure, the sudden jump from vibrational behavior in $^{150}$Sm to rotational behavior in $^{152}$Sm \cite{Iachello2001,Casten2001} being a characteristic example. The neutron number is used as a control parameter in these transitions \cite{Casten1999}, while various collective quantities are used as the order parameter \cite{Werner2002,Iachello2004,Iachello2006,Bonatsos2008}. 

The Ehrenfest classification of nuclear QPTs started in 1981 \cite{Feng1981}, with the realization that within the parameter space of the Interacting Boson Model 
\cite{Arima1976,Arima1978,Arima1979,Iachello1987} a second-order QPT takes place between U(5) (spherical) \cite{Arima1976}  and O(6) ($\gamma$-unstable, i.e., soft towards triaxial deformation) \cite{Arima1979} nuclei, while a first-order QPT occurs between U(5) (spherical) and SU(3) (deformed) \cite{Arima1978} nuclei. The critical point symmetries E(5) \cite{Iachello2000,Caprio2007} and X(5) \cite{Iachello2001,Bijker2003,Caprio2005} have been introduced, in 2000 and 2001 respectively, for the description of these QPTs within the framework of the Bohr collective Hamiltonian \cite{Bohr1952,Bohr1998a,Bohr1998b} providing parameter independent (up to overall scales) predictions for spectra and  transition rates at the critical point of these QPTs. The $N=90$ isotones $^{150}$Nd \cite{Krucken2002}, $^{152}$Sm \cite{Casten2001}, $^{154}$Gd \cite{Tonev2004}, $^{156}$Dy \cite{Caprio2002} have been suggested as the best experimental manifestations of the X(5) CPS. A $\gamma$-rigid version of X(5), called X(3), in which the $\gamma$ value has been fixed to zero, has also been introduced \cite{Bonatsos2006}.

Even before the introduction of the concept of CPS, the mechanism of the onset of deformation within microscopic models was investigated in a series of papers by Federman and Pittel \cite{Federman1977,Federman1978,Federman1979a,Federman1979b}. They pointed out the crucial role played by the proton-neutron interaction in the creation of deformation, in particular stressing the major role played by intruder neutron orbitals in the development of deformation across the nuclear chart. In addition, the concept of quasi-dynamical symmetries \cite{Rowe2004b}, i.e., symmetries persisting in the presence of strong symmetry-breaking interactions, has been introduced for both first-order \cite{Rosensteel2005} and second-order \cite{Rowe2004c,Turner2005} QPTs.  

Recently, the connection between CPSs linked to QPTs and the effect of shape coexistence (SC) \cite{Heyde1983,Wood1992,Heyde2011,Garrett2022,Atoms2023} attracted considerable interest \cite{GRamos2018,GRamos2019,GRamos2020,GRamos2022}. Shape coexistence \cite{Heyde1983,Wood1992,Heyde2011,Garrett2022,Atoms2023}
is said to occur in nuclei in which  the ground state band and an excited $K=0$ band lie close in energy and simultaneously exhibit radically different structures and symmetries, for example one of them being spherical and the other one deformed, or both of them being deformed, but one of them exhbiting a prolate (rugby-ball-like) shape and the other one showing an oblate (pancake-like) one. The $N=60$ isotones $^{100}$Zr \cite{GRamos2018,GRamos2019,GRamos2020} and $^{98}$Sr \cite{GRamos2022} have been found as the best nuclei in which the connection between CPS and QPT is observed. The relevant studies have been carried out in the framework of the IBM with configuration mixing (IBM-CM), which takes particle-hole (p-h) excitations into account \cite{DeCoster1996,Lehmann1997,DeCoster1997,DeCoster1999} and allows for mixing of the ground state configuration of $N$ bosons with the excited 2p-2h configuration, described by $N+2$ bosons \cite{Duval1981,Duval1982}. It should be noticed that the occurence of phase coexistence in the region of transition between spherical and deformed shapes has been pointed out in the IBM framework already in 1998 \cite{Iachello1998}, while the structural similarity between the $N=90$ and $N=60$ regions has been pointed out already in 1981 \cite{Casten1981}, implying a common microscopic origin of the QPT from spherical to deformed shapes in these two regions. Furthermore, it has been recently suggested \cite{Gavrielov2019,Gavrielov2020,Gavrielov2022} that $^{100}$Zr is the critical point of two intertwined QPTs, the one from spherical to deformed shapes already mentioned and another one from normal configurations (without particle-hole excitations)  to intruder configurations (involving particle-hole excitations). 

In addition, a recent study \cite{JPG2023} of the systematics of energy levels and $B(E2)$ transition rates in the $N=40$ region around $^{74}$Se pointed out the structural similarity of the $N=40$ region to the above mentioned $N=60$ and $N=90$ regions.  It has been shown that in all these three regions intruder orbitals start participating to the onset of deformation. In particular, neutron particle-hole excitations are found to lead to both SC and QPT from spherical to prolate deformed shapes taking place in parallel.

The above findings suggest that the three regions with $N=90$, $N=60$, and $N=40$ exhibit in a similar way a QPT from spherical to deformed shapes, based on the microscopic picture of the onset of deformation due to increased proton-neutron interaction, with SC triggered at the same time by the creation of neutron particle-hole excitations (called proton-induced particle-hole excitations in the framework of the dual shell mechanism \cite{Martinou2021} for SC developed within the proxy-SU(3) approximation
\cite{Bonatsos2017a,Bonatsos2017b,Bonatsos2023}, corroborated through covariant density functional theory calculations \cite{Bonatsos2022a,Bonatsos2022b}). Theoretical calculations in these three regions, related to the presence of SC, have been recently reviewed in \cite{Atoms2023}. 

However, despite the similar microscopic origins of the three regions, only the critical isotones with $N=90$ can be described in terms of the parameter-free predictions of the X(5) CPS \cite{Krucken2002,Casten2001,Tonev2004,Caprio2002}, while for $N=60$ and $N=40$ the data for spectra and $B(E2)$ transition rates are far away from the X(5) predictions. It is therefore of interest to look for a more flexible theoretical framework, possibly able to accommodate these three critical regions simultaneously.  

Therefore, in the present paper we investigate the three CPS regions within different microscopic and macroscopic approaches, attempting to shed more light on the nature of the CPS and possibly find a common framework for all of them. In Sec.~\ref{emp} empirical systematics of different spectral and transitional signatures of CPS are discussed in the three regions of interest. In Sec.~\ref{MMFC} we turn to microscopic calculations based on the Skyrme-Hartree-Fock + Bardeen-Cooper-Schrieffer (SHF + BCS) model to study potential energy curves (PEC) as functions of quadrupole, octupole and hexadecapole deformations, and introduce microscopic position and occupation indices for 
positive-parity and negative-parity proton and neutron single-quasiparticle states around the Fermi level. In Sec.~\ref{ACM} we briefly introduce the Algebraic Collective Model (ACM), which is then applied to $^{74}$Se, $^{102}$Mo and $^{150}$Nd in Sec.~\ref{Num_ACM}.

\section{Empirical systematics}  \label{emp} 

We are going to use the energy ratios 
\begin{equation}
R_{4/2} = {E(4_1^+) \over E(2_1^+)}, \quad R_{2/0} = {E(2_1^+) \over E(0_2^+)}, \quad R_{2/2} = {E(2_\gamma^+) \over E(2_1^+)},
\end{equation}
as well as the rates of change  with the neutron number $N$ (which possess twice the value of relevant mathematical derivative), 
\begin{eqnarray}
{\mathrm{d}R_{4/2}\over \mathrm{d}N}(N)=  R_{4/2}(N)-R_{4/2}(N-2), \\ \nonumber
{\mathrm{d}R_{2/2}\over \mathrm{d}N}(N)=  R_{2/2}(N)-R_{2/2}(N-2). \nonumber
\end{eqnarray}
In addition we are going to use the transition rates $B(E2; 2_1^+ \to 0_1^+)$, and their rate of change with the neutron number $N$ (which possesses twice the value of relevant mathematical derivative), 
\begin{equation}
{\mathrm{d}B(E2; 2_1^+ \to 0_1^+)\over \mathrm{d}N}(N) = B(E2; 2_1^+ \to 0_1^+)(N)-B(E2; 2_1^+ \to 0_1^+)(N-2),
\end{equation}

The ratio $R_{4/2}$ is a well known indicator of collectivity \cite{Casten1990}, obtaining values 2.0--2.4 for near-spherical nuclei, 2.4--3.0 for $\gamma$-unstable nuclei, and 3.0--3.33 for deformed nuclei. 

The transition rate $B(E2; 2_1^+ \to 0_1^+)$ is known \cite{Raman2001} to be proportional to the square of the quadrupole deformation parameter $\beta$, expressing the deviation of the nuclear shape from sphericity. 

The ratio $R_{2/0}$ is known \cite{JPG2023} to exhibit a maximum in the region of the QPT from spherical to deformed nuclei. 

\begin{table*}

\caption{Energy ratios $R_{4/2}$, $R_{2/2}$, and transition rates $B(E2; 2_1^+ \to 0_1^+)$, as well as their rates of change as a function of the neutron number $N$, 
${\mathrm{d}R_{4/2} \over \mathrm{d}N}(N)$,  ${\mathrm{d}R_{2/2} \over \mathrm{d}N}(N)$, and ${\mathrm{d}B(E2)\over \mathrm{d}N}(N)$,  are listed for nuclei at $N=90$, 60, 42 and 40, lying near the maxima appearing in Figs. \ref{FR42}-\ref{FR22}. In addition, the energy ratios $R_{2/0}$ are shown for the isotones listed, labeled as  $R_{2/0}(N)$. The parameter-independent predictions of the critical point symmetries X(5) \cite{Iachello2001,Bonatsos2004} and X(3) \cite{Bonatsos2006} are also shown for comparison.  
See Sec. \ref{emp} for further discussion. }

\bigskip

\begin{tabular}{ r r r r r r  r r }

\hline
nuc. & $R_{4/2}$ & ${\mathrm{d}R_{4/2} \over \mathrm{d}N}(N)$ & $B(E2)$ & ${\mathrm{d}B(E2)\over \mathrm{d}N}(N)$ & $R_{2/0}(N)$ & $R_{2/2}$ & ${\mathrm{d}R_{2/2} \over \mathrm{d}N}(N)$\\
     &           &                       &  W.u.   & W.u.   &  &  &  \\

 \hline
  
$^{150}$Nd$_{90}$ & 2.927 & 0.434 & 114.4 & 56.9 & 0.193 & 8.156 & 4.017 \\
$^{152}$Sm$_{90}$ & 3.009 & 0.693 & 143.7 & 86.8 & 0.178 & 8.916 & 5.786 \\
$^{154}$Gd$_{90}$ & 3.015 & 0.821 & 158.0 & 89.3 & 0.181 & 8.095 & 4.873 \\
$^{156}$Dy$_{90}$ & 2.934 & 0.700 & 149.2 & 50.4 & 0.204 & 6.464 & 3.392 \\
 
 $^{98}$Sr$_{60}$ & 3.006 & 0.806 &  92.4 & 74.7 & 0.669 &10.674 & 8.676 \\
$^{100}$Zr$_{60}$ & 2.656 & 0.982 &  80.5 &      & 0.642 & 5.628 & 3.808 \\
$^{102}$Mo$_{60}$ & 2.507 & 0.386 &  69.0 & 30.6 & 0.425 & 2.859 & 0.873 \\

 $^{72}$Zn$_{42}$ & 2.297 & 0.278 &  21.1 &  3.5 & 0.432 & 2.539 & 0.551 \\
 $^{74}$Ge$_{42}$ & 2.457 & 0.385 &  33.1 &  9.6 & 0.402 & 2.021 & 0.266 \\
 $^{76}$Se$_{42}$ & 2.380 & 0.232 &  45.1 &  6.4 & 0.498 & 2.175 & 0.176 \\
 $^{74}$Se$_{40}$ & 2.148 & 0.249 &  38.7 & 17.4 & 0.743 & 1.999 & 0.472 \\
 
 X(5)             & 2.907 &       &       &      & 0.177 &       &       \\
 X(3)             & 2.440 &       &       &      & 0.350 &       &       \\

\hline

\end{tabular} \label{Tmax}
\end{table*}


\begin{figure*} [htb]

    {\includegraphics[width=75mm]{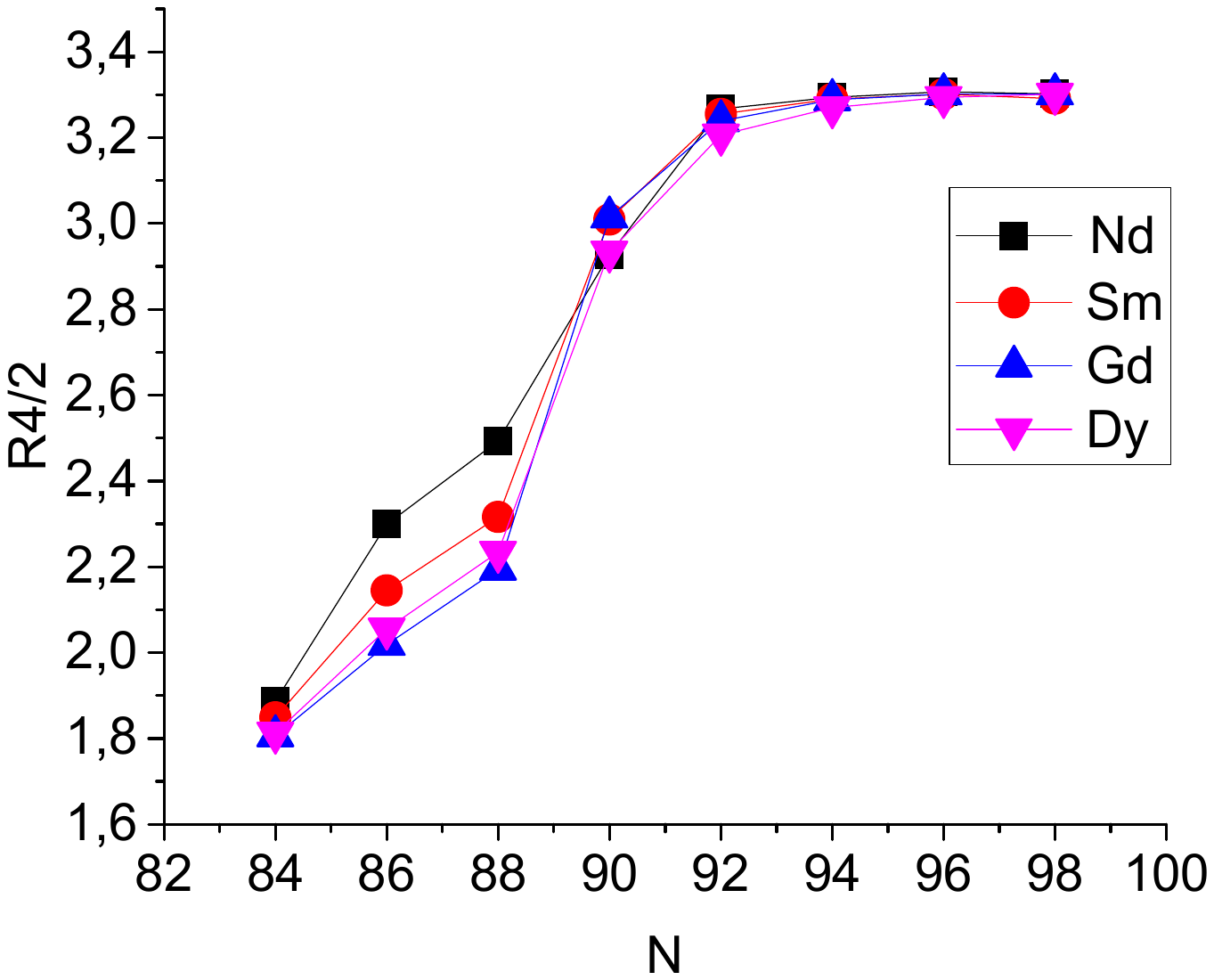}\hspace{5mm}
    \includegraphics[width=75mm]{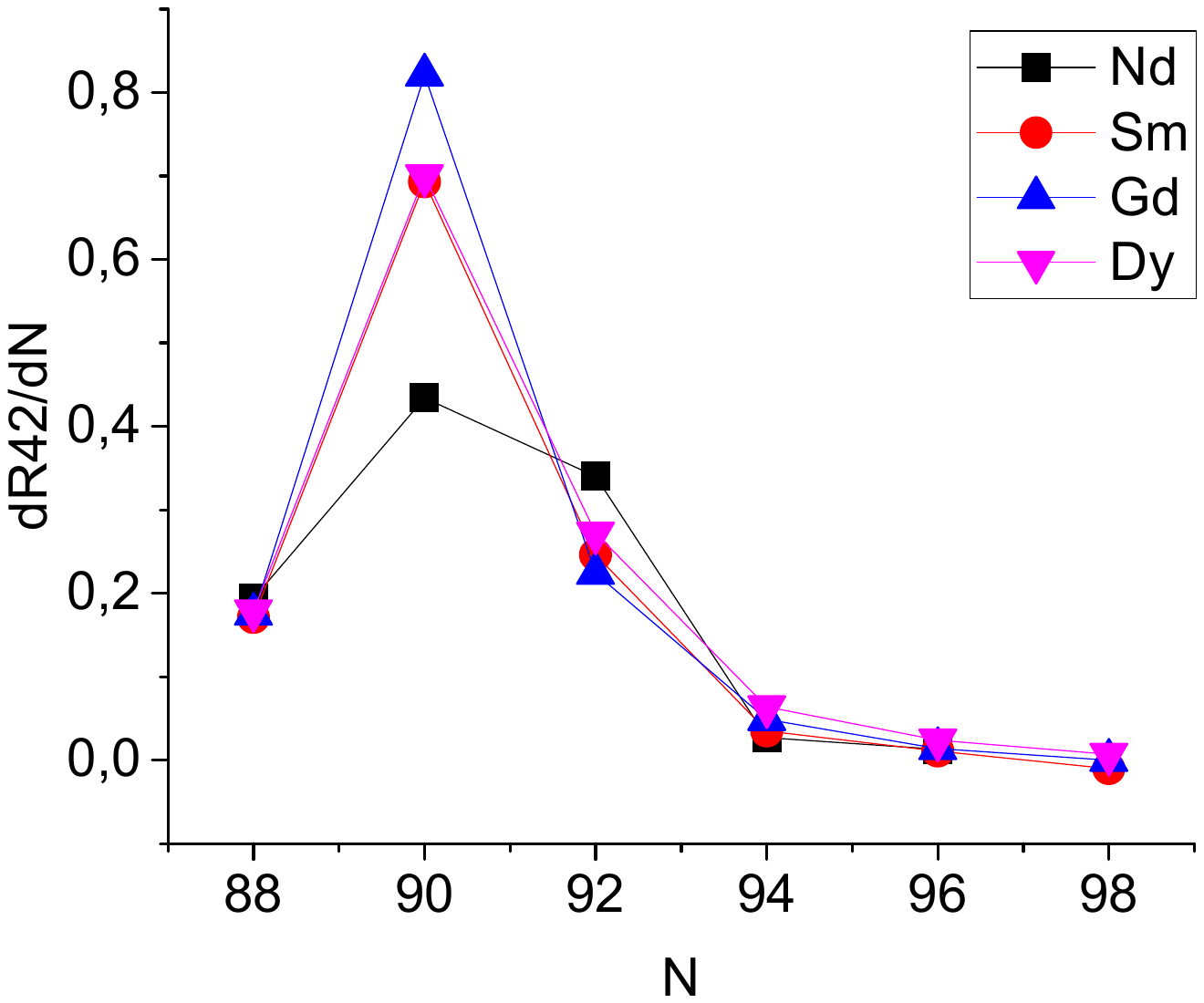}}
      {\includegraphics[width=75mm]{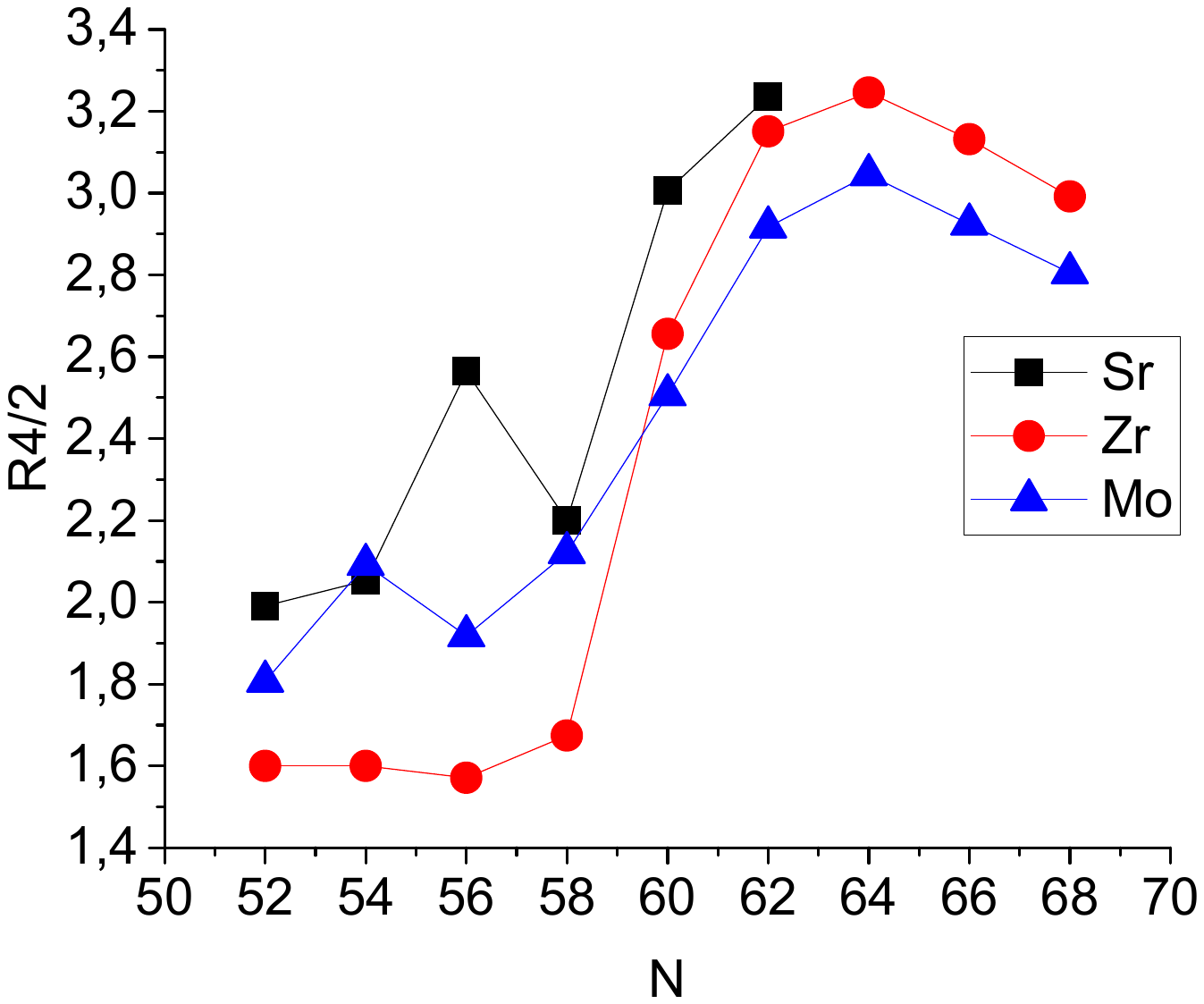}\hspace{5mm}
    \includegraphics[width=75mm]{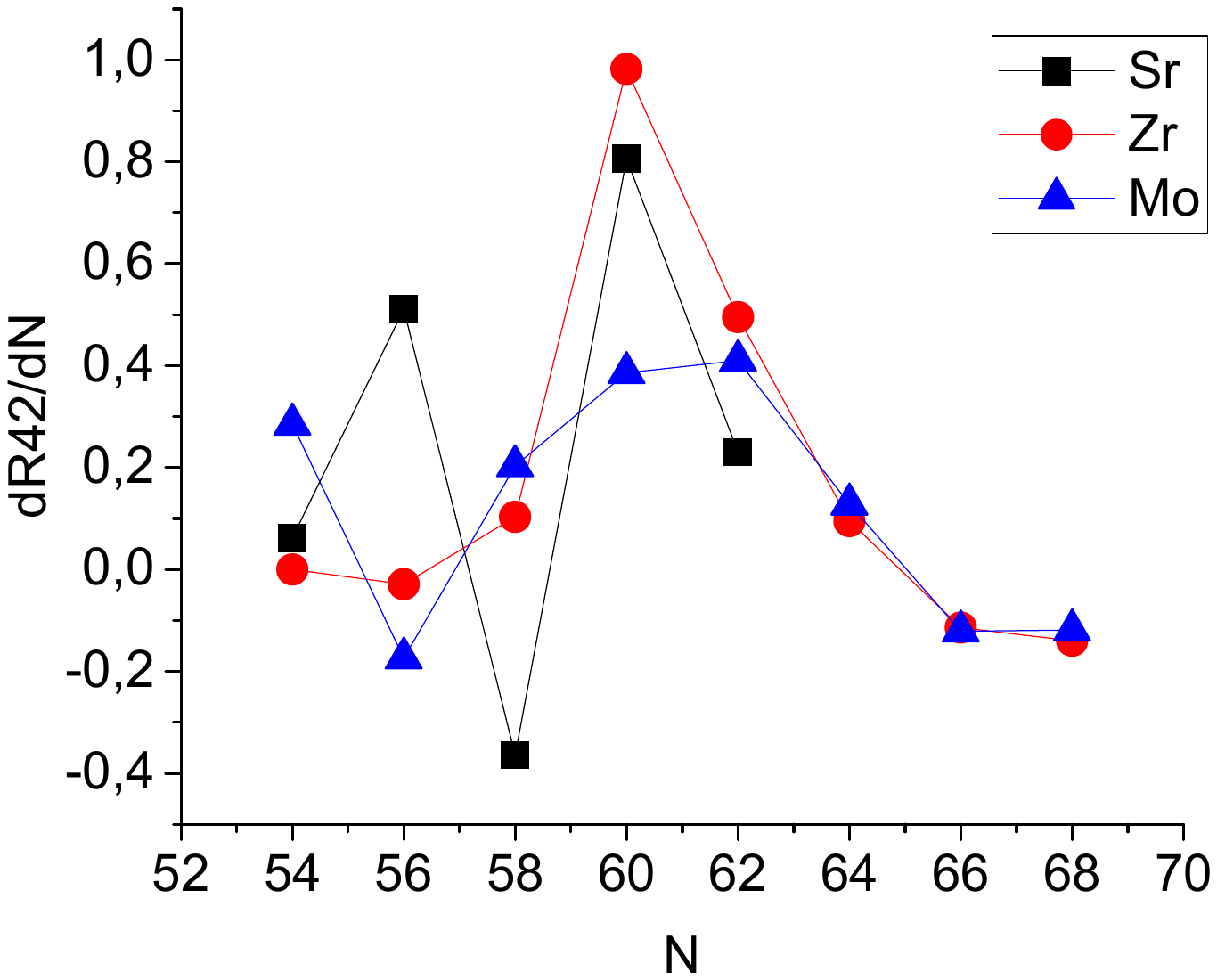}}
      {\includegraphics[width=75mm]{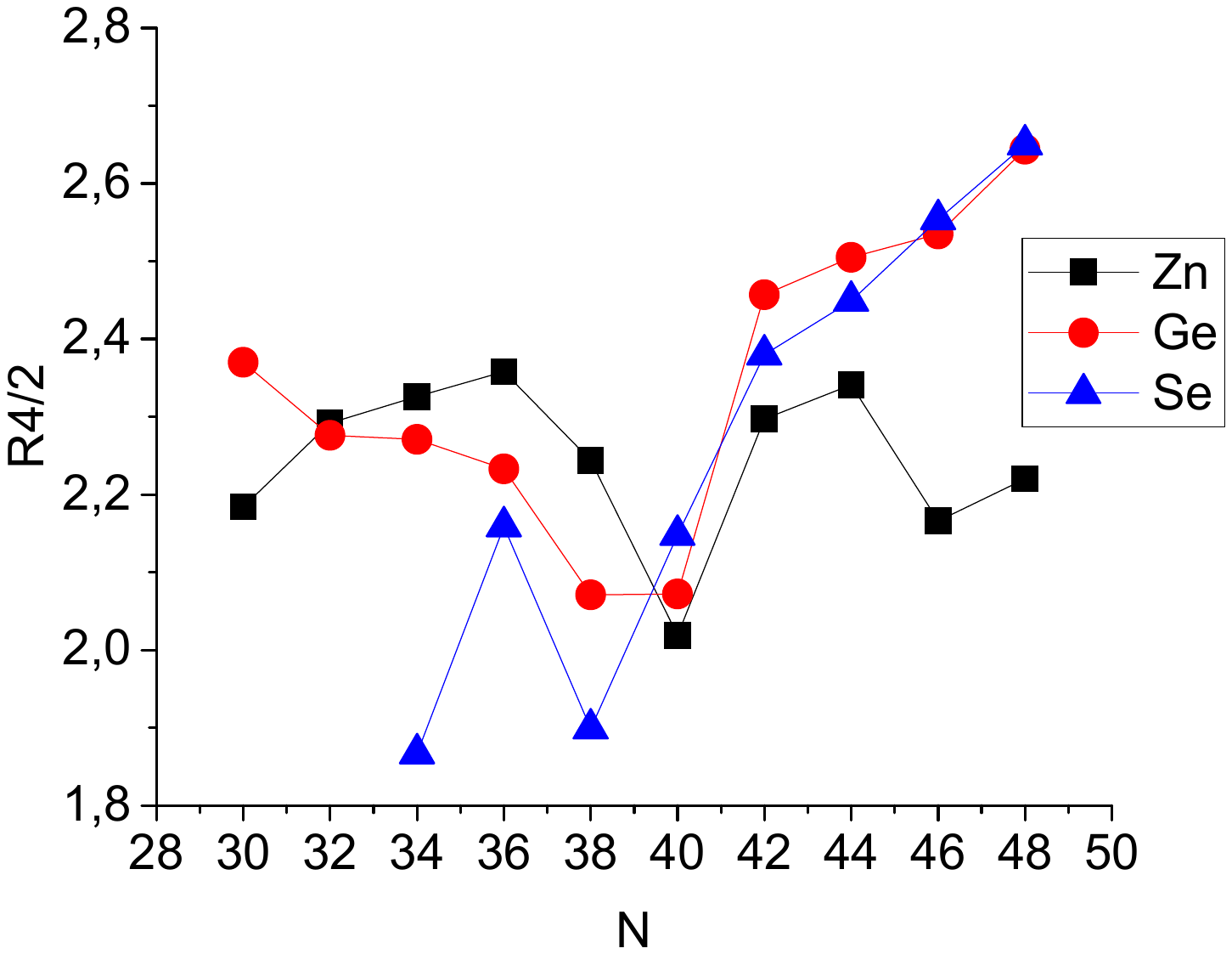}\hspace{5mm}
    \includegraphics[width=75mm]{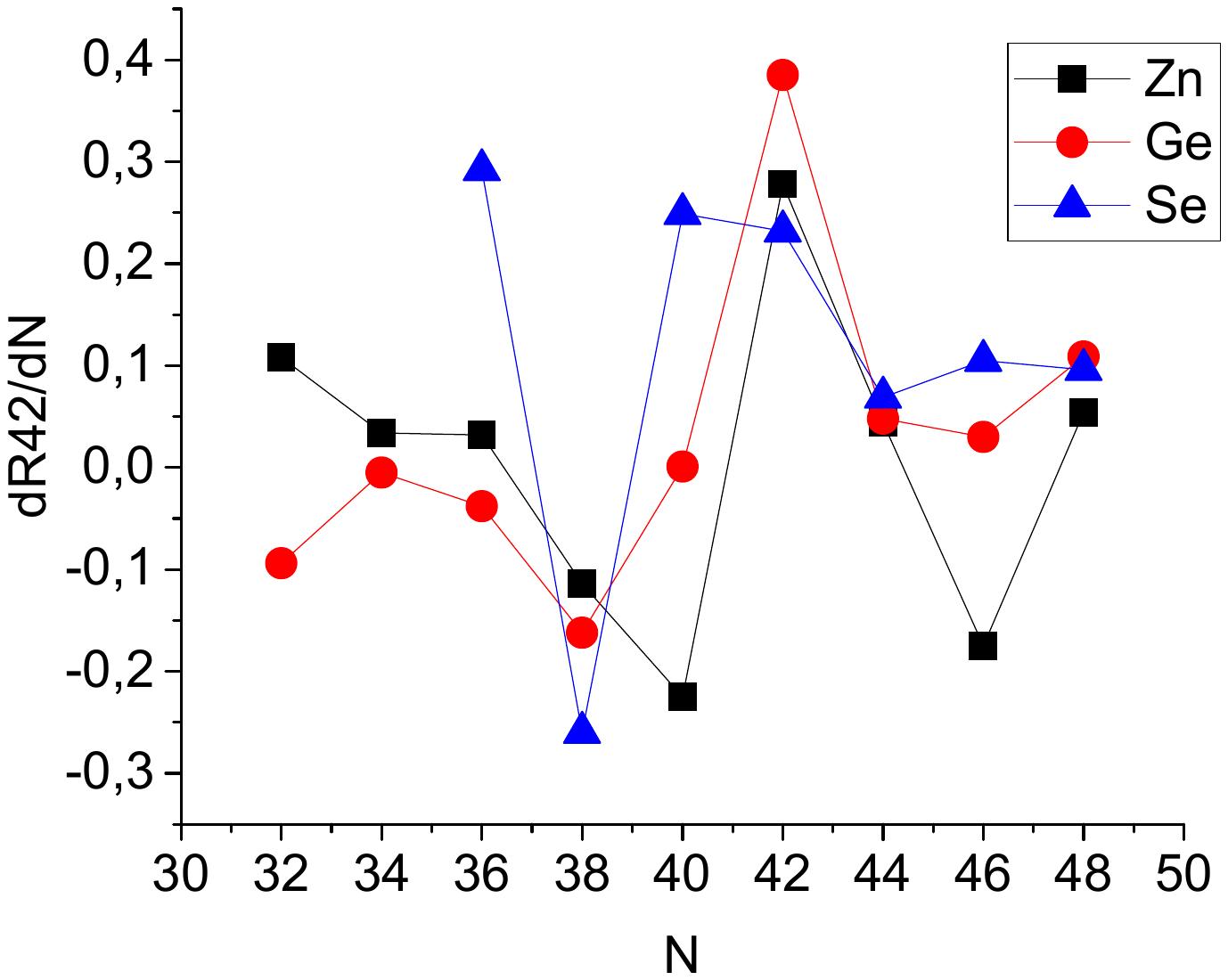}}
   
    \caption{Experimental \cite{ensdf} energy ratios $R_{4/2}$ and their rate of change $\mathrm{d}R_{4/2}/\mathrm{d}N$ with respect to the neutron number $N$ in the $N=90$, 60, 40 regions. See Section \ref{emp} for further discussion.} 
    \label{FR42}
\end{figure*}


\begin{figure*} [htb]

    {\includegraphics[width=75mm]{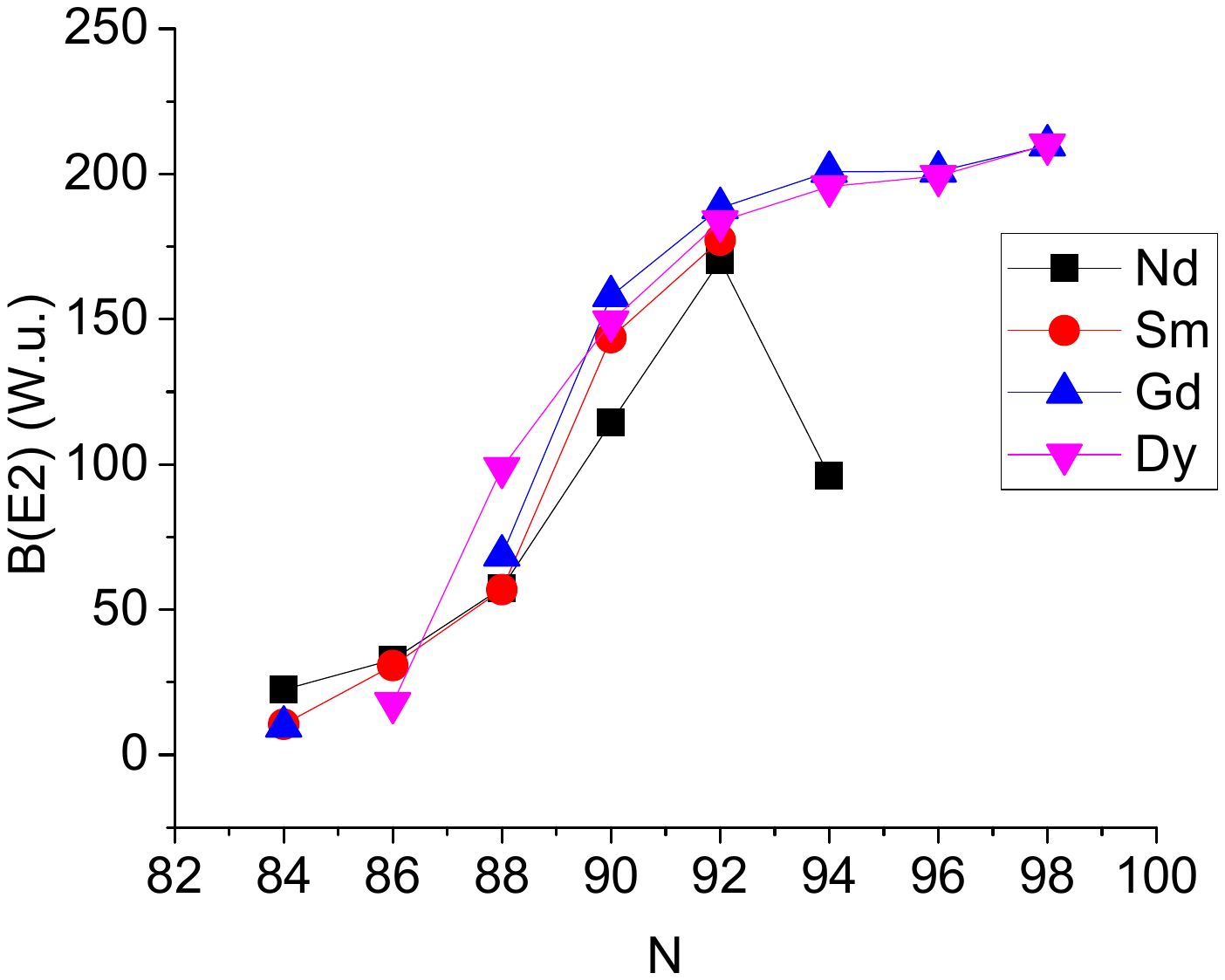}\hspace{5mm}
    \includegraphics[width=75mm]{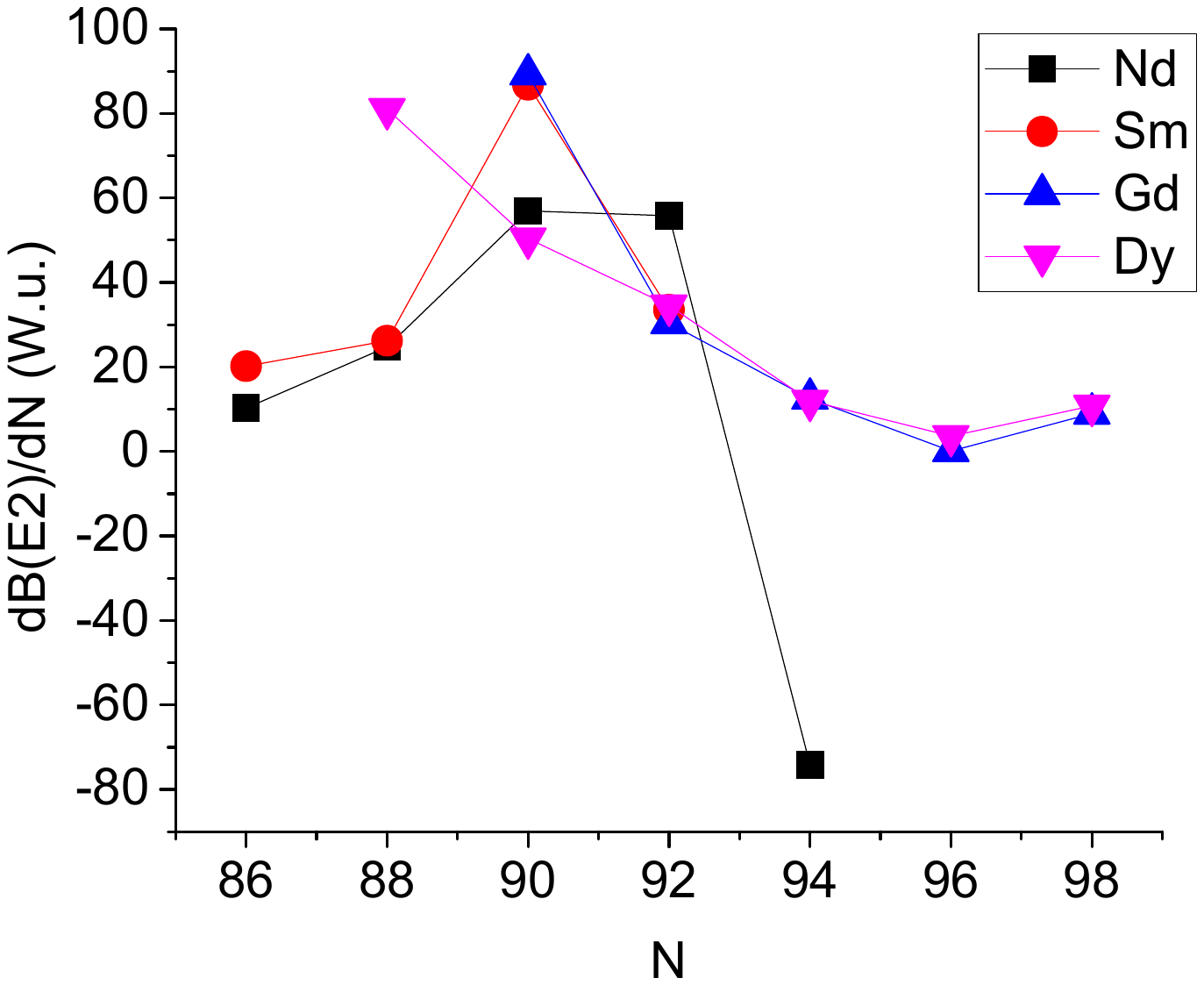}}
      {\includegraphics[width=75mm]{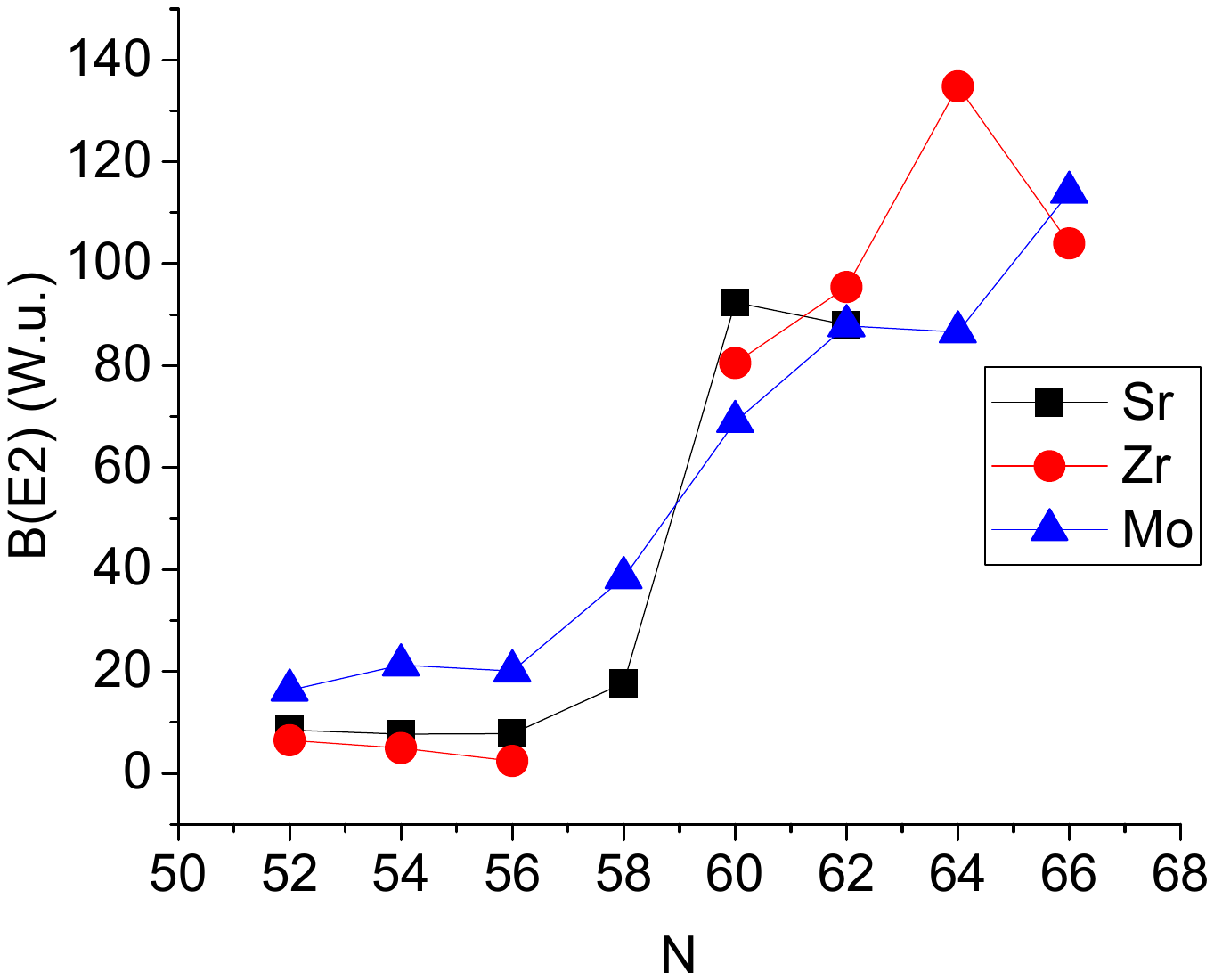}\hspace{5mm}
    \includegraphics[width=75mm]{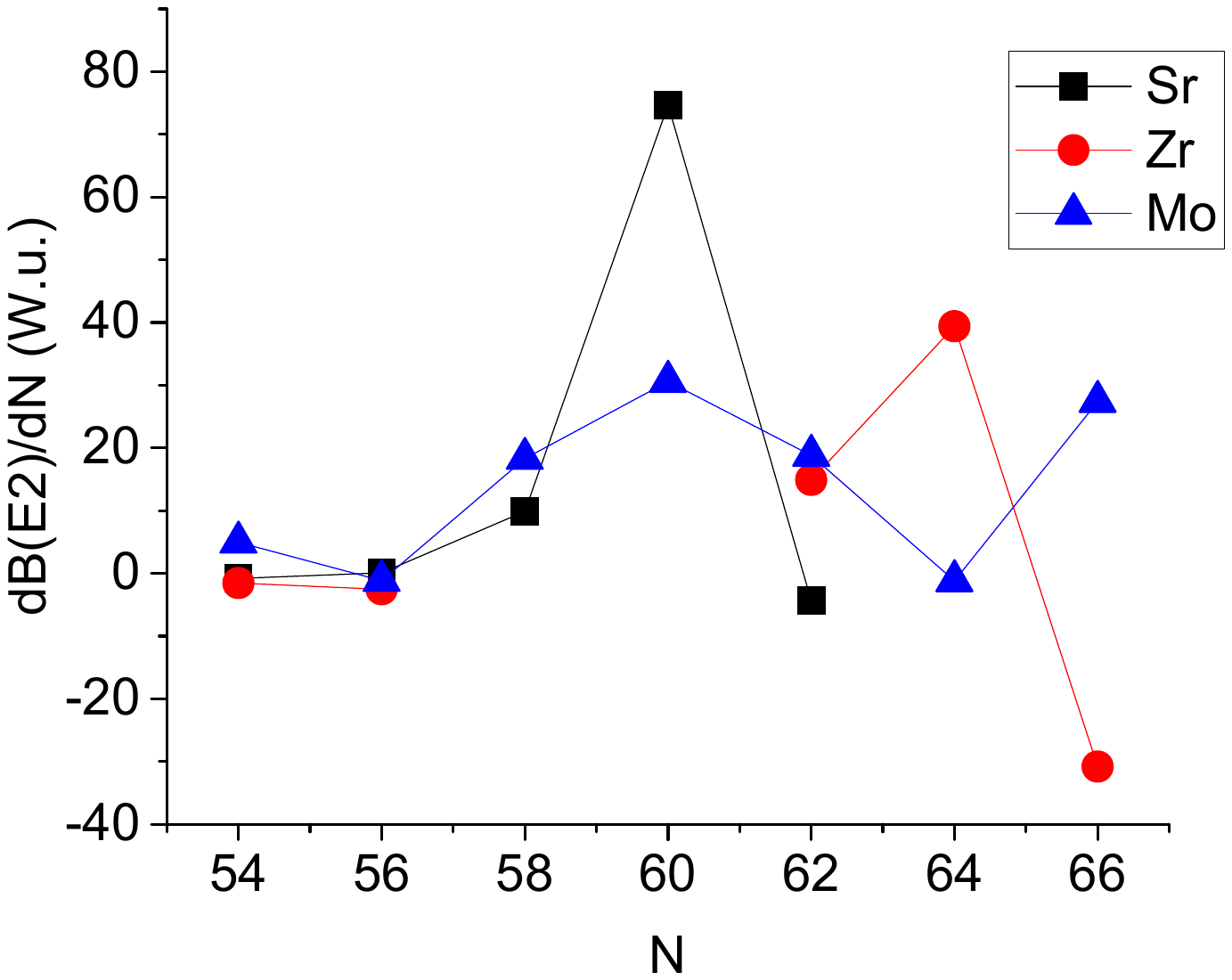}}
      {\includegraphics[width=75mm]{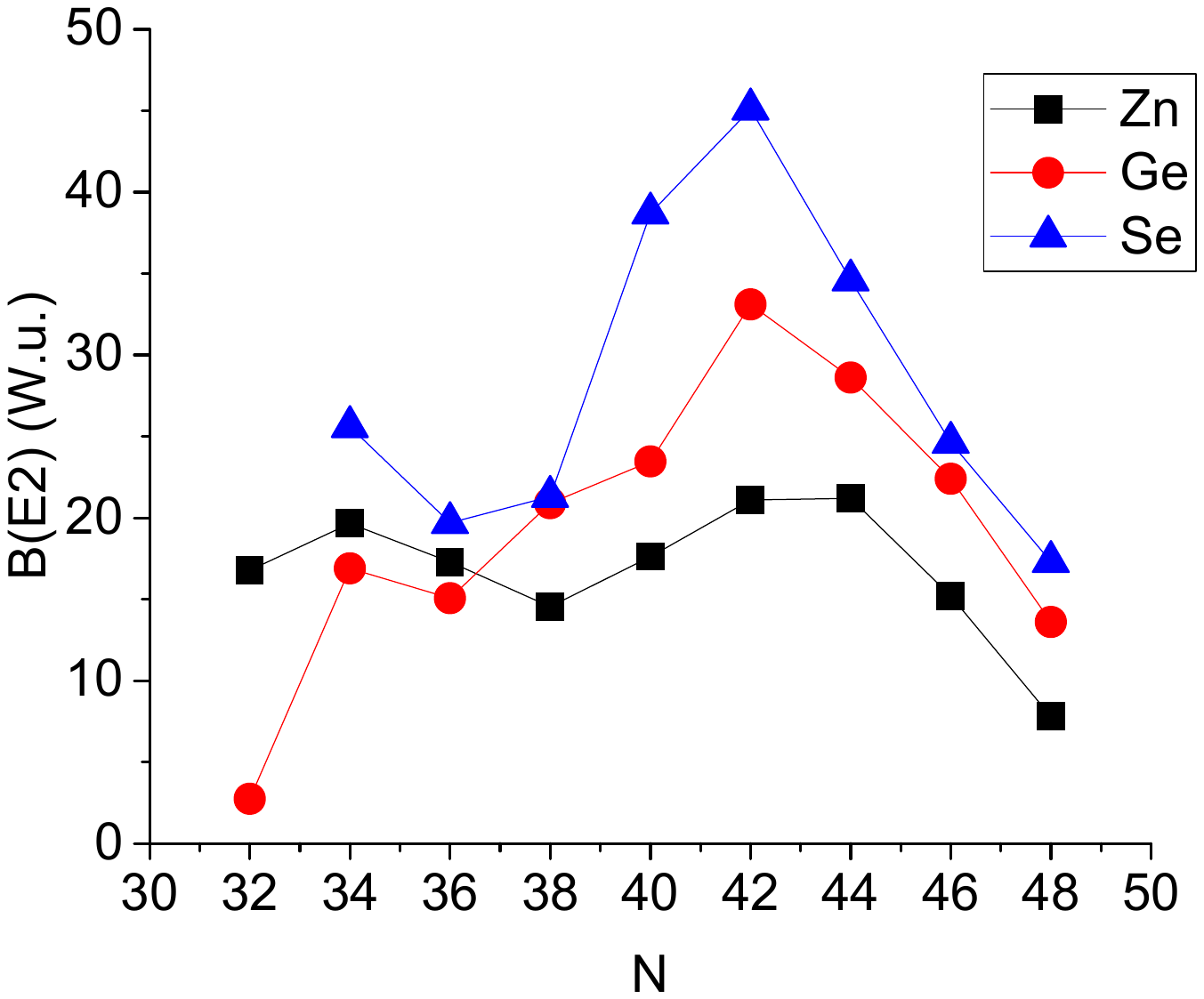}\hspace{5mm}
    \includegraphics[width=75mm]{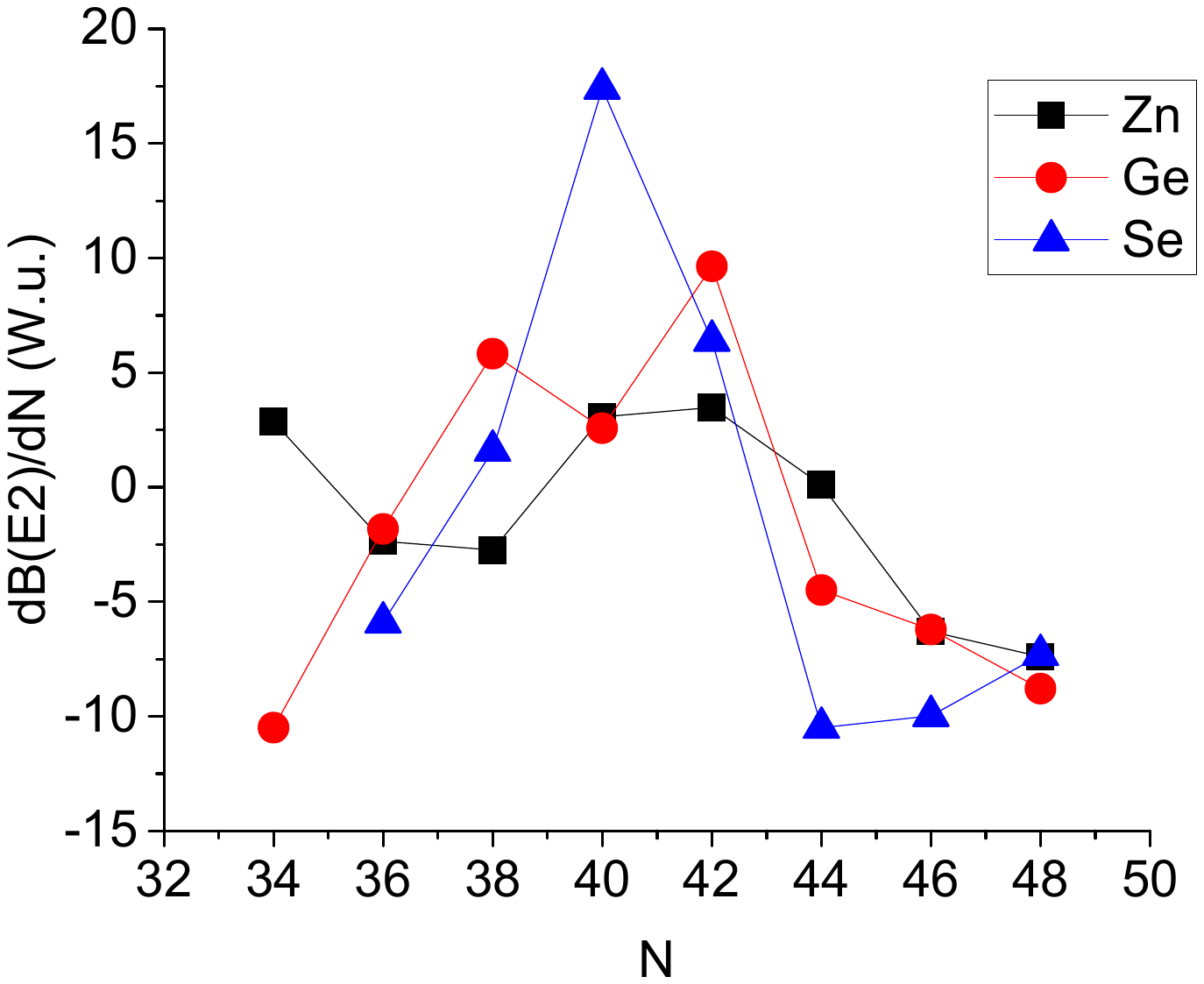}}
    
    \caption{ Experimental $B(E2; 2_1^+\to 0_1^+)$ transition rates \cite{Pritychenko} and their rate of change $\mathrm{d}B(E2; 2_1^+\to 0_1^+)/\mathrm{d}N$ with respect to the neutron number $N$ in the $N=90$, 60, 40 regions. See Section \ref{emp} for further discussion. }
    \label{FB}
\end{figure*}


\begin{figure} [htb]

    \includegraphics[width=75mm]{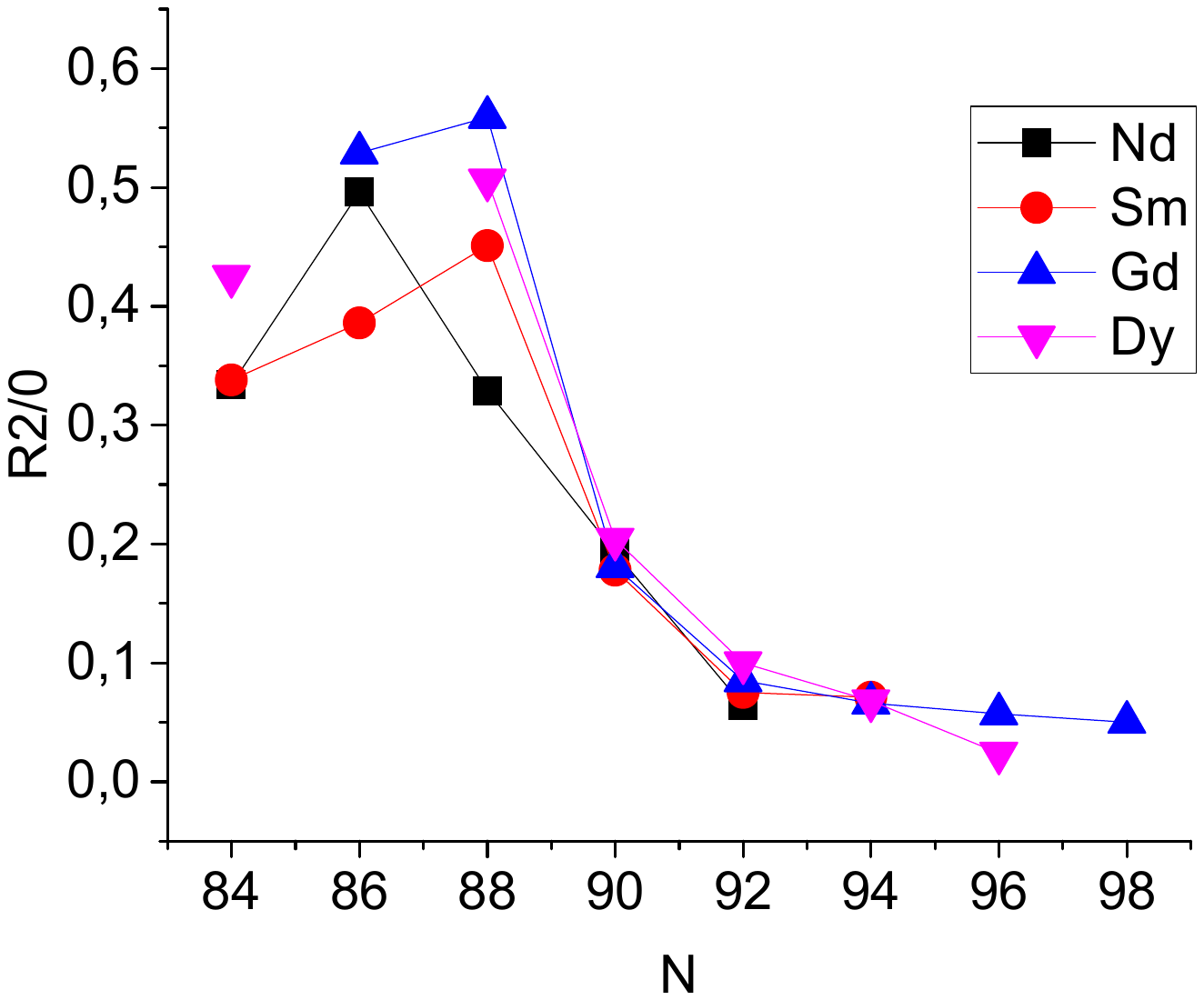}
    \includegraphics[width=75mm]{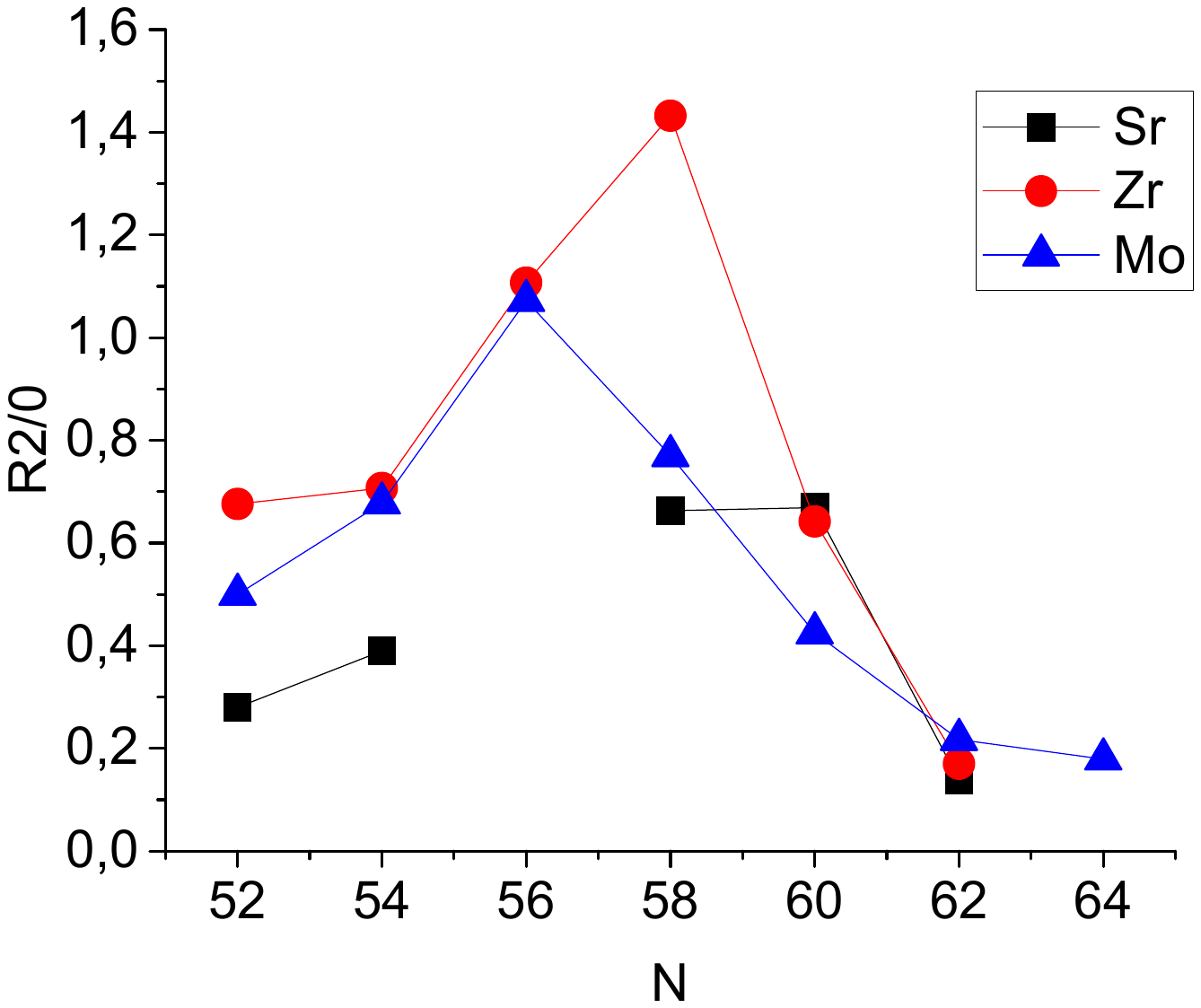}
      \includegraphics[width=75mm]{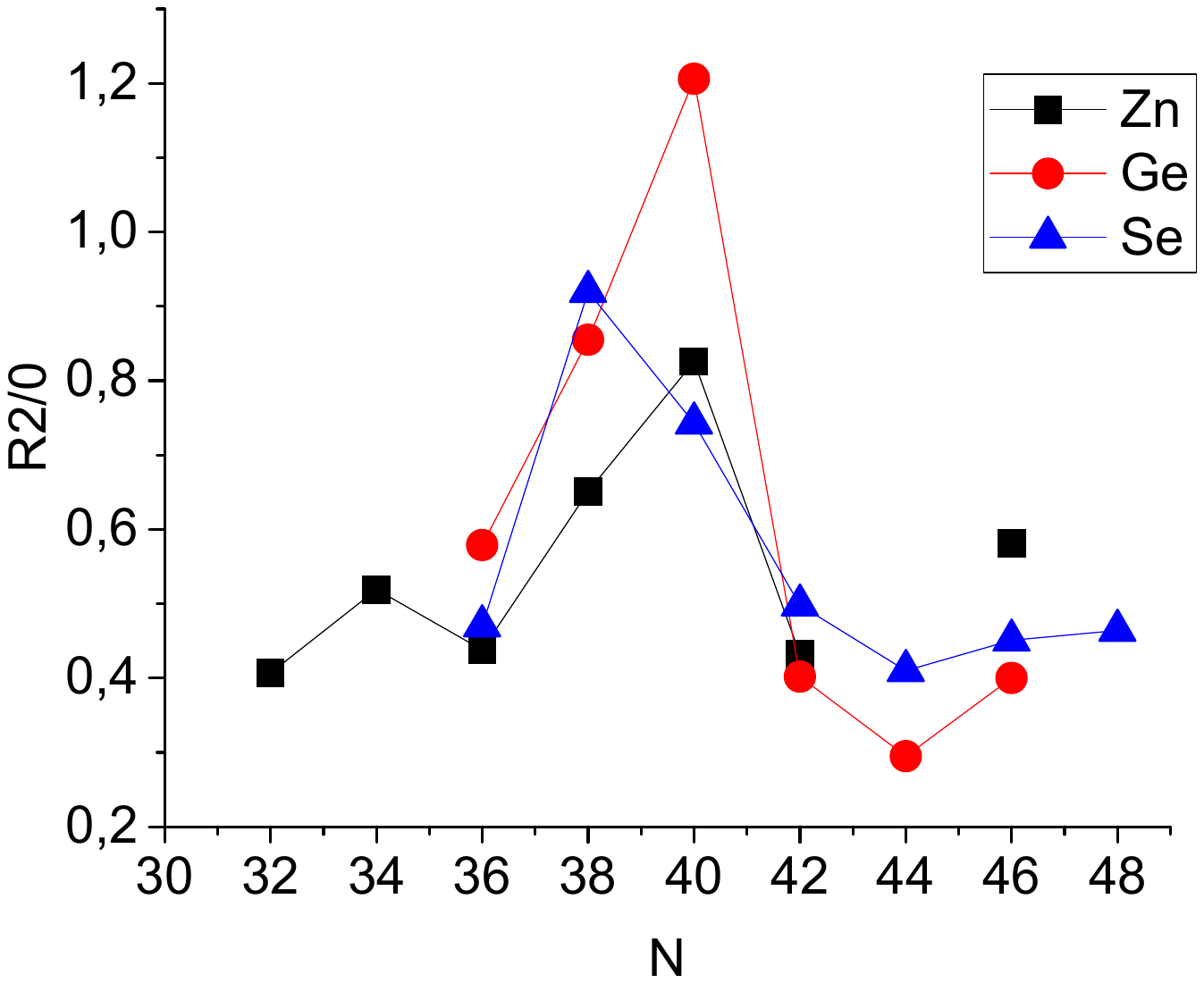}

    \caption{ Experimental \cite{ensdf} energy ratios $R_{2/0}$ in the $N=90$, 60, 40 regions. See Section \ref{emp} for further discussion. }
    \label{FR20}
\end{figure}


\begin{figure*} [htb]

    {\includegraphics[width=75mm]{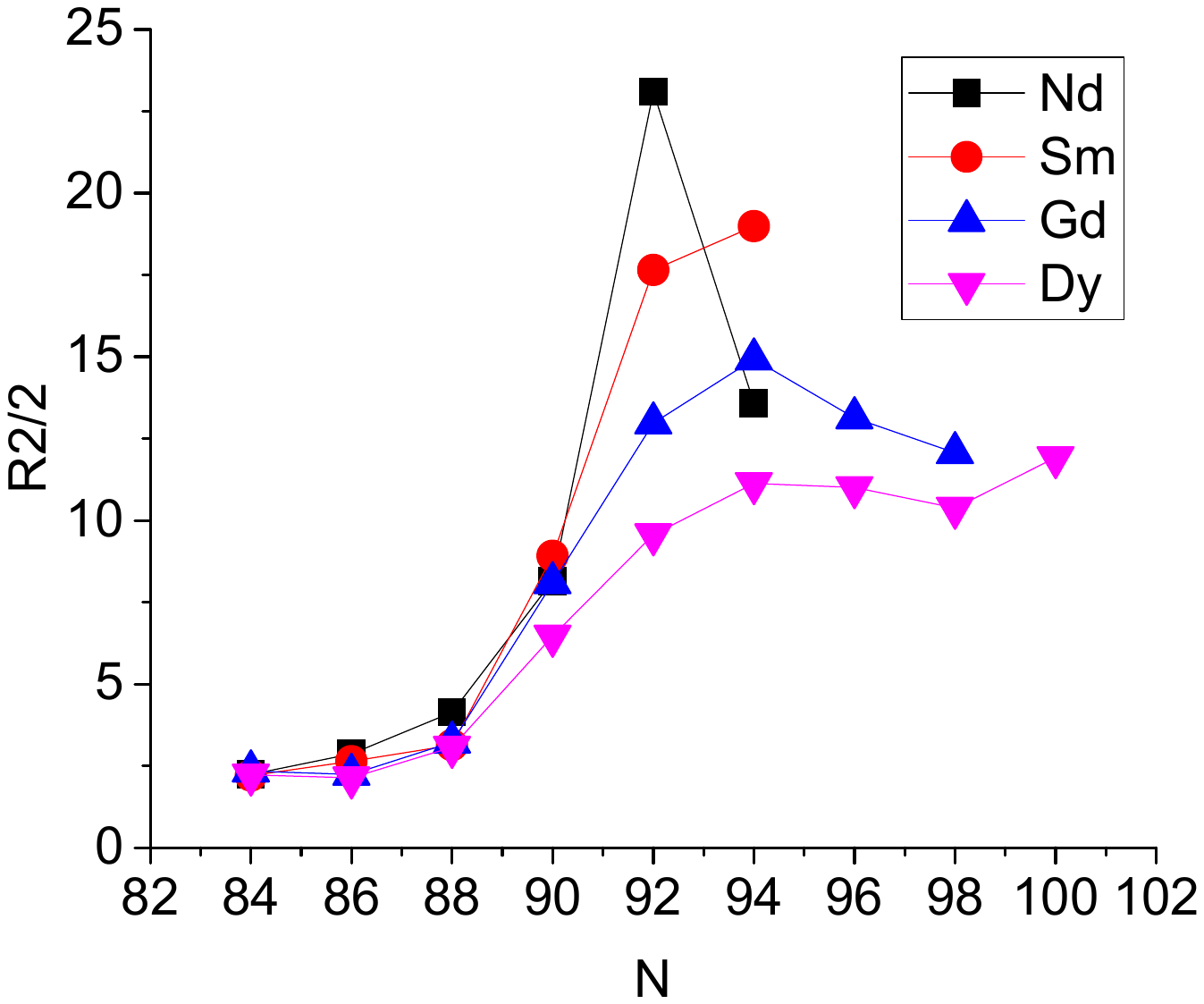}\hspace{5mm}
    \includegraphics[width=75mm]{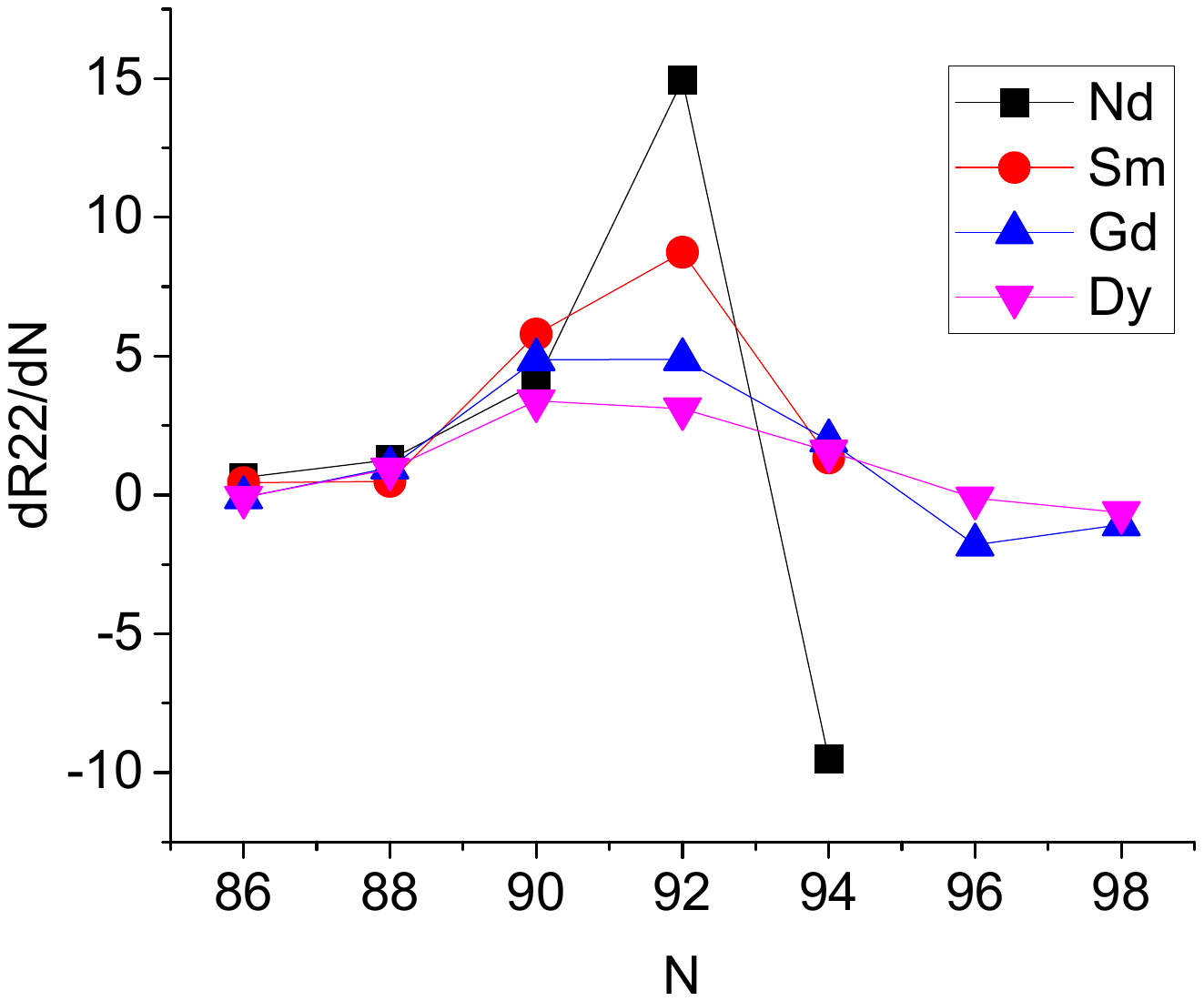}}
      {\includegraphics[width=75mm]{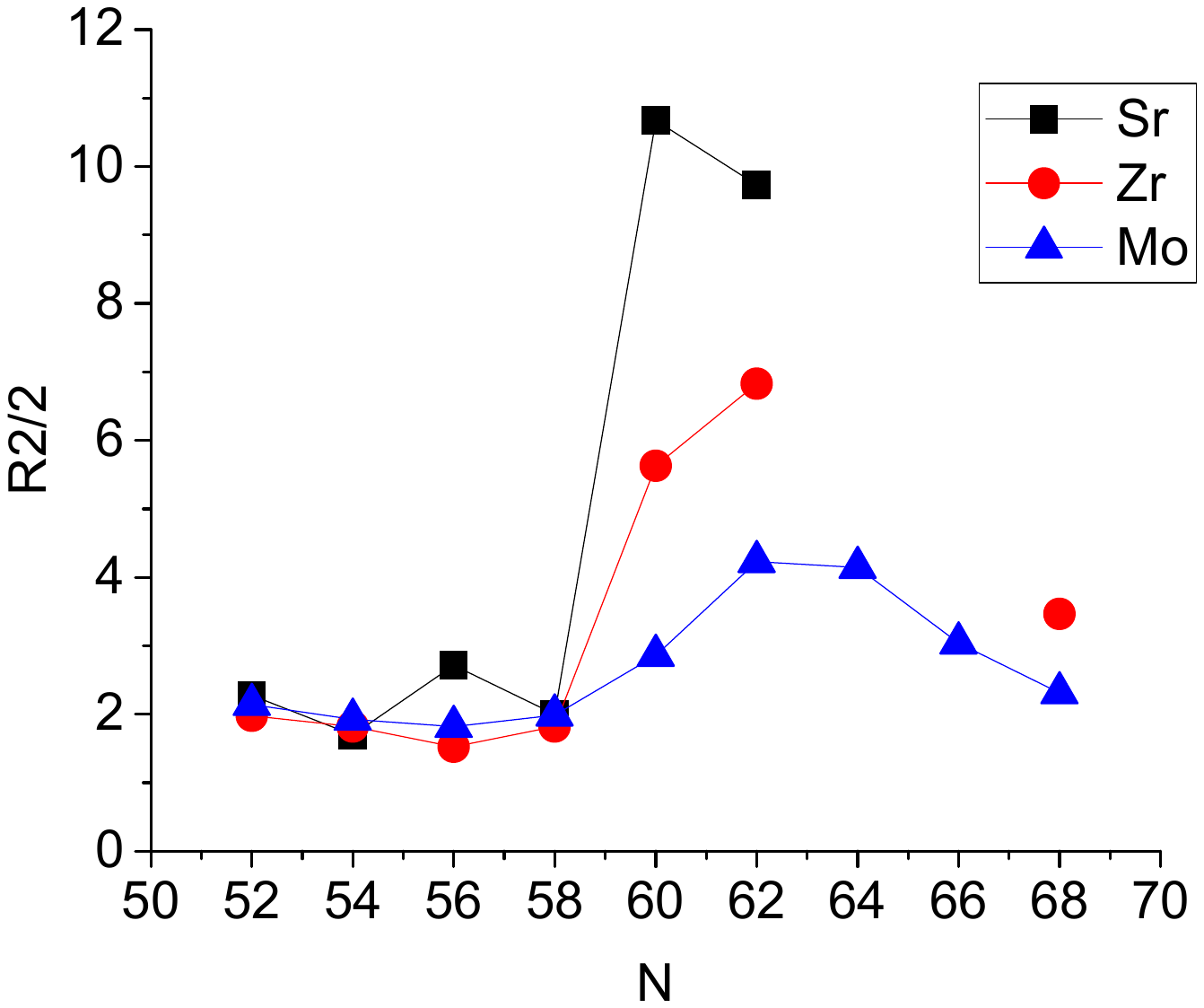}\hspace{5mm}
    \includegraphics[width=75mm]{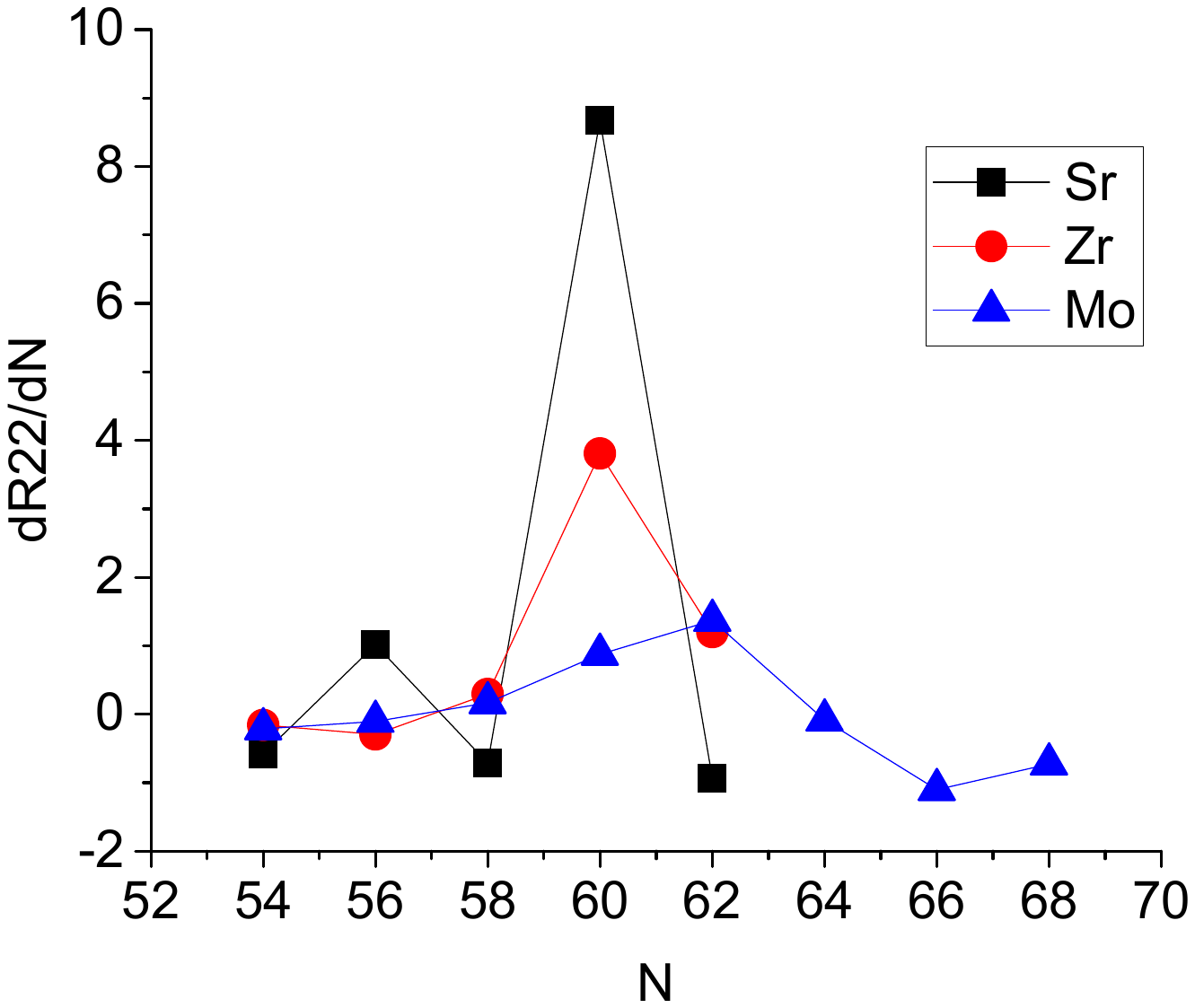}}
      {\includegraphics[width=75mm]{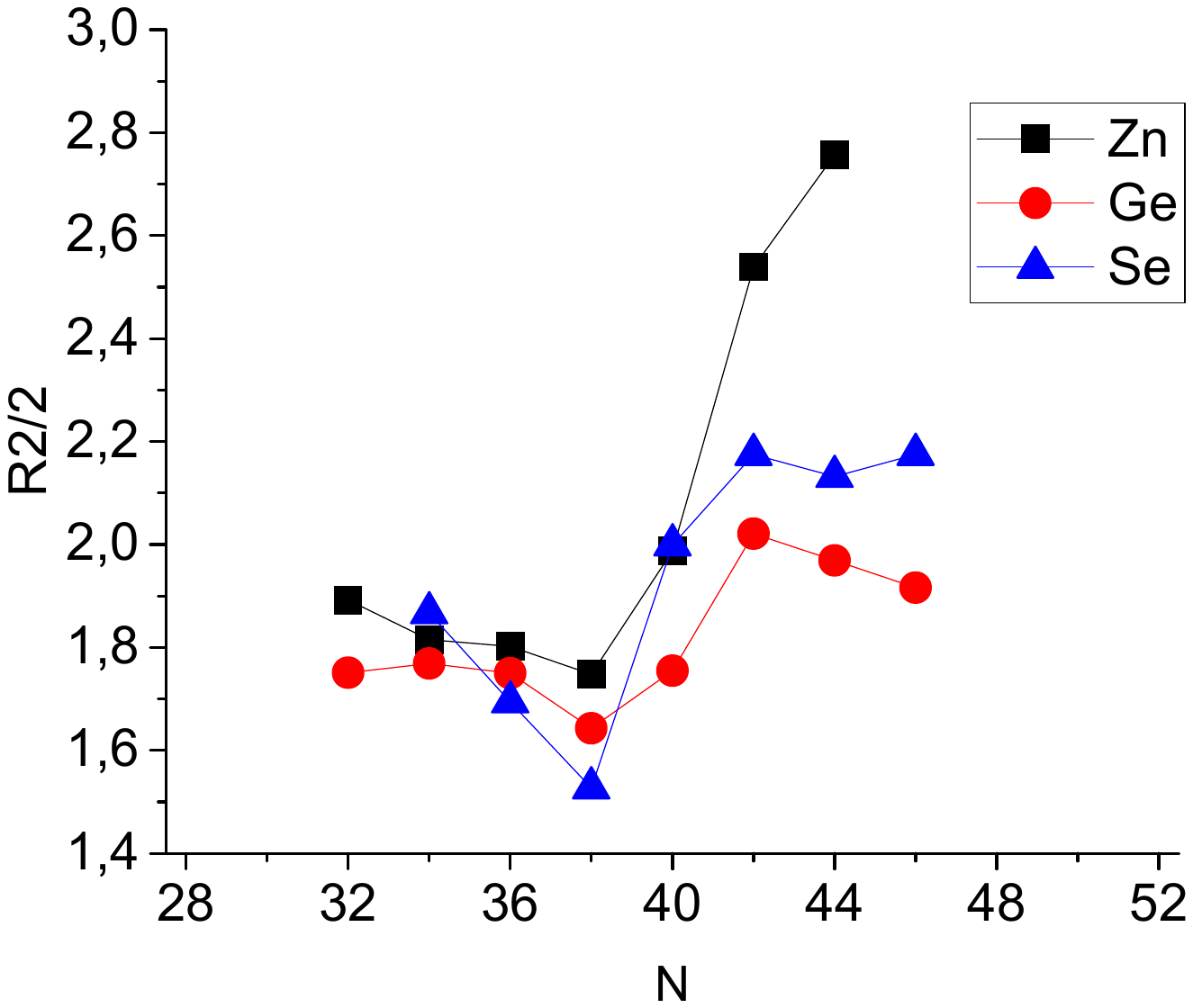}\hspace{5mm}
    \includegraphics[width=75mm]{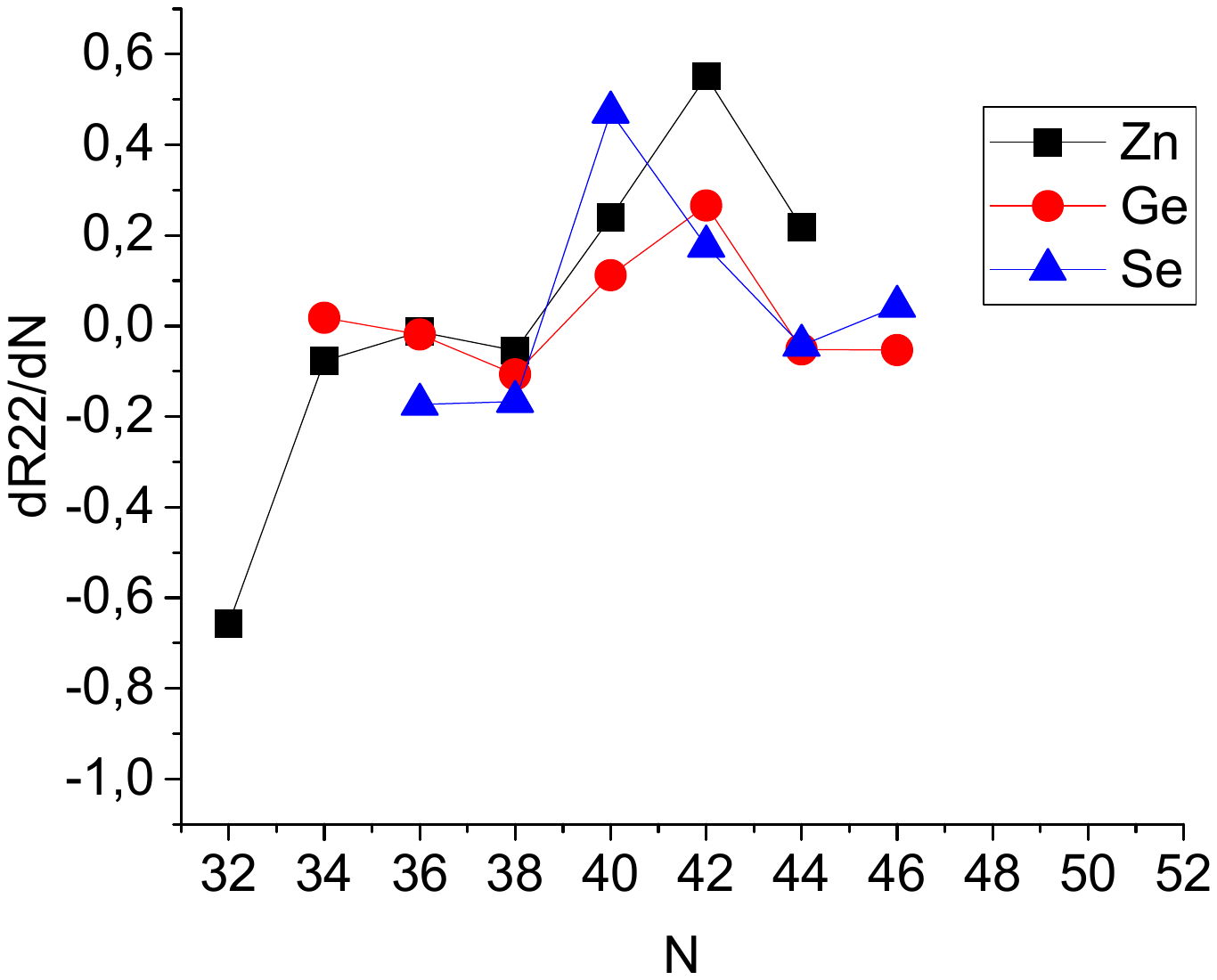}}
   
    \caption{Experimental \cite{ensdf} energy ratios $R_{2/2}$ and their rate of change $\mathrm{d}R_{2/2}/\mathrm{d}N$ with respect to the neutron number $N$ in the $N=90$, 60, 40 regions. See Section \ref{emp} for further discussion.} 
    \label{FR22}
\end{figure*}

In Fig. \ref{FR42} an abrupt increase of $R_{4/2}$ is seen at $N=90$ in the Nd-Sm-Gd-Dy region, at $N=60$ in the Sr-Zr-Mo region, and at $N=42$ in the Zn-Ge-Se region. This is corroborated by the rate of change of $R_{4/2}$ vs. $N$ \cite{Werner2002} within these series of isotopes. 

In Fig. \ref{FB} an abrupt increase of $B(E2; 2_1^+ \to 0_1^+)$ is seen around $N=90$ in the Nd-Sm-Gd-Dy region, around $N=60$ in the Sr-Zr-Mo region, and around $N=42$ in the Zn-Ge-Se region. This is corroborated by the rate of change of $B(E2; 2_1^+ \to 0_1^+)$ vs. $N$ \cite{Werner2002} within these series of isotopes. 

In Fig. \ref{FR20} maxima of the energy ratio $R_{2/0}$ are seen around $N=88$ in the Nd-Sm-Gd-Dy region, around $N=58$ in the Sr-Zr-Mo region, and around $N=40$ in the Zn-Ge-Se region. In other words, the maxima appear two neutrons earlier in relation to the maximal changes of $R_{4/2}$ and $B(E2; 2_1^+ \to 0_1^+)$ seen in Figs. \ref{FR42} and \ref{FB}. 

In Fig. \ref{FR22} an abrupt increase of $R_{2/2}$ is seen around $N=90$ in the Nd-Sm-Gd-Dy region, around $N=60$ in the Sr-Zr-Mo region, and around $N=42$ in the Zn-Ge-Se region. This is corroborated by the rate of change of $R_{2/2}$ vs. $N$ \cite{Werner2002} within these series of isotopes. 

In  Table \ref{Tmax} the values of these quantities for the nuclei close to the maxima are reported. We see that while the three regions exhibit in Figs. 
\ref{FR42}-\ref{FR22} very similar behavior, the numerical values in each region differ. In particular

a) The $R_{4/2}$ ratio exhibits near-rotational values in the $N=90$ region, intermediate values in the $N=60$ region, and much lower, near-vibrational values in the $N=40$ region. 

b) In qualitative agreement with a), the $B(E2; 2_1^+ \to 0_1^+)$ transition rate exhibits higher values in the $N=90$ region, intermediate values in the $N=60$ region, and lower values in the $N=40$ region. This is expected, since the $B(E2; 2_1^+ \to 0_1^+)$ transition rate is proportional to the square of the quadrupole deformation $\beta$ \cite{Raman2001}, despite the fact that part of the difference is due to the factor $A^{4/3}$ accompanying $\beta^2$ in the relevant equation \cite{Raman2001}.

c) The $R_{2/0}$ values exhibit the opposite trend, namely lower values in the $N=90$ region, evolving to higher values in the $N=40$ region, in agreement to the well known fact \cite{ensdf} that the $0_2^+$ state raises to very high values as a nucleus approaches the rotational limit of $R_{4/2}=10/3$.   

d) The $R_{2/2}$ ratio exhibits higher values in the $N=90$ region, intermediate values in the $N=60$ region, and lower values in the $N=40$ region, corroborating the structural similarity between the ground state band and the $\gamma$-band \cite{Bonatsos2021}.  

These observations indicate that the parameter-independent values of $R_{4/2}=2.904$ and $R_{2/0}= 0.177$ \cite{Iachello2001,Bonatsos2004} characterizing the X(5) critical point symmetry \cite{Iachello2001} apply very well to the $N=90$ isotones, but they do not apply in the $N=60$ and $N=40$ regions, for which a more flexible model appears to be required. The values seen in the $N=40$ region, in particular, are quite similar to the parameter-independent values of  $R_{4/2}=2.44$ and $R_{2/0}= 0.35$ seen in X(3) \cite{Bonatsos2006b}, the $\gamma$-rigid analog of X(5).  

The $N=90$ isotones $^{150}$Nd \cite{Krucken2002}, $^{152}$Sm \cite{Casten2001}, $^{154}$Gd \cite{Tonev2004}, $^{156}$Dy \cite{Caprio2002} are well established examples of the X(5) critical point symmetry. 

In the $N=60$ region, $^{104}$Mo$_{62}$ has initially been suggested \cite{Bizzeti2002} as an X(5) candidate, based on its ratio $R_{4/2}=2.917$, which is very close to the X(5) value of 2.904. In addition, its ratio $R_{2/0}=0.217$, is also very close to the X(5) value of 0.177. However, it was later disregarded, since its $B(E2)$s within the ground state band have been found \cite{Hutter2003} to exhibit a deformed behavior. Figs. \ref{FR42}-\ref{FR22} suggest that the $N=60$ isotones $^{98}$Sr, $^{100}$Zr, $^{102}$Mo are better candidates for the critical point of the spherical to deformed QPT in the $Z=40$ region.  

It should be noticed that Figs. \ref{FR42}-\ref{FR22} shed light on the nature of ground state QPTs \cite{Iachello2004,Iachello2006,Cejnar2010}. According to the Ehrenfest classification, a first order phase transition occurs when the first derivative of a physical quantity (serving as the order parameter) exhibits a discontinuity with respect to the control parameter \cite{Iachello2004,Iachello2006}. 
The ground state spectra and $B(E2)$ transition rates shown in Figs. \ref{FR42} and \ref{FB} suggest the occurrence of a first order ground state QPT within the ground state band for nuclei around $N=90$, 60, 42, since the rate of change of these quantities vs. the neutron number $N$ exhibits a discontinuity, while the quantities themselves show an increase with $N$, which becomes abrupt at the critical point of the QPT. The same behavior is seen in Fig. \ref{FR22}, suggesting a similar behavior for the $\gamma$-band. However, this is not the case for the ratio  $R_{2/0}$ (Fig. \ref{FR20}), which presents itself a maximum around the critical point.    
This difference is rooted to the fact that the ground state band and the $\gamma$-band tend to have similar structures \cite{Bonatsos2021}, while the nature of the 
$\beta$-band has been a point of controversy \cite{SharpeySchafer2008,SharpeySchafer2010,SharpeySchafer2011,SharpeySchafer2019}, with the first excited $0^+$ state being  able to correspond to several different physical situations and not necessarily to the band-head of the $\beta$-band \cite{Garrett2001}.

\section{Microscopic mean field calculations}
\label{MMFC}

The regions of interest have been investigated within several different mean-field frameworks. Non-relativistic calculations using the Skyrme interaction have been performed for the Nd-Sm-Gd-Dy \cite{RGuzman2007} and Zr \cite{Skalski1993,Reinhard1999} isotopes, while the Gogny interaction has been used in the Nd-Sm-Gd-Dy \cite{Robledo2008,Rodriguez2008,Rodriguez2009}, Zr \cite{Rodriguez2011} and Se \cite{Peru2014} regions. Relativistic mean field calculations have been performed for the Nd-Sm-Gd-Dy \cite{Meng2005,Sheng2005,Naz2018} and  Sr-Zr-Mo \cite{Abusara2017a,Abusara2017b,Kumar2021,Thakur2021} isotopes. A 5-dimensional quadrupole collective Hamiltonian with parameters determined from relativistic mean field calculations has been used in the Nd-Sm-Gd-Dy \cite{Li2009,Li2016,Majola2019} and Sr-Zr-Mo \cite{Xiang2012} isotopes. In addition, an IBM Hamiltonian with parameters determined by covariant density functional theory calculations has been applied to the Nd-Sm-Gd-Dy \cite{Fossion2006,Niksic2007,Nomura2010,Nomura2019}, Sr-Zr-Mo \cite{Nomura2016}, and Se \cite{Nomura2017} regions. 

In the present section we are trying to accommodate all three regions of $N=90$, $N=60$ and $N=40$ within the same theoretical framework. 

Self-consistent mean-field methods based e.g. on the most widely used Skyrme-Hartree-Fock + Bardeen-Cooper-Schrieffer (SHF + BCS) calculations
 \cite{Bender2003} represent a microscopic possibility to investigate phase transitions in finite nuclei. One would expect to observe a typical bump
 structure in the potential energy curves (PEC) calculations as a function of the quadrupole deformation parameter $\beta_2$ \cite{Fossion2006,Niksic2007} that
 manifests itself in a wider flat region around the minimum of the PEC compared to neigboring nuclei. In addition, the presence of opposite-parity
 intruder states in both proton and neutron single-quasiparticle spectra close to the Fermi level represents another microscopic signature of the phase
 transition \cite{Federman1977,Federman1978,Federman1979a,Federman1979b}. 
 
 Because there exist plenty of SHF functional parametrizations we investigate here the whole family of parametrizations based on the SV-bas one 
 (15 parametrizations) \cite{Klupfel2009}. In calculations using the axial SHF code SKYAX \cite{Reinhard2021} with
 a density dependent $\delta$-force interaction in the pairing channel
\cite{Terasaki1996} single-particle levels up to 75 MeV were taken into account (8
oscillator shells). We investigated PECs as a function of quadrupole $\beta_2$, octupole $\beta_3$ and hexadecapole $\beta_4$ deformations for four isotopic
chains around $N=40$, 60 and 90 ($^{68}$Se -- $^{80}$Se, $^{92}$Zr -- $^{104}$Zr, $^{94}$Mo -- $^{106}$Mo, $^{144}$Nd -- $^{156}$Nd).

\begin{figure}[!htb]
\begin{minipage}[t]{.45\textwidth}
\includegraphics[width=0.95\textwidth]{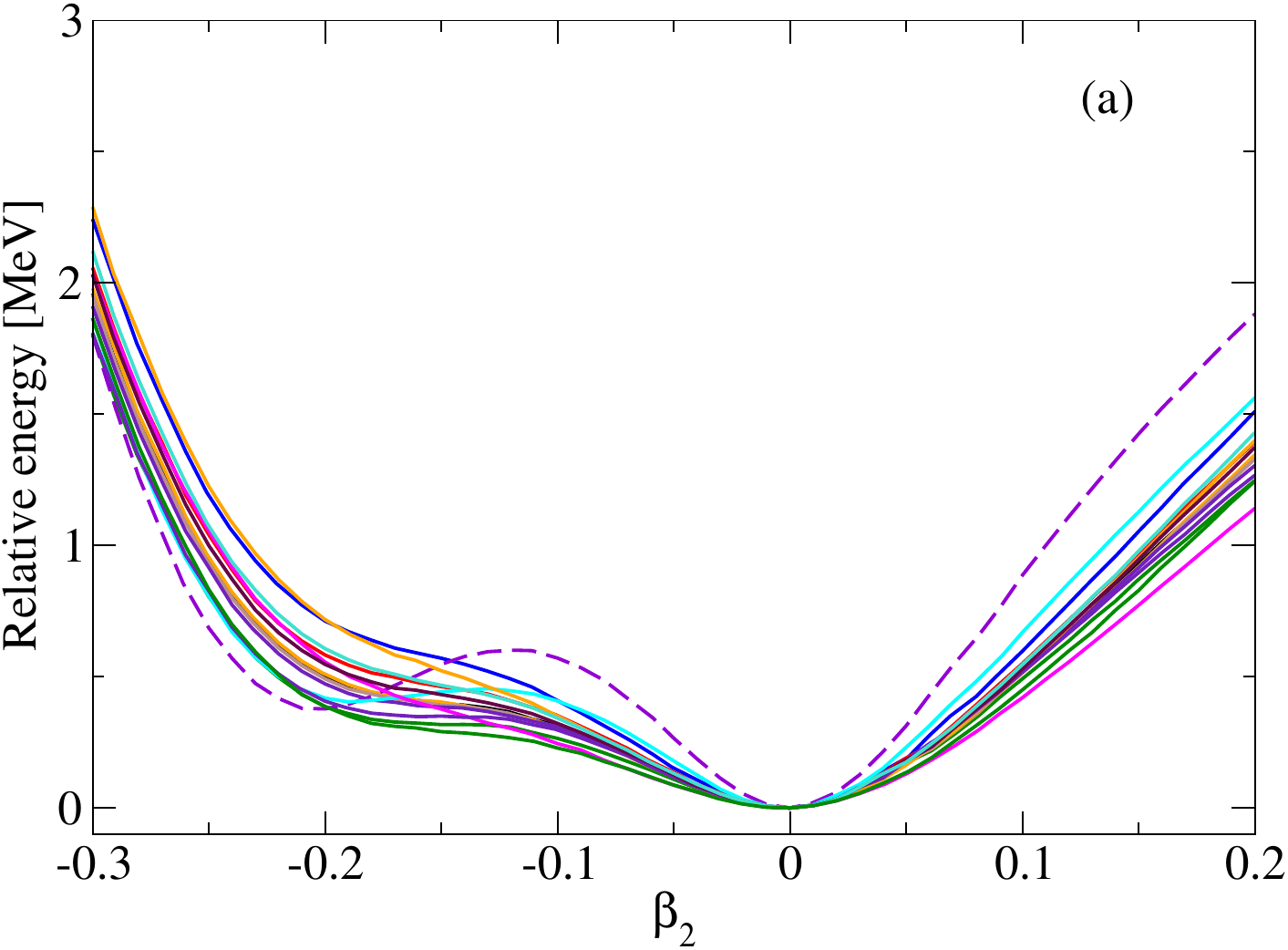}
\end{minipage}
\begin{minipage}[t]{.45\textwidth}
\includegraphics[width=0.95\textwidth]{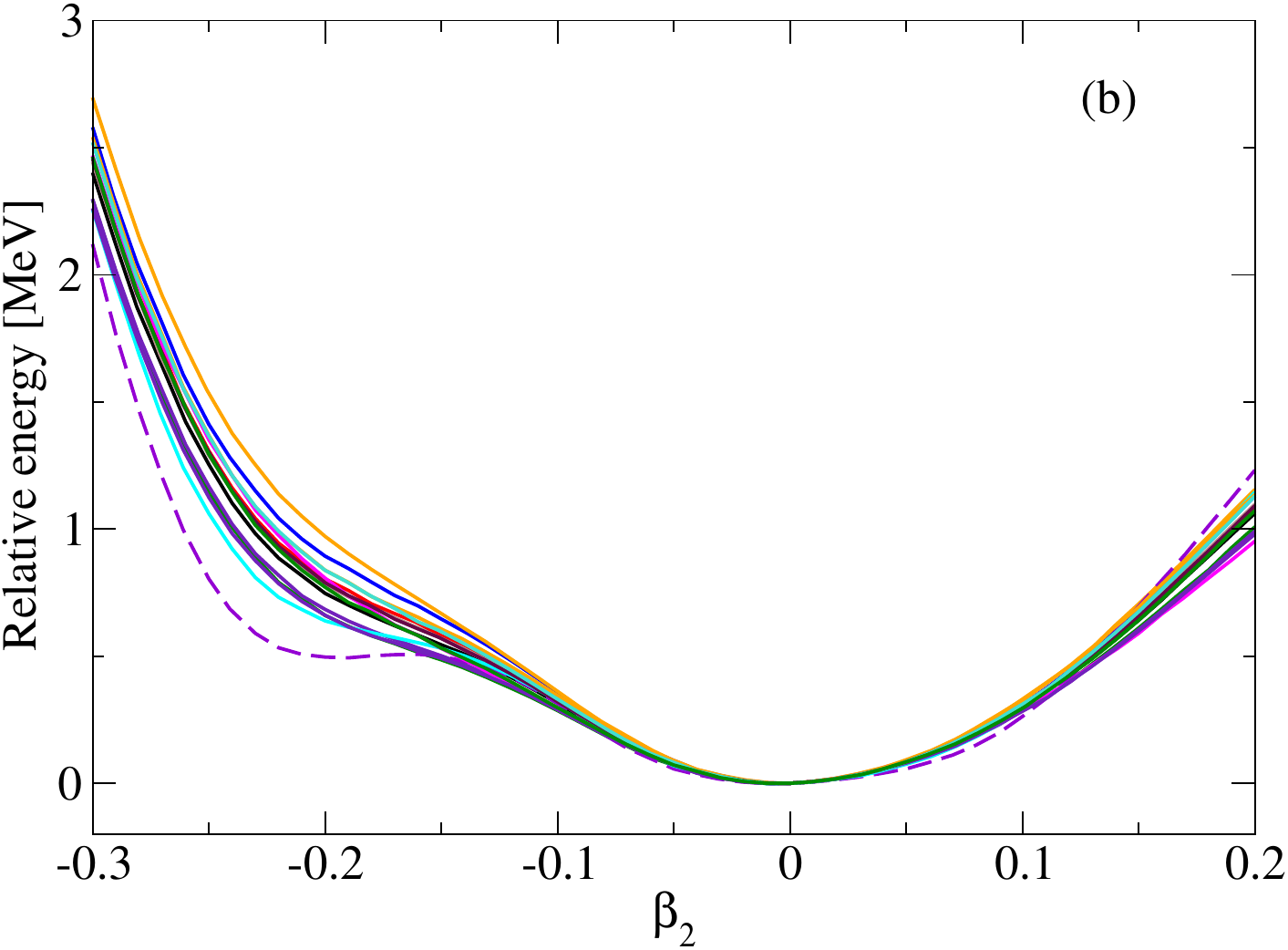}
\end{minipage}
\caption{PEC as a function of $\beta_2$ for $^{74}$Se (a) and $^{76}$Se (b) for all investigated parametrizations. The parametrization SV-mas07 (violet dashed) gives a slightly different PEC compared to the others.}
\label{Se-b2-all}
\end{figure}

\begin{figure}[!htb]
\begin{minipage}[t]{.45\textwidth}
\includegraphics[width=0.95\textwidth]{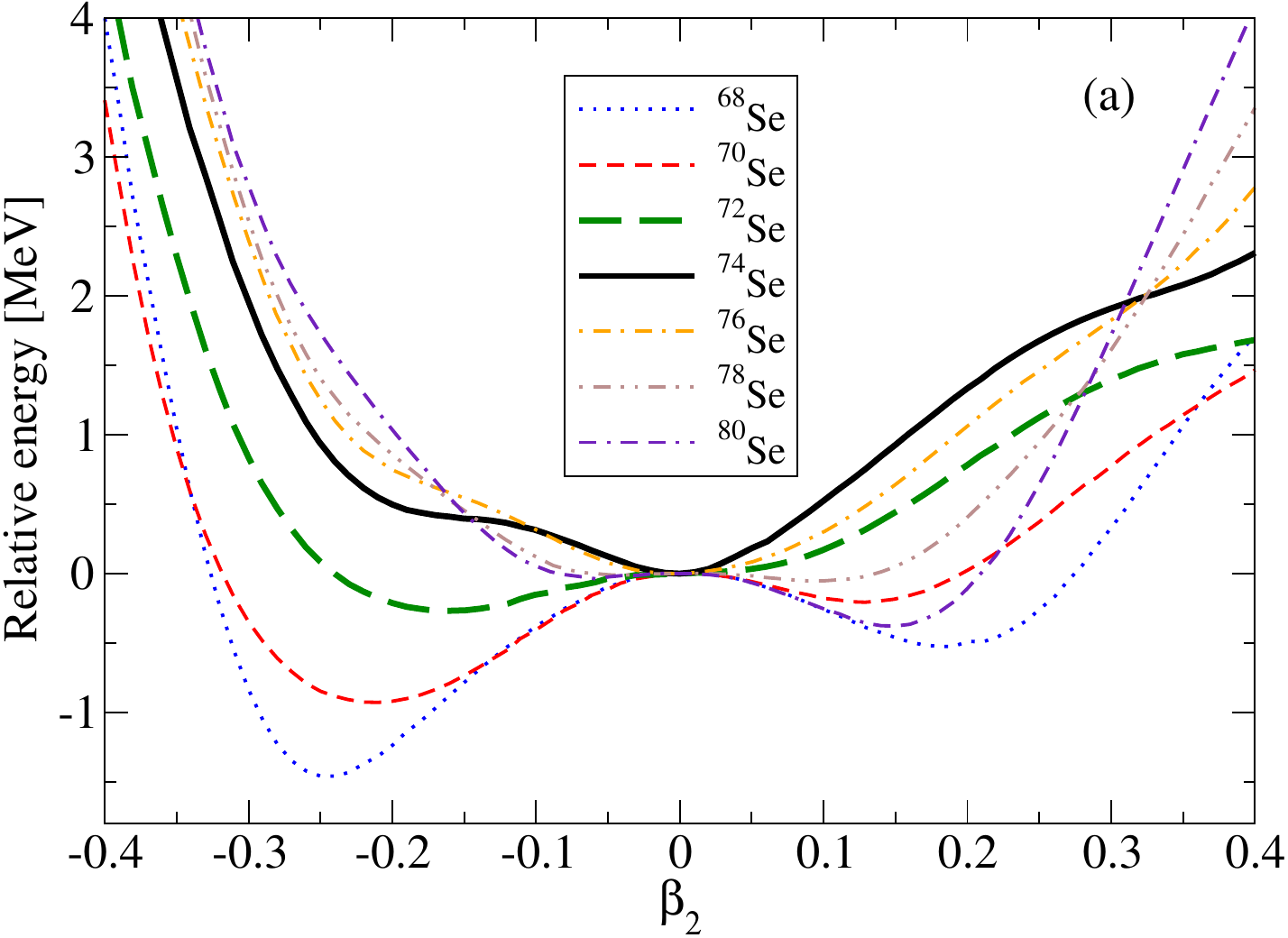}
\end{minipage}
\begin{minipage}[t]{.45\textwidth}
\includegraphics[width=0.95\textwidth]{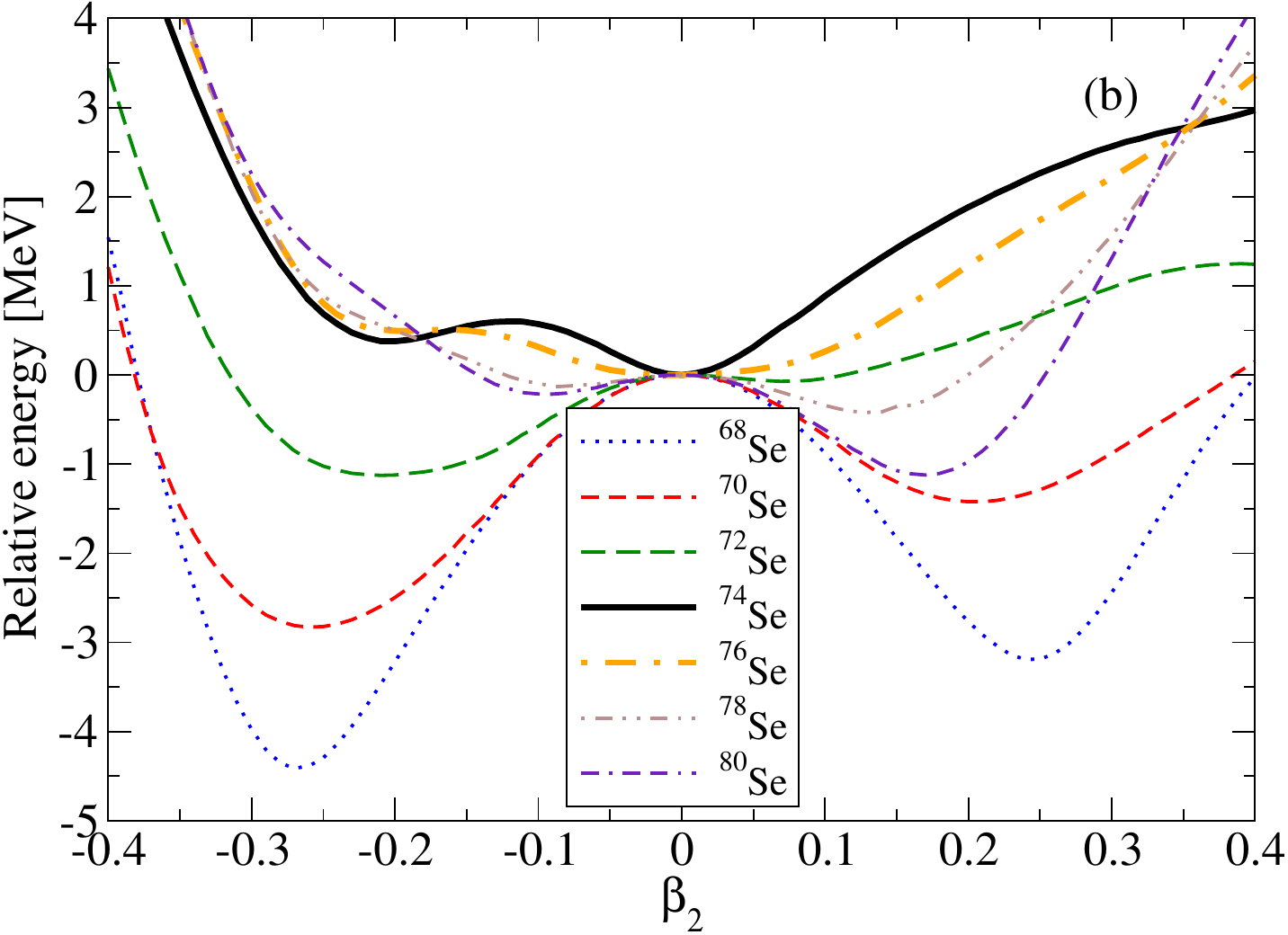}
\end{minipage}
\caption{PEC as a function of $\beta_2$ for the selenium isotopes for two SHF parametrizations, SV-bas (a) and SV-mas07 (b).
Flat-bottomed PECs are indicated by thicker curves.}
\label{Se-b2}
\end{figure}

In the selenium chain all parametrizations predict similar PECs, the parametrization SV-mas07 gives slightly different results (see Fig.~\ref{Se-b2-all}). 
In the quadrupole deformation PECs one observes phase transition from 
oblate to prolate shape at $N = 38$-44 (see Fig.~\ref{Se-b2}). For $N=40$ SC is predicted with one spherical and one oblate minimum close in energy. A wider flat region is observed for $N=38$ and 40 (SV-bas) and for $N=40$ and 42
(SV-mas07).
Octupole deformation does not play any significant role, as expected. Hexadecapole deformation remains positive and close to zero ($\beta_4 =0$ for $N=40$ and 42).

\begin{figure}[!htb]
\begin{minipage}[t]{.45\textwidth}
\includegraphics[width=0.95\textwidth]{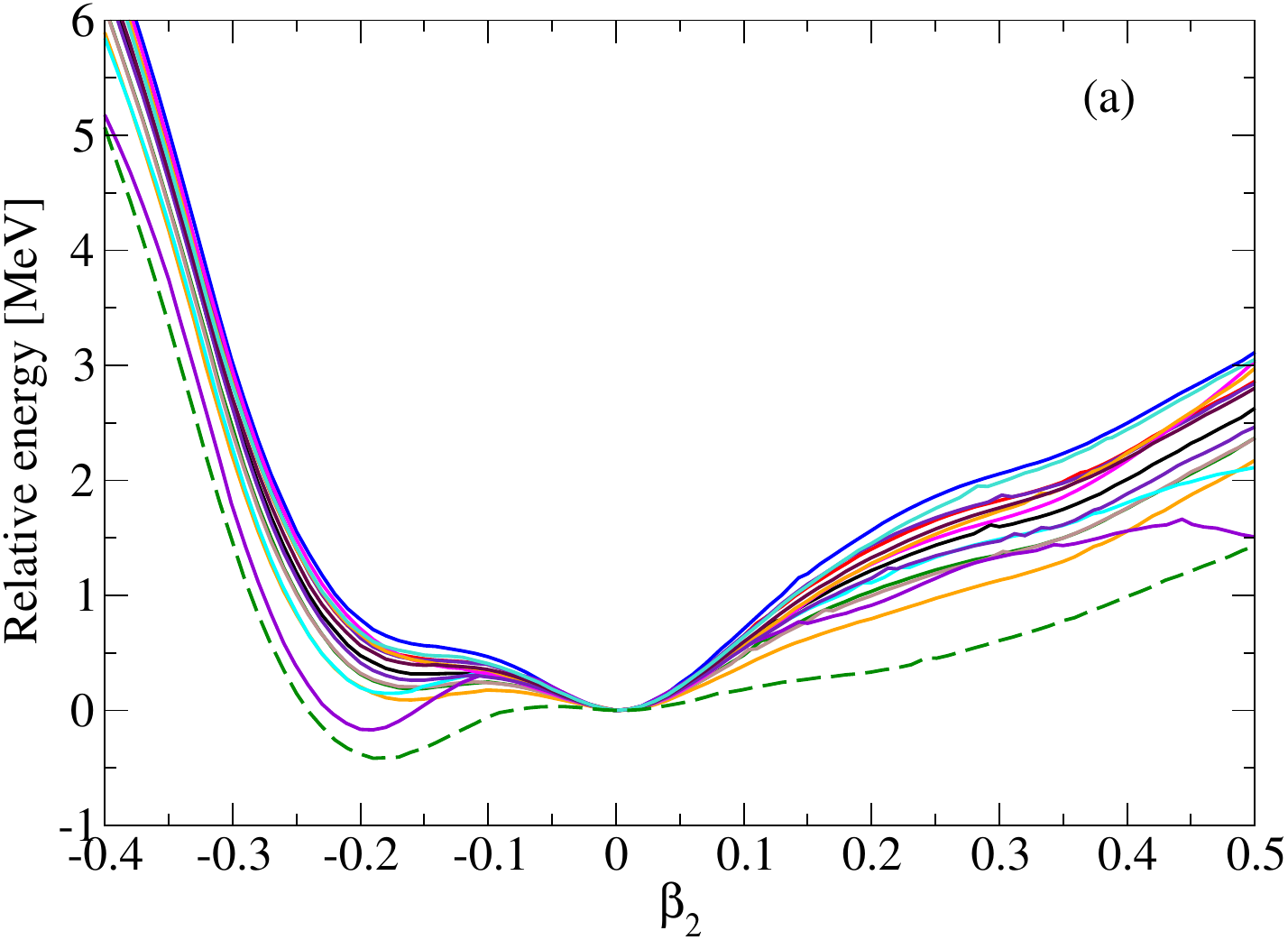}
\end{minipage}
\begin{minipage}[t]{.45\textwidth}
\includegraphics[width=0.95\textwidth]{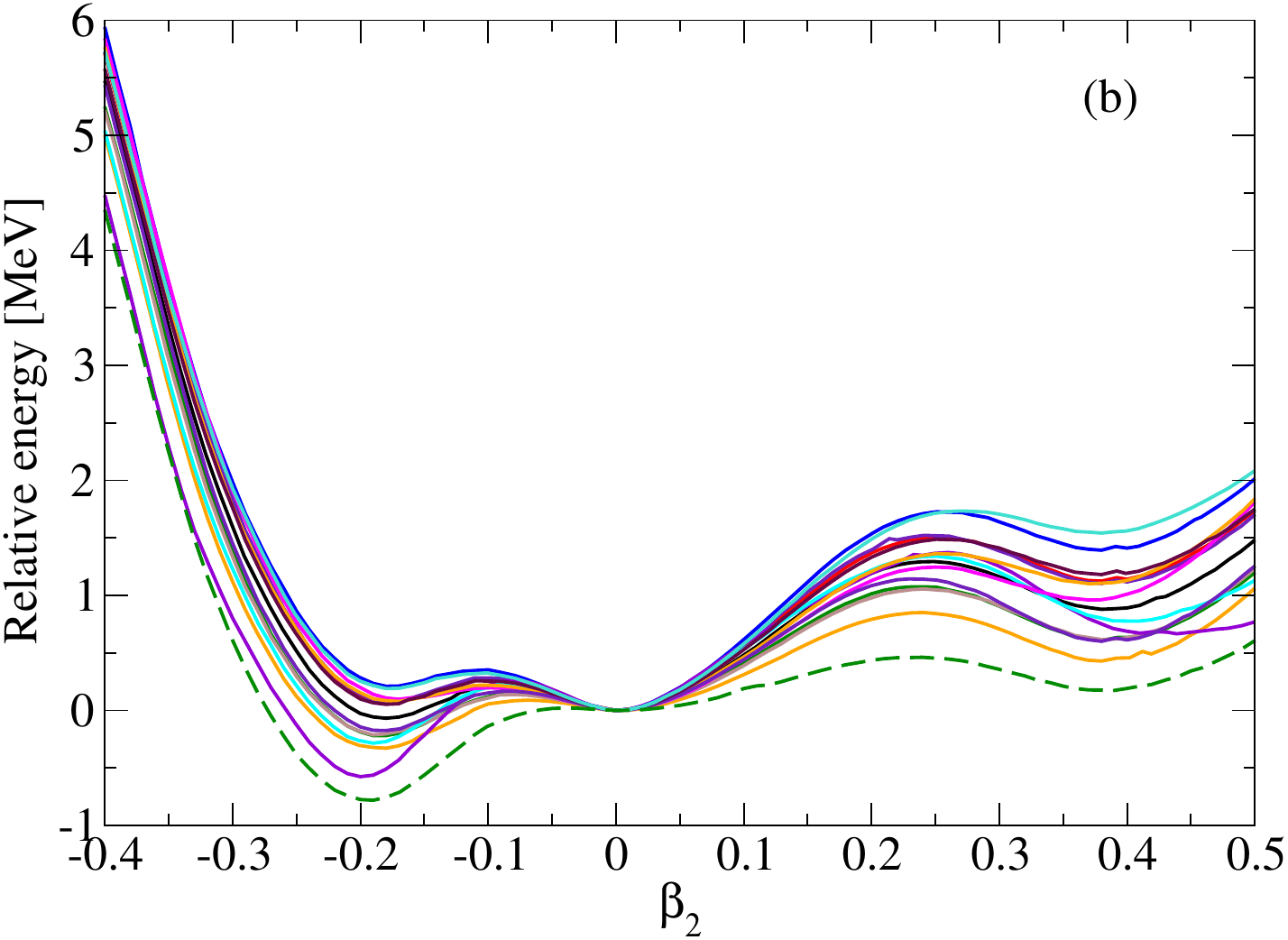}
\end{minipage}
\caption{PEC as a function of $\beta_2$ for $^{98}$Zr (a) and $^{100}$Zr (b) for all investigated parametrizations. The parametrization SV-tls (dark green dashed) gives a more pronounced oblate minimum and the lowest energy differences for different $\beta_2$ values.}
\label{Zr-b2-all}
\end{figure}

\begin{figure}[!htb]
\begin{minipage}[t]{.45\textwidth}
\includegraphics[width=0.95\textwidth]{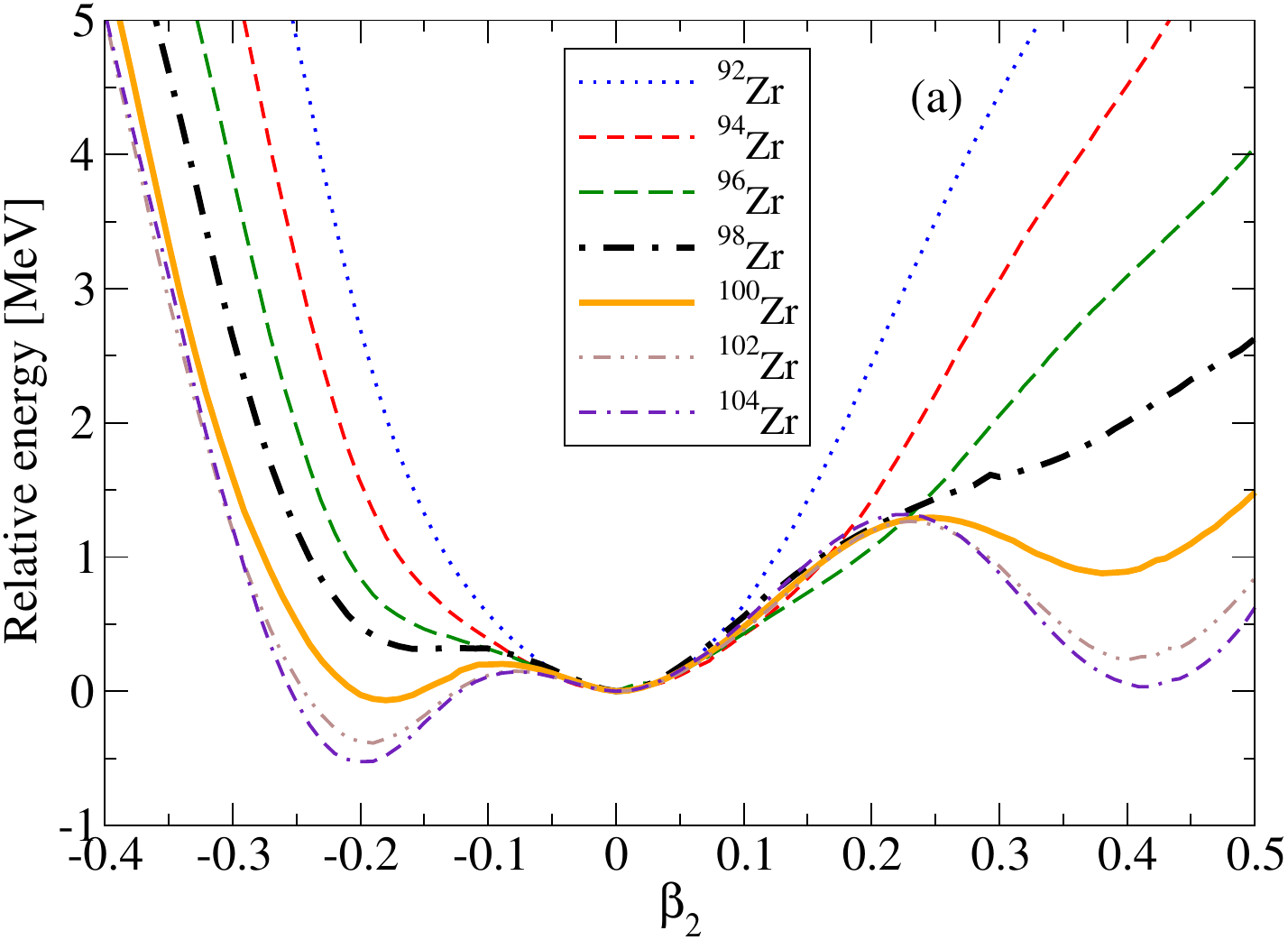}
\end{minipage}
\begin{minipage}[t]{.45\textwidth}
\includegraphics[width=0.95\textwidth]{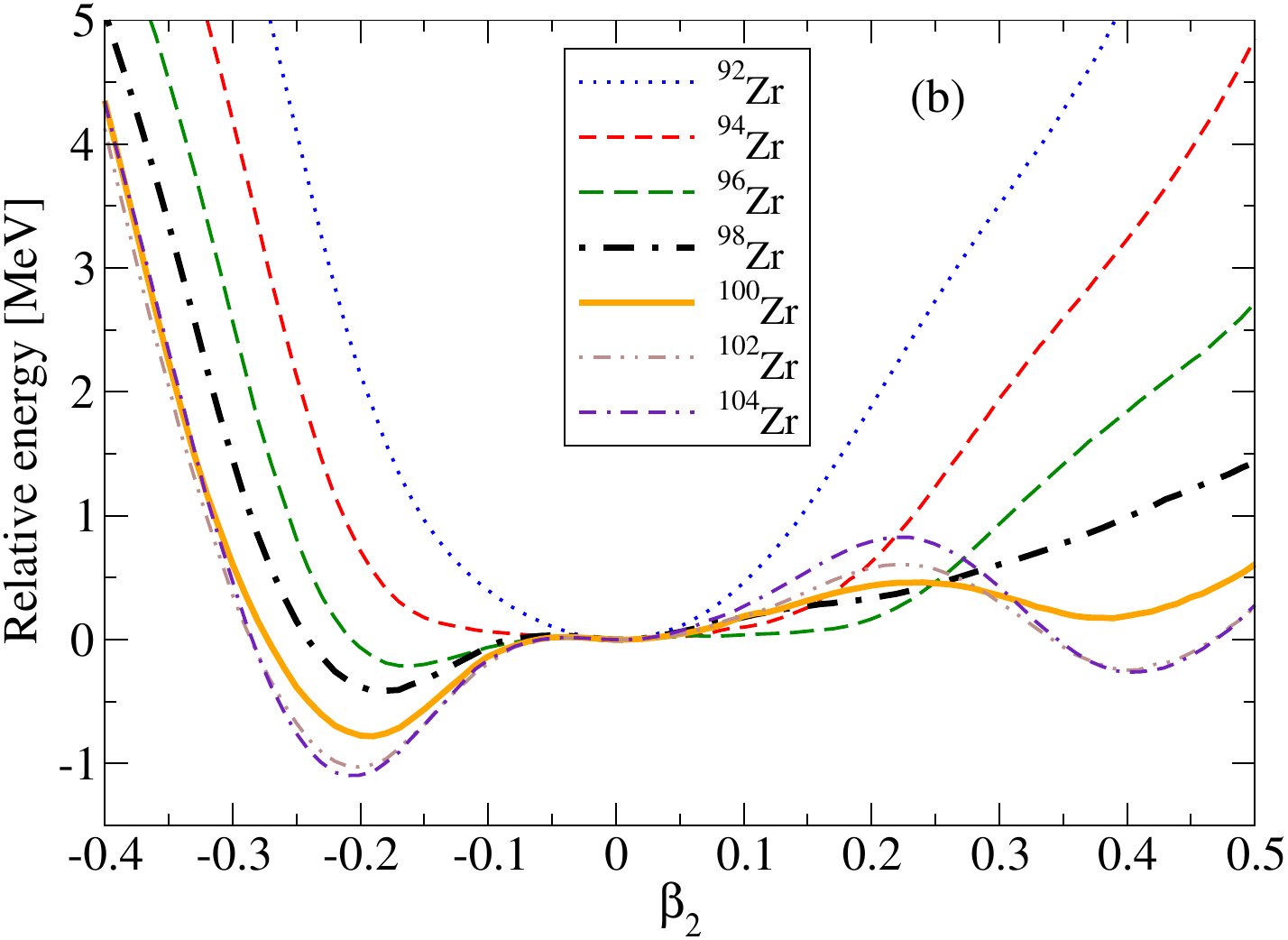}
\end{minipage}
\caption{PEC as a function of $\beta_2$ for the zirconium isotopes for two SHF parametrizations, SV-bas (a) and SV-tls (b).
Flat-bottomed PECs are indicated by thicker curves.}
\label{Zr-b2}
\end{figure}

In the zirconium chain all parametrizations again predict similar quadrupole-deformation PECs with a more pronounced minima for the SV-tls parametrization (see Figs.~\ref{Zr-b2-all} and \ref{Zr-b2}). Interestingly, for the SV-bas one observes three minima, oblate, spherical and prolate for $N\ge 58$, the oblate one and the spherical one being almost degenerate for $N=60$ (the flattest PEC). 
For SV-tls, the flat-bottomed PECs were found for $N = 58$-60.  Octupole deformation is again not important. Hexadecapole deformation is around $\beta_4 =0$.

\begin{figure}[!htb]
\begin{minipage}[t]{.45\textwidth}
\includegraphics[width=0.95\textwidth]{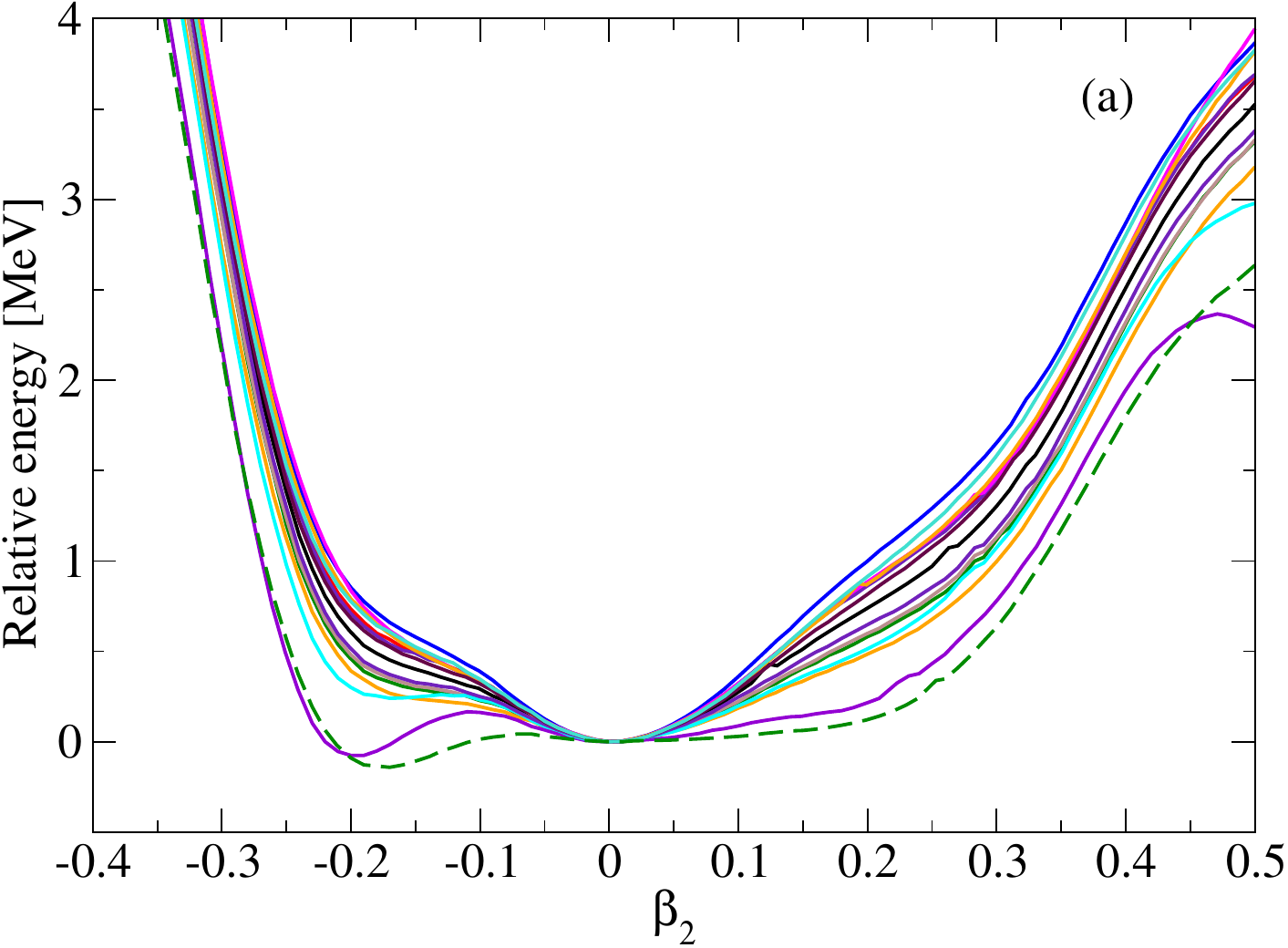}
\end{minipage}
\begin{minipage}[t]{.45\textwidth}
\includegraphics[width=0.95\textwidth]{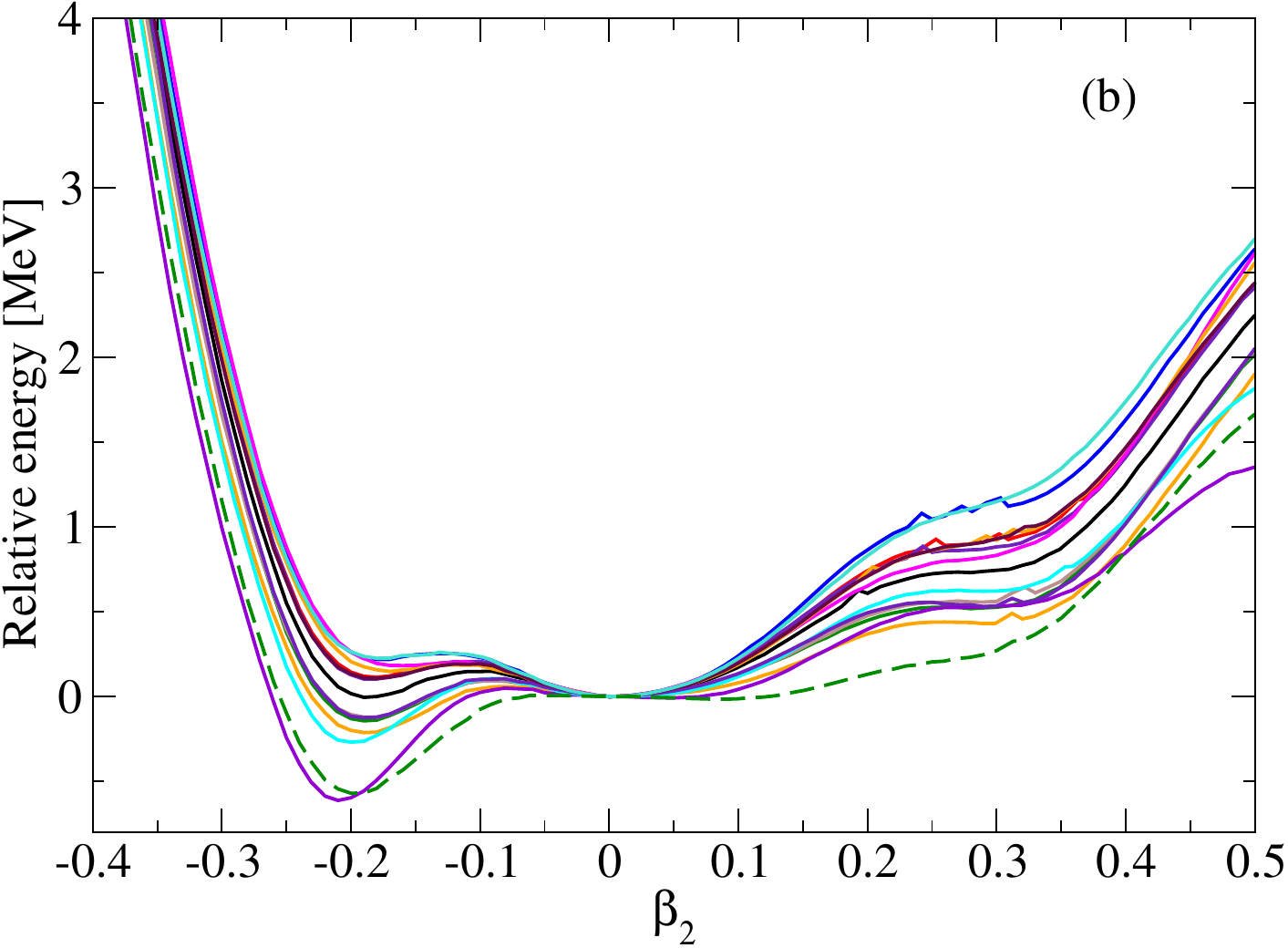}
\end{minipage}
\caption{PEC as a function of $\beta_2$ for $^{100}$Mo (a) and $^{102}$Mo (b) for all investigated parametrizations. The parametrization SV-tls (dark green dashed) gives a more pronounced oblate minimum and the lowest energy differences for different $\beta_2$ values.}
\label{Mo-b2-all}
\end{figure}

\begin{figure}[!htb]
\begin{minipage}[t]{.45\textwidth}
\includegraphics[width=0.95\textwidth]{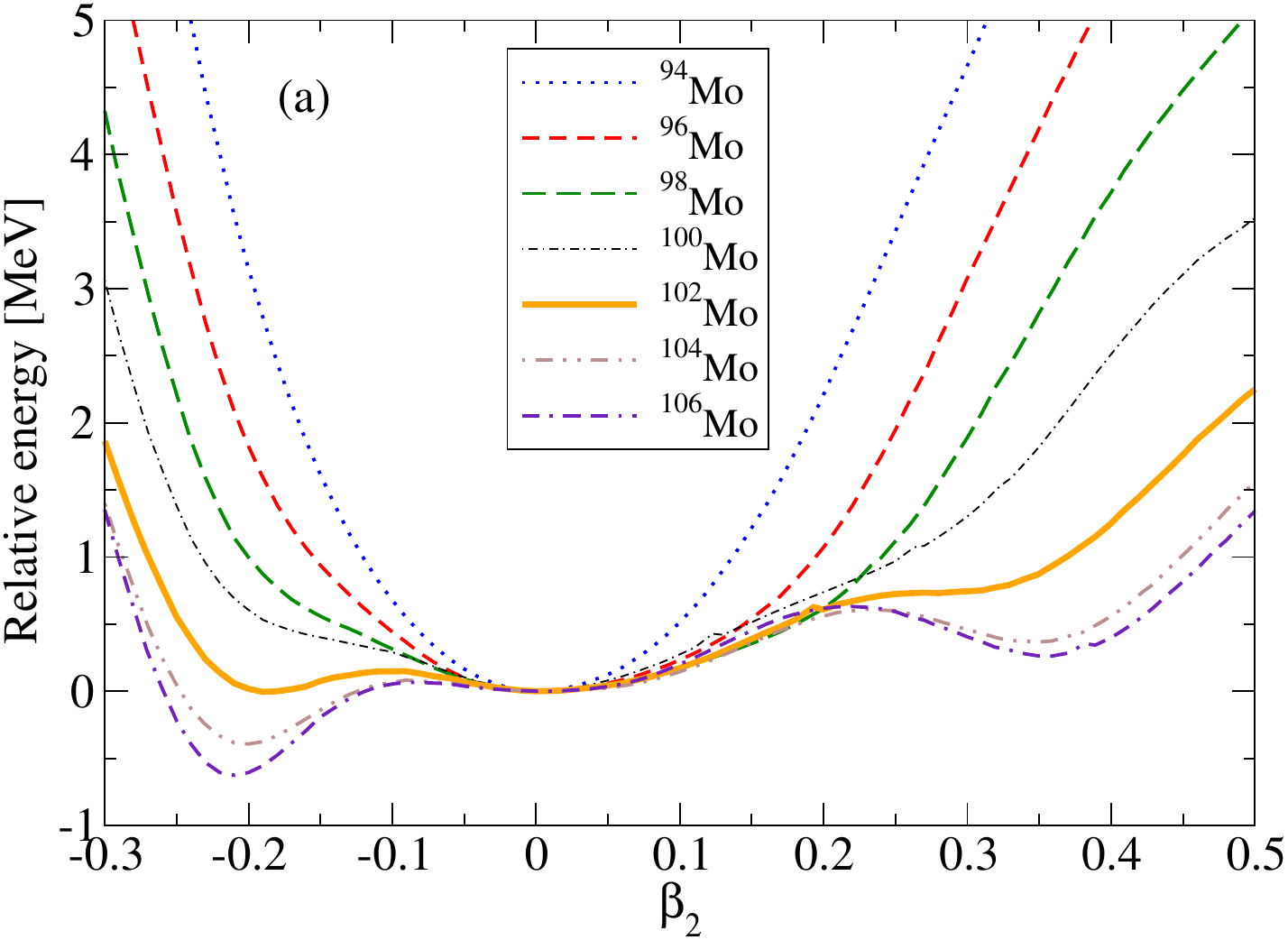}
\end{minipage}
\begin{minipage}[t]{.45\textwidth}
\includegraphics[width=0.95\textwidth]{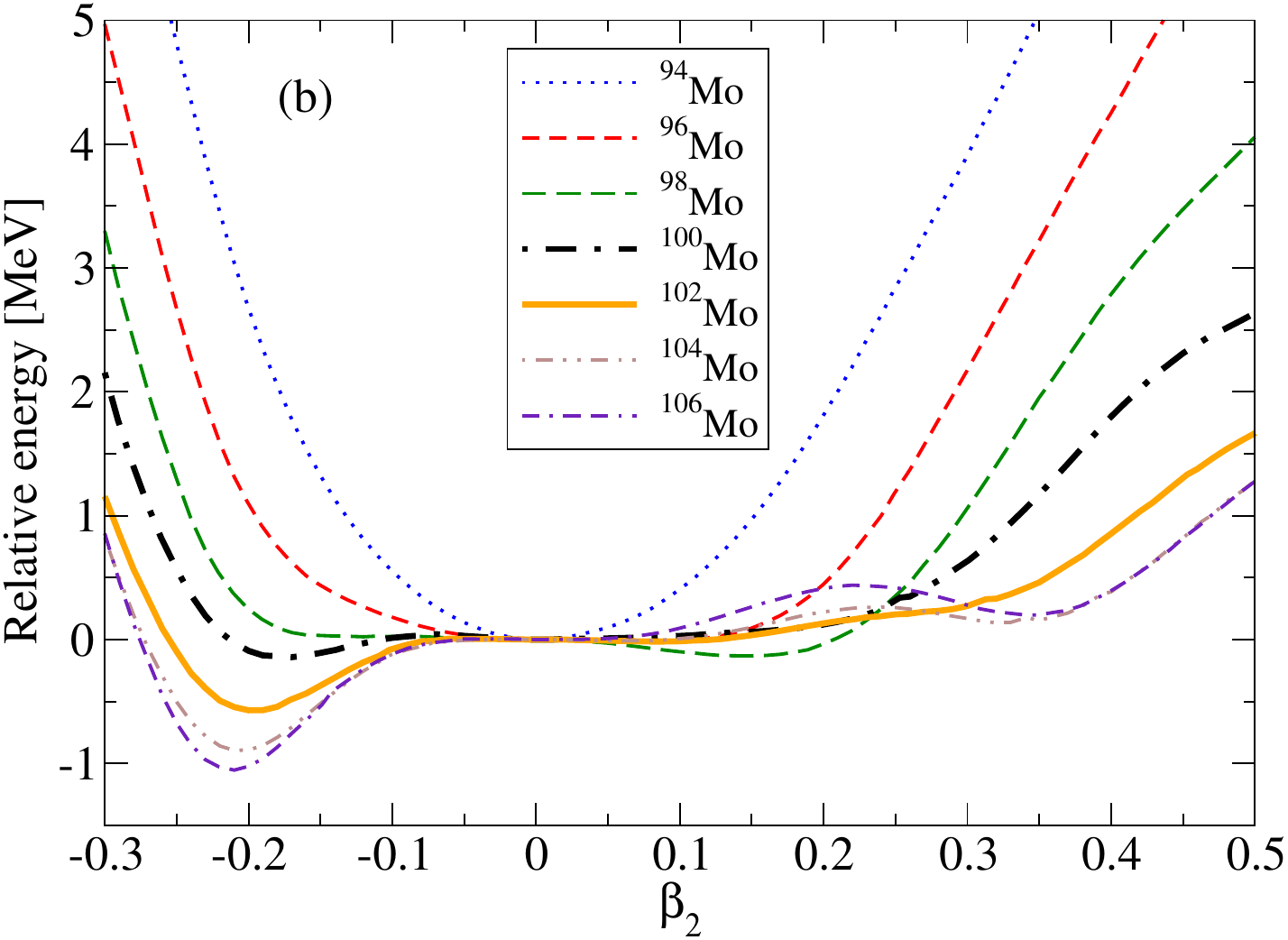}
\end{minipage}
\caption{PEC as a function of $\beta_2$ for the molybdenum isotopes for two SHF parametrizations, SV-bas (a) and SV-tls (b).
Flat-bottomed PECs are indicated by thicker curves.}
\label{Mo-b2}
\end{figure}

The molybdenum chain behaves similarly (see Figs.~\ref{Mo-b2-all} and \ref{Mo-b2}). The flat-bottomed PECs were found
for $N=60$ (SV-bas) and $N=58$ (SV-tls).

\begin{figure}[!htb]
\begin{minipage}[t]{.45\textwidth}
\includegraphics[width=0.95\textwidth]{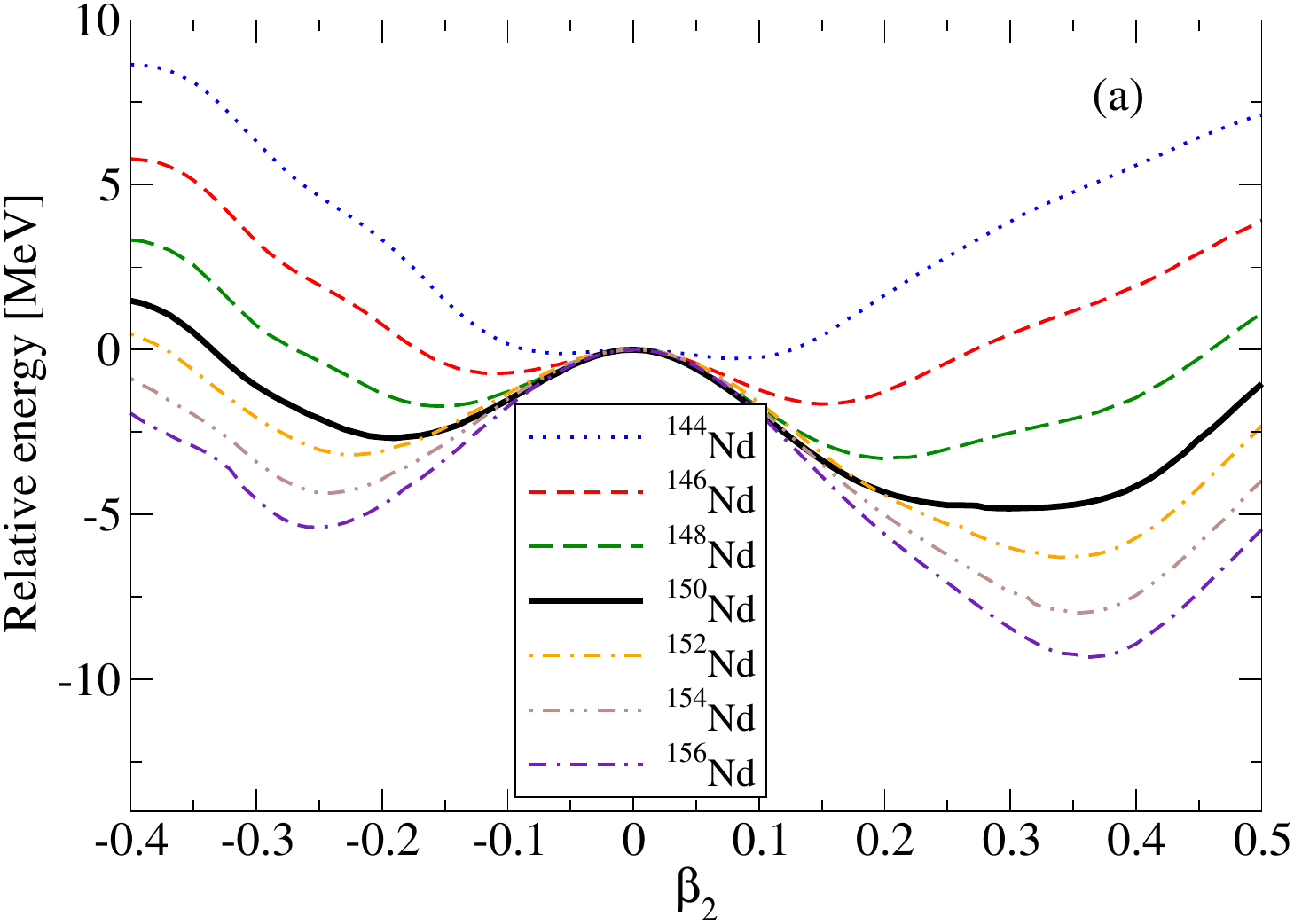}
\end{minipage}
\begin{minipage}[t]{.45\textwidth}
\includegraphics[width=0.95\textwidth]{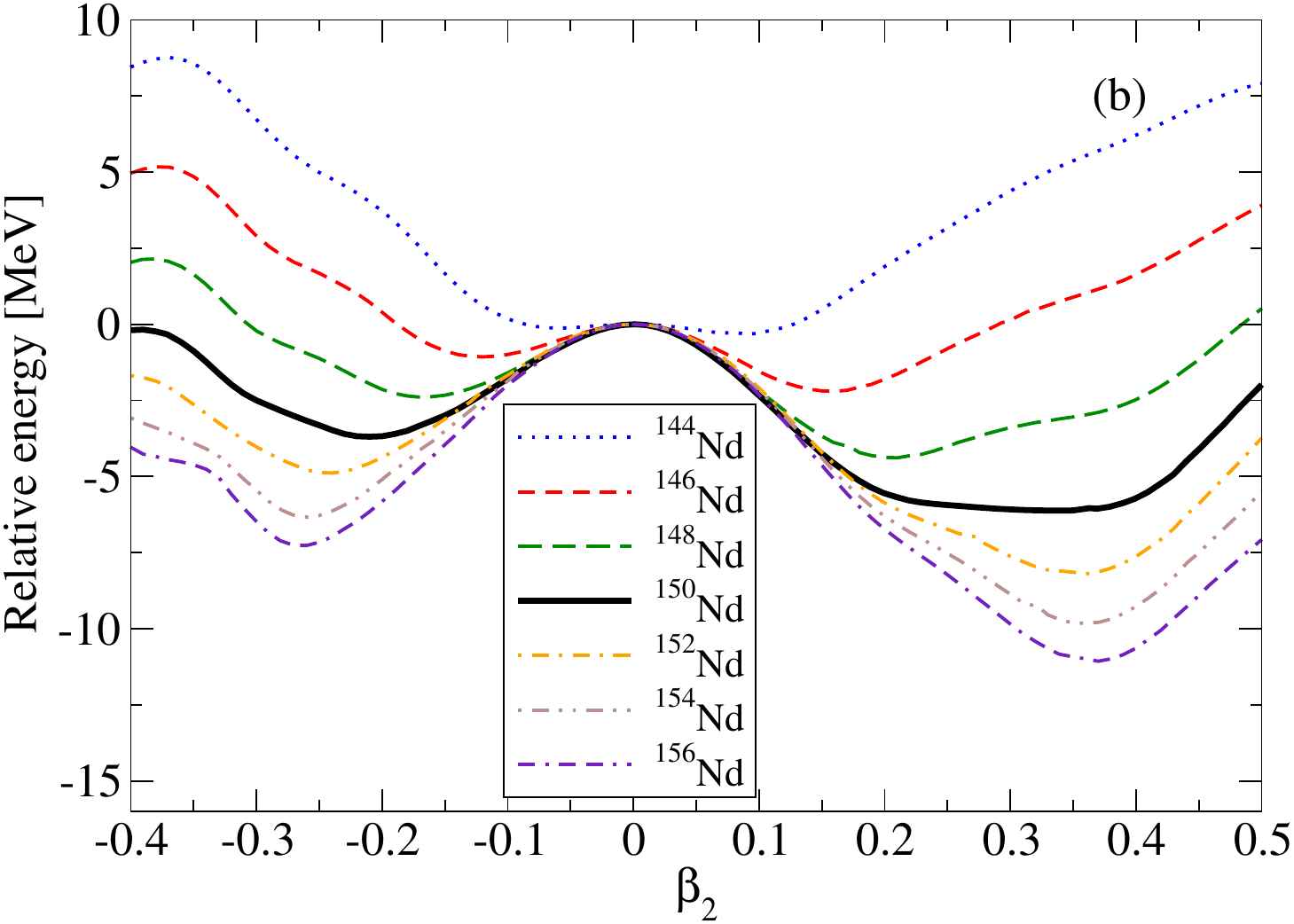}
\end{minipage}
\caption{PEC as a function of $\beta_2$ for the neodymium isotopes for two SHF parametrizations, SV-bas (a) and SV-mas07 (b).
Flat-bottomed PECs are indicated by thicker curves.}
\label{Nd-b2}
\end{figure}

\begin{figure}[!htb]
\begin{minipage}[t]{.45\textwidth}
\includegraphics[width=0.95\textwidth]{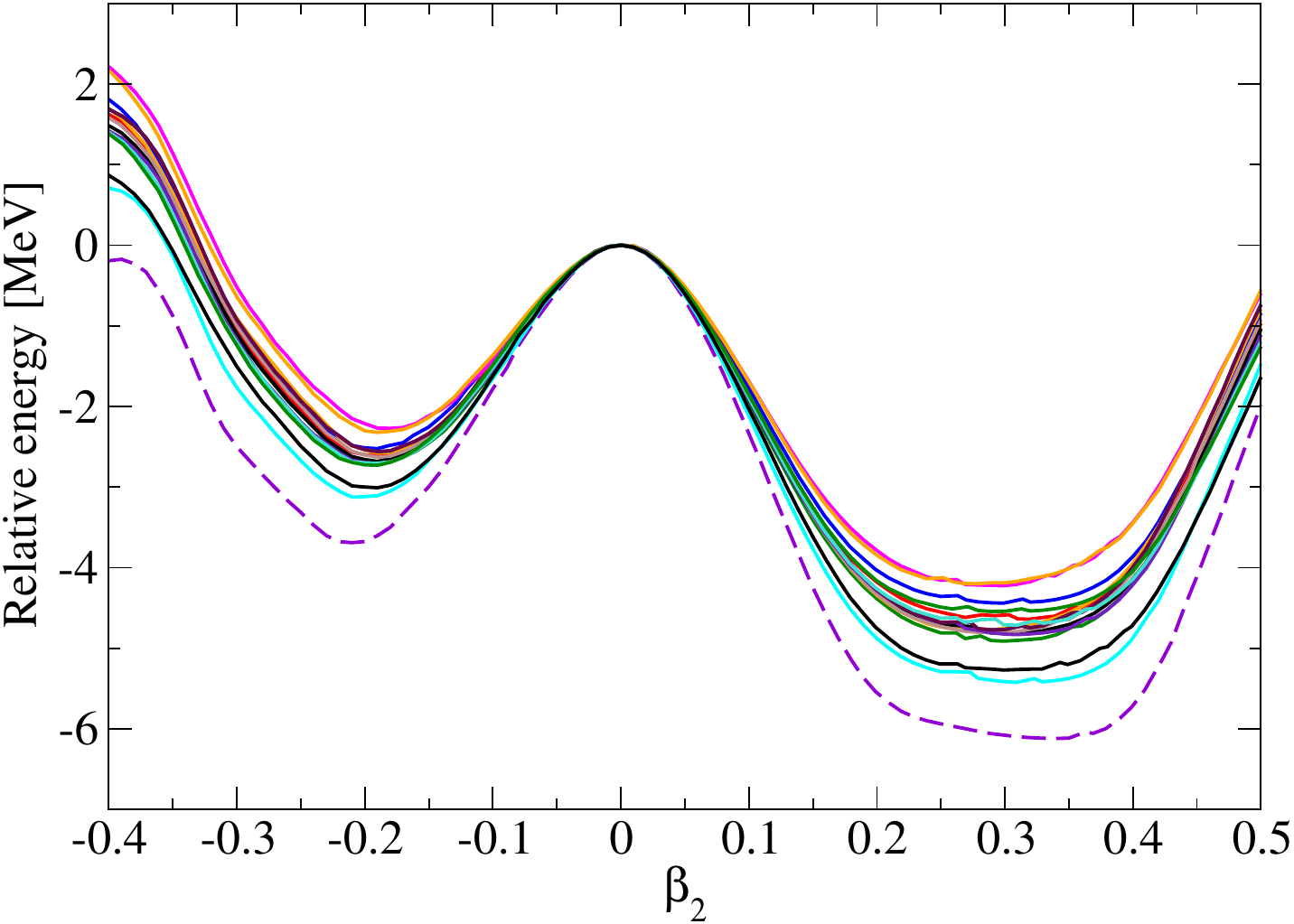}
\caption{PEC as a function of $\beta_2$ for $^{150}$Nd for all investigated parametrizations. The parametrization SV-mas07 (violet dashed) gives more pronounced minima.}
\label{150Nd-all}
\end{minipage}
\end{figure}

\begin{figure}[!htb]
\begin{minipage}[t]{.45\textwidth}
\includegraphics[width=0.95\textwidth]{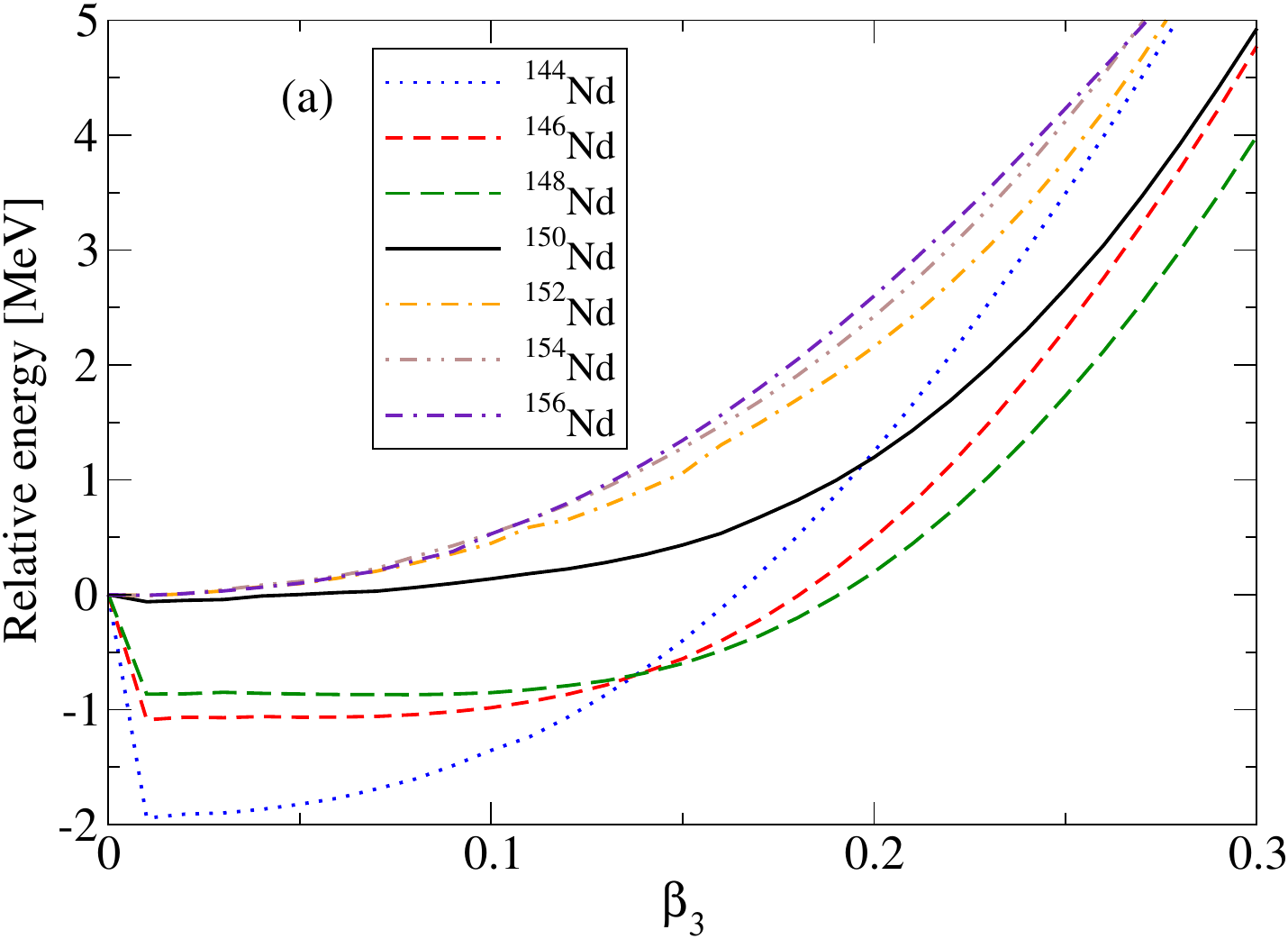}
\end{minipage}
\begin{minipage}[t]{.45\textwidth}
\includegraphics[width=0.95\textwidth]{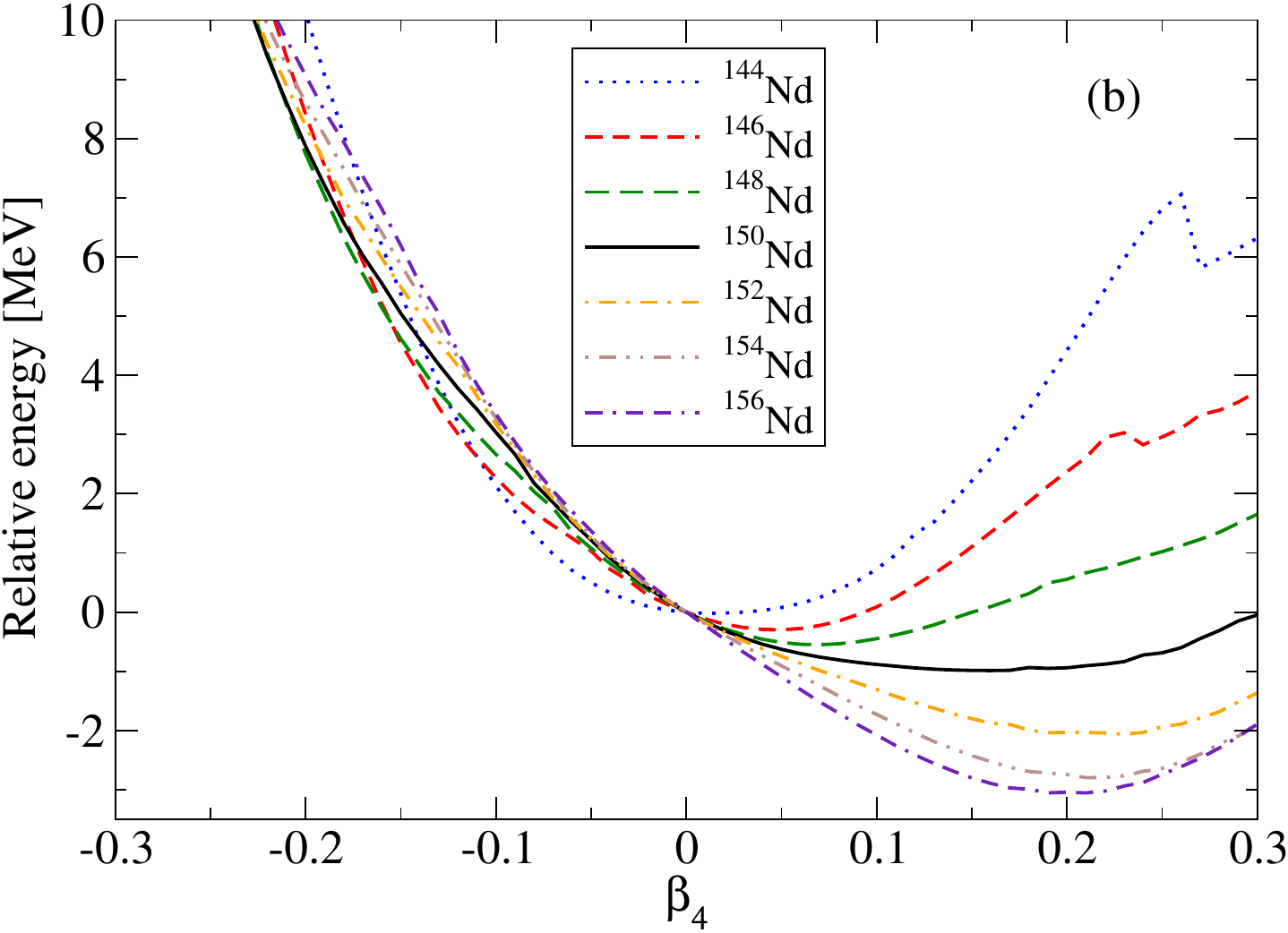}
\end{minipage}
\caption{PEC as a function of $\beta_3$ (a) and $\beta_4$ (b) for the neodymium isotopes for SV-bas parametrization.}
\label{Nd-b3b4}
\end{figure}

In the neodymium chain one observes a gradual development of prolate deformation (see Fig.~\ref{Nd-b2}) with a wider flat region around the prolate quadrupole deformation minimum for $N=90$ for all parametrizations (see Fig.~\ref{150Nd-all}). Octupole deformation does not play any significant role, in the region $N \le 90$
octupole deformation softness occurs. Hexadecapole deformation increases from $\beta_4=0$ to $\beta_4=0.22$ along the chain with a PEC wider flat region for $N=60$ (see Fig.~\ref{Nd-b3b4}).

To check if triaxiality plays an important role in the investigated regions we have to use Sky3D code \cite{Maruhn2014}. Results for the parametrization SV-bas are presented
in Table~\ref{Sky3D}. Because the codes use different definitions of $\beta_2$, $\beta_{\mathrm{2SKYAX}}$ is recalculated according to the $\beta_2$ definition in the Sky3D code.
One can see that both values of $\beta_2$ coincide and the triaxiality is not important in the investigated regions.

\begin{table}

\caption{Comparison of quadrupole deformations, $\beta_{\mathrm{2SKYAX}}$ and $\beta_{\mathrm{2Sky3D}}$, $\gamma_\mathrm{{Sky3D}}$, calculated using SKYAX and Sky3D codes for the parametrization SV-bas.}
\begin{tabular}{ c | c| c c}
\hline
nucleus &  $\beta_{\mathrm{2SKYAX}}$ &   $\beta_{\mathrm{2Sky3D}}$ & $\gamma_{\mathrm{Sky3D}}$ \\
\hline
 $^{68}$Se & -0.150 & 0.147 & 60.0 \\
 $^{70}$Se & -0.126 & 0.132 & 60.0 \\
 $^{72}$Se & -0.102 & 0.108 & 60.0 \\
 $^{74}$Se &  0.000 & 0.000 & 22.4 \\
 $^{76}$Se &  0.000 & 0.003 & 52.4 \\
 $^{78}$Se &  0.053 & 0.063 &  0.1 \\
 $^{80}$Se &  0.090 & 0.086 &  0.0 \\
 $^{92}$Zr &  0.000 & 0.002 & 17.8 \\
 $^{94}$Zr &  0.000 & 0.004 &  7.2 \\
 $^{96}$Zr &  0.000 & 0.007 & 10.5 \\
 $^{98}$Zr &  0.000 & 0.004 & 53.2 \\
$^{100}$Zr & -0.108 & 0.005 & 56.8 \\
$^{102}$Zr & -0.114 & 0.122 & 59.6 \\
$^{104}$Zr & -0.121 & 0.003 & 57.0 \\
 $^{94}$Mo &  0.000 & 0.003 &  5.4 \\
 $^{96}$Mo &  0.000 & 0.009 &  2.8 \\
 $^{98}$Mo &  0.000 & 0.012 &  4.6 \\
$^{100}$Mo &  0.000 & 0.009 & 41.3 \\
$^{102}$Mo & -0.114 & 0.118 & 57.1 \\
$^{104}$Mo & -0.120 & 0.126 & 57.9 \\
$^{106}$Mo & -0.126 & 0.129 & 59.4 \\
$^{144}$Nd &  0.048 & 0.033 &  7.8 \\
$^{146}$Nd &  0.089 & 0.089 &  0.2 \\
$^{148}$Nd &  0.119 & 0.125 &  0.0 \\
$^{150}$Nd &  0.179 & 0.203 &  0.0 \\
$^{152}$Nd &  0.204 & 0.214 &  0.0 \\
$^{154}$Nd &  0.210 & 0.219 &  0.0 \\
$^{156}$Nd &  0.216 & 0.225 &  0.0 \\
\hline
\end{tabular}
\label{Sky3D}
\end{table}

\begin{figure}[!htb]
\begin{minipage}[t]{.45\textwidth}
\includegraphics[width=0.95\textwidth]{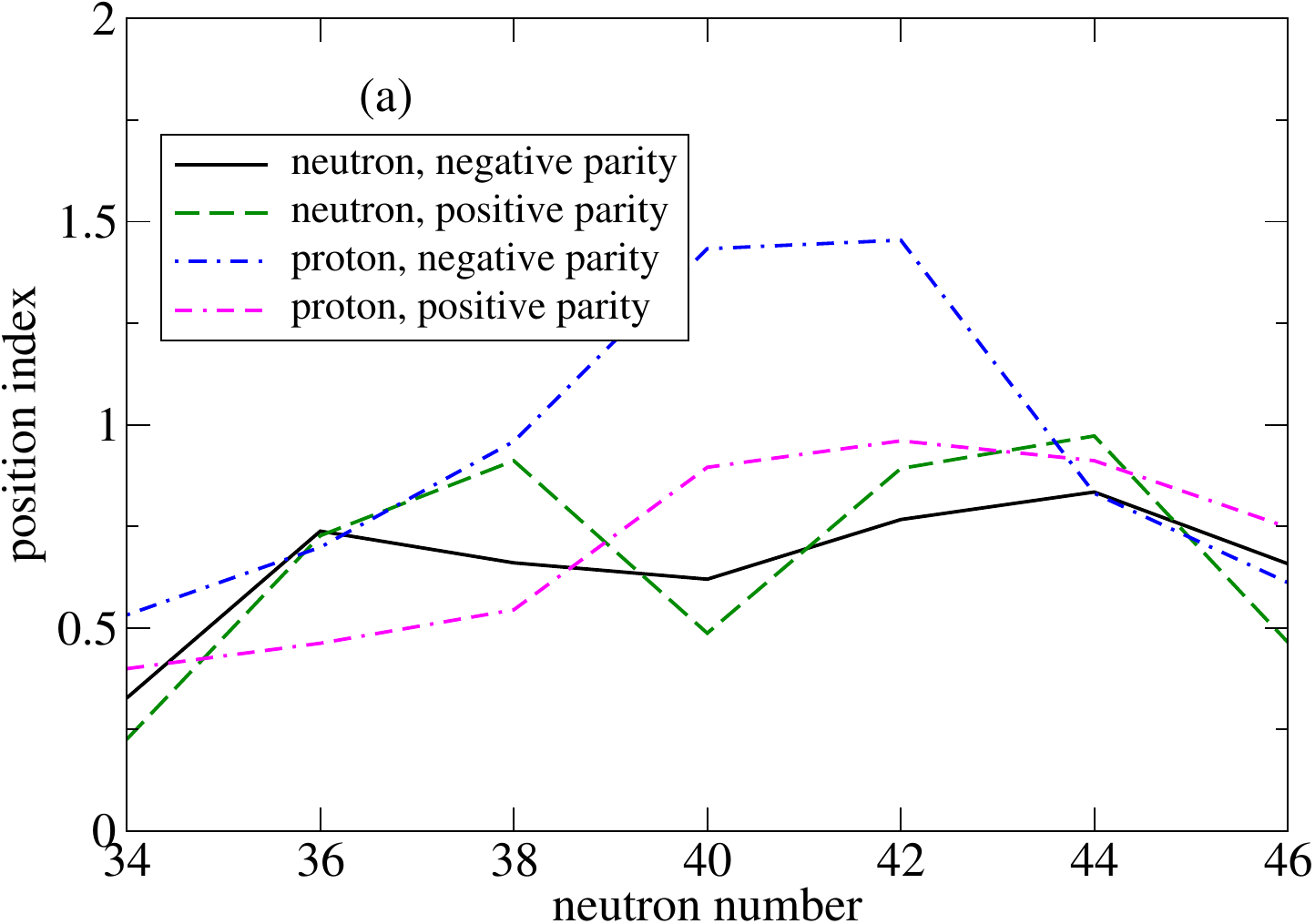}
\end{minipage}
\begin{minipage}[t]{.45\textwidth}
\includegraphics[width=0.95\textwidth]{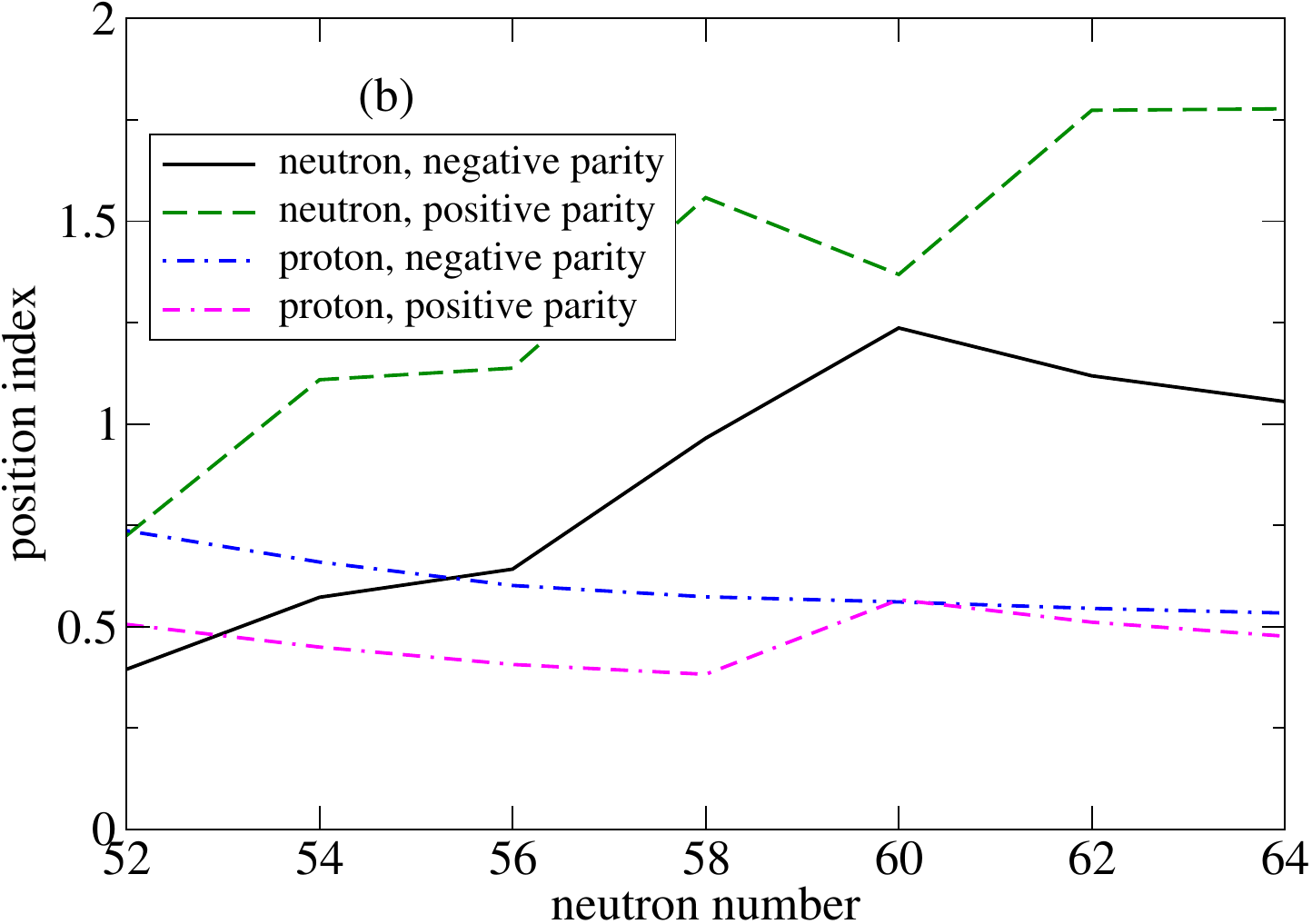}
\end{minipage}
\caption{Position index $I$ for the selenium (a) and zirconium (b) chains (SV-bas parametrization).}
\label{SeZr-index}
\end{figure}

\begin{figure}[!htb]
\begin{minipage}[t]{.45\textwidth}
\includegraphics[width=0.95\textwidth]{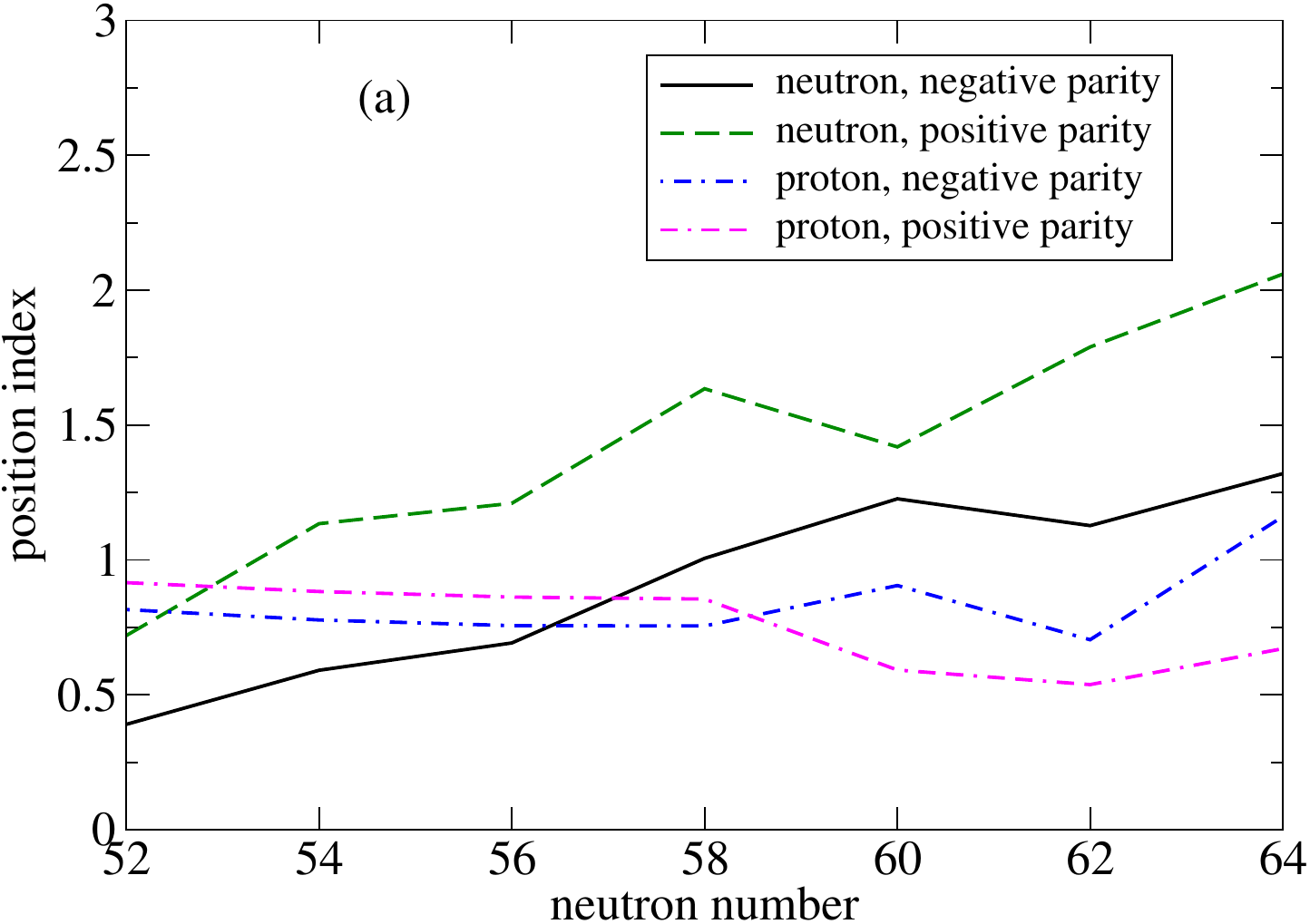}
\end{minipage}
\begin{minipage}[t]{.45\textwidth}
\includegraphics[width=0.95\textwidth]{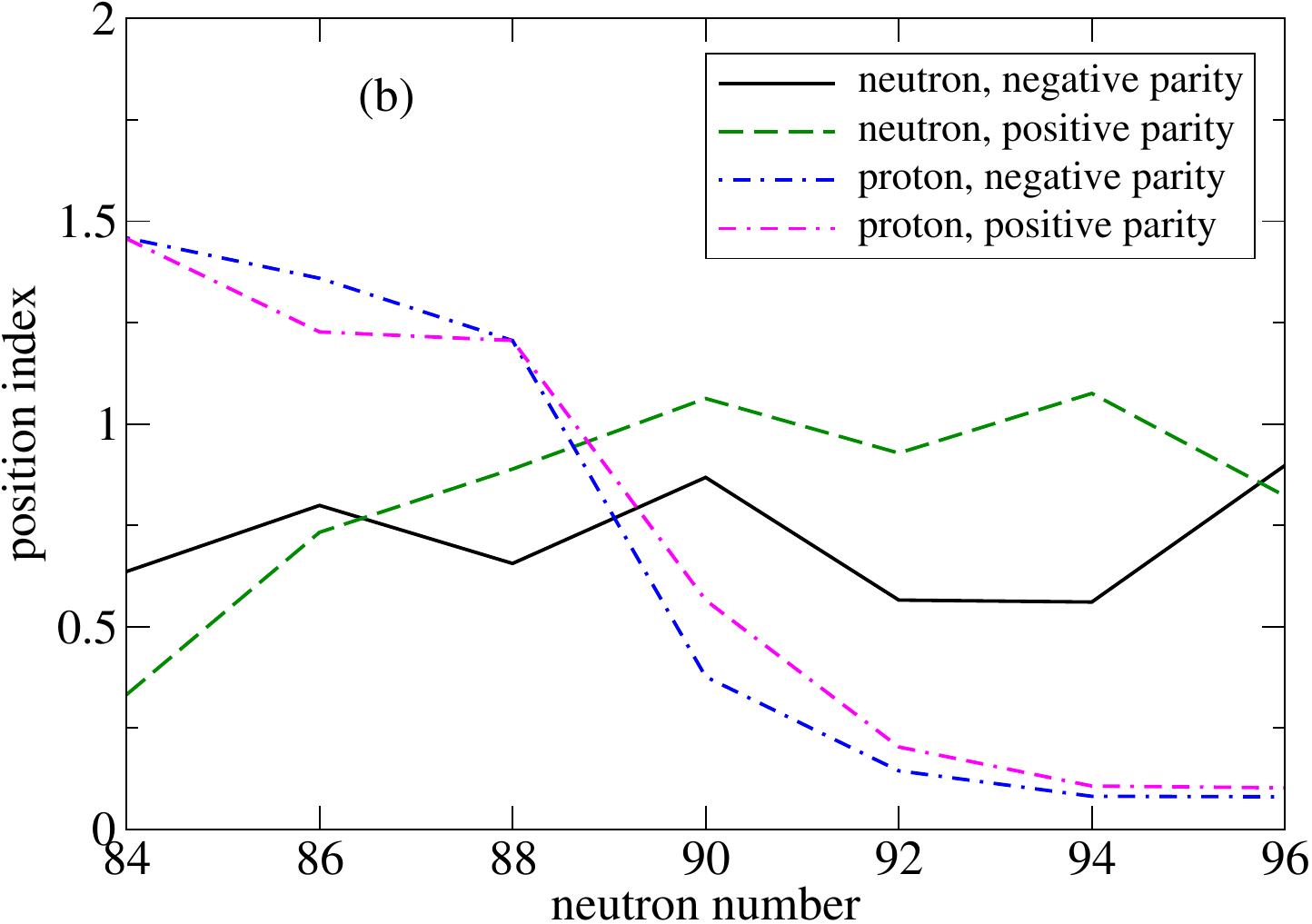}
\end{minipage}
\caption{Position index $I$ for the molybdenum (a) and neodymium (b) chains (SV-bas parametrization).}
\label{MoNd-index}
\end{figure}

To track the phase transitions on a single-quasiparticle level, one has to investigate a relative position of single-quasiparticle positive and negative
parity proton and neutron states with respect to the Fermi level. We, therefore, define new four position indices $I(\nu,\pi)$ for proton positive ($\nu =p$, $\pi=+$), proton negative ($\nu =p$, $\pi=-$), neutron positive ($\nu =n$, $\pi=+$), 
and neutron negative ($\nu =n$, $\pi=-$) parity states. Values of $I(\nu,\pi)$ are found from $I(\nu,\pi) = \sum_{i(\nu,\pi)}
(0.5 - |v^2_i - 0.5|)$, where $v^2_i$ is the pairing occupation probability of the state $i$ and we sum over all model states of defined
$\nu$ and $\pi$. Because $v^2=0.5$ exactly at the Fermi level, the contribution of the levels close to the Fermi level to the position
 index approaches the maximum value of 0.5, whereas the contribution of the levels 
far away from the Fermi level approaches to zero. The position index also depends on the strength of the pairing force. If 
it is low, the indices approach zero, too. In the limit of zero pairing $v^2_i =0$ or 1 for all states $i$ and $I(\nu,\pi) =0$. Parity enables to separate intruder states: In the selenium isotopes ($Z=34$) both proton and neutron intruder states have positive parity, 
while in the zirconium ($Z=40$) and molybdenum ($Z=42$) isotopes proton intruder states have positive parity and neutron intruder states have negative parity. In the neodymium ($Z=60$) isotopes, instead, proton intruder states have negative parity and neutron intruder states have positive parity. Results for selenium, zirconium, molybdenum and neodymium chains are presented in Figs.~\ref{SeZr-index} and
\ref{MoNd-index} for the SV-bas parametrization.

In selenium isotopes protons and neutrons occupy the same shell and $I(p,+)=I(n,+)$ between $N=38$ and 40 and around $N=44$. $I(p,-)=I(n,-)$ symmetrically around $N=40$ for 
$N=36$ and $N=44$. 
In zirconium isotopes $I(p,-)=I(n,-)$ for $N=56$ and neutron intruder position index $I(n,-)$ reaches its maximum for $N=60$, i.e., these states are closer to the Fermi level.
In molybdenum isotopes $I(p,-)=I(n,-)$ again close to $N=56$ and the neutron intruder position index $I(n,-)$ gradually increases up to $N=60$.

In neodymium isotopes, both $I(p,+)=I(n,+)$ and $I(p,-)=I(n,-)$ between $N=88$ and 90 where the phase transition should occur. The position indices for intruder proton 
negative-parity states and for proton positive-parity states are almost the same. This means that almost equal number of proton negative and positive parity states lie close to the Fermi level, with a slight preference for the intruder states (higher position index) before the phase transition point. Close to the phase transition point a crossing of both indices occurs and after it, the proton intruder states are more distant from the Fermi level and, simultaneously, the proton pairing gets weaker. A common feature of
both regions,
$N=60$ and $N=90$, namely a gradual increase of $I(n,+)$ and $I(n,-)$ and a gradual decrease of $I(p,+)$ and $I(p,-)$ is more pronounced in the case of neodymium. In addition, in both regions, the intruder-neutron position index approaches its maximum for
$N=60$ and 90, respectively, indicating a maximum proximity of neutron intruder states to the Fermi level.

In the case of Nd, the position indices for intruder negative- and normal positive-parity states are almost the same. This means that almost equal number of proton negative and positive parity states lie close to the Fermi level, with a slightly preferable intruder states (higher position index) before the phase transition point. At the phase transition point a crossing of the both indices occurs and after it, normal proton states are close to the Fermi level and the intruder states are more distant and, simultaneously, the proton pairing gets weaker. Similar situation occurs for Mo, except the weakening of the pairing interaction.

\begin{figure}[!htb]
\begin{minipage}[t]{.45\textwidth}
\includegraphics[width=0.95\textwidth]{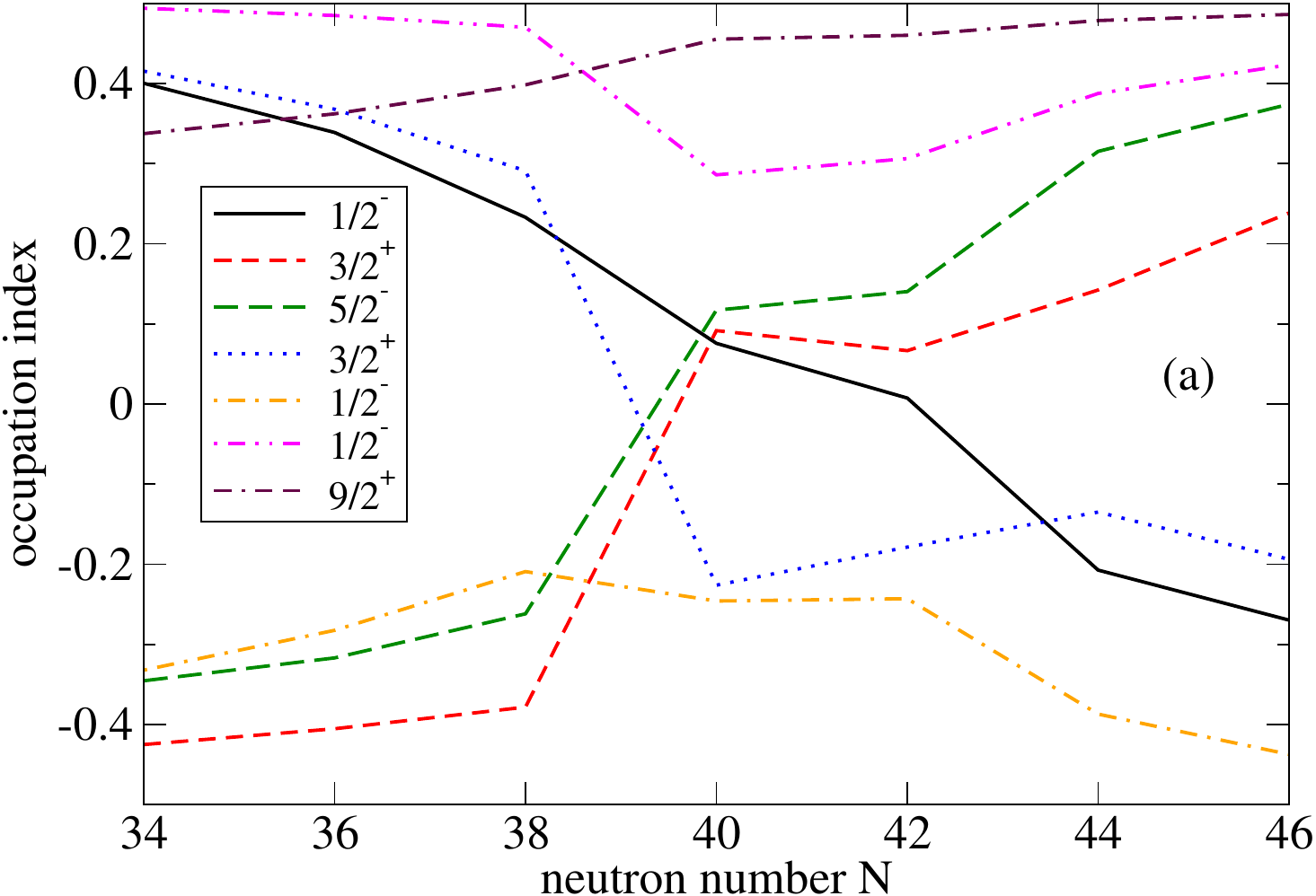}
\end{minipage}
\begin{minipage}[t]{.45\textwidth}
\includegraphics[width=0.95\textwidth]{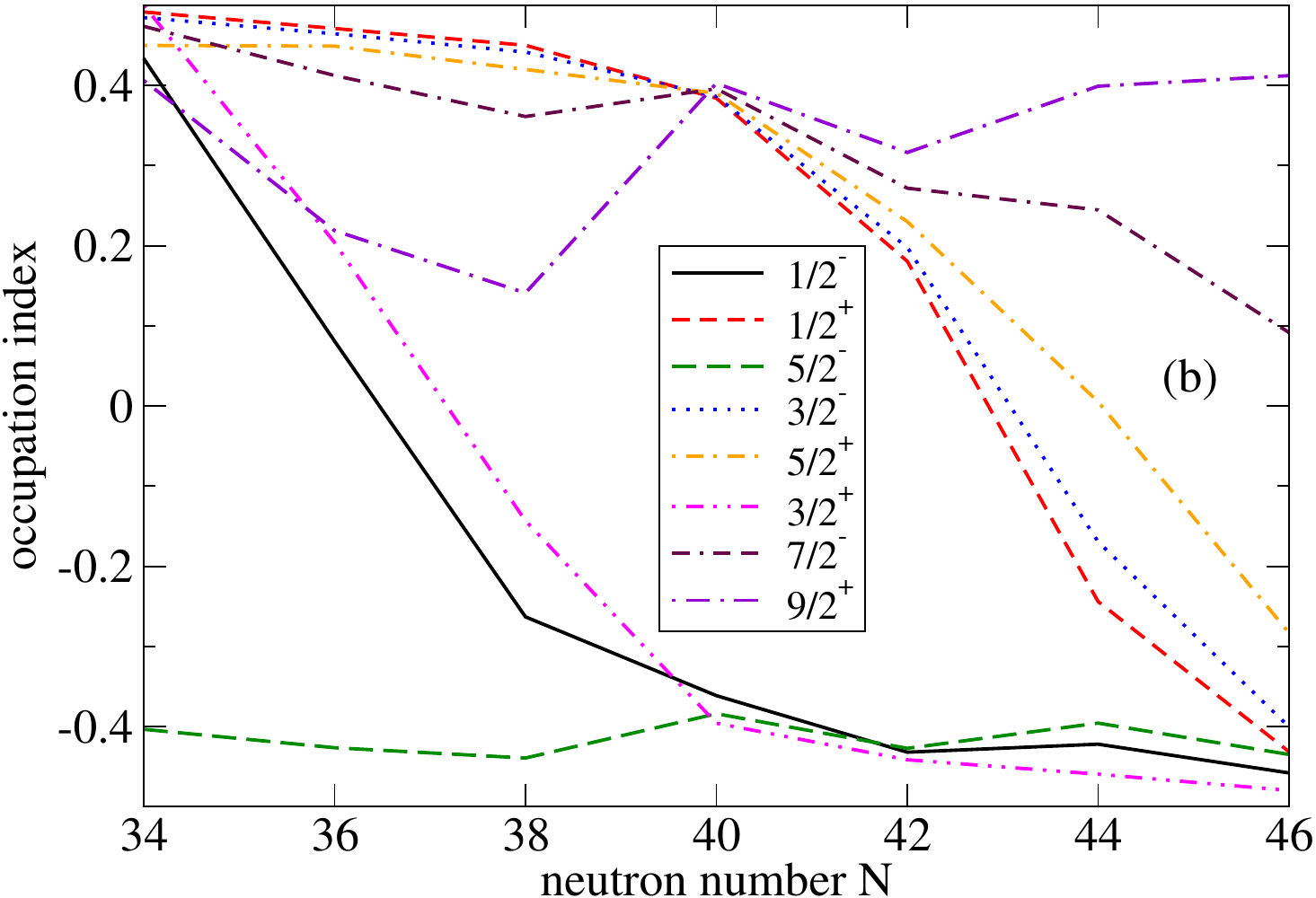}
\end{minipage}
\caption{Occupation index $i$ for the proton (a) and neutron (b) states closest to the Fermi level in the selenium chain (SV-bas parametrization).}
\label{Se-oindex}
\end{figure}

\begin{figure}[!htb]
\begin{minipage}[t]{.45\textwidth}
\includegraphics[width=0.95\textwidth]{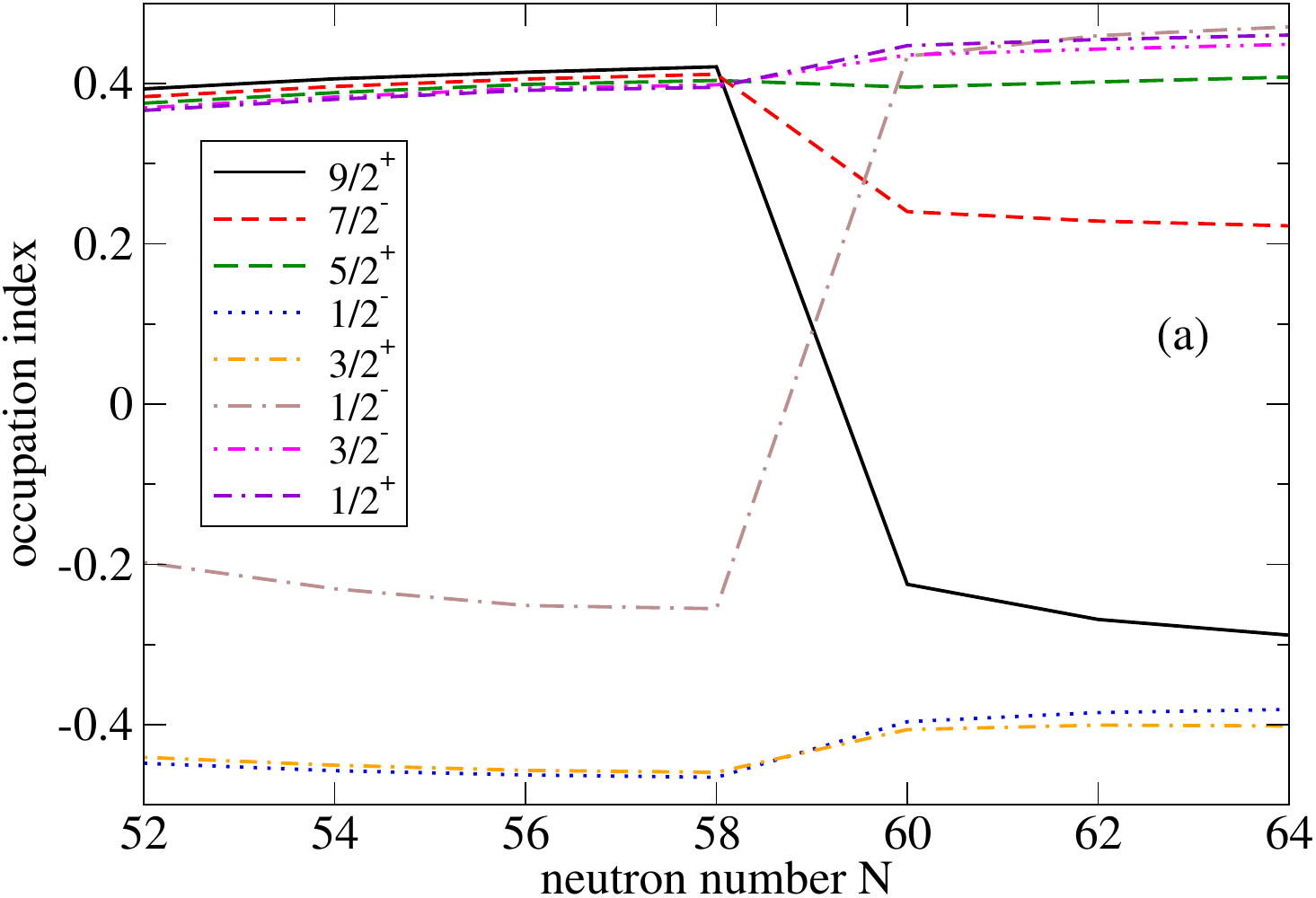}
\end{minipage}
\begin{minipage}[t]{.45\textwidth}
\includegraphics[width=0.95\textwidth]{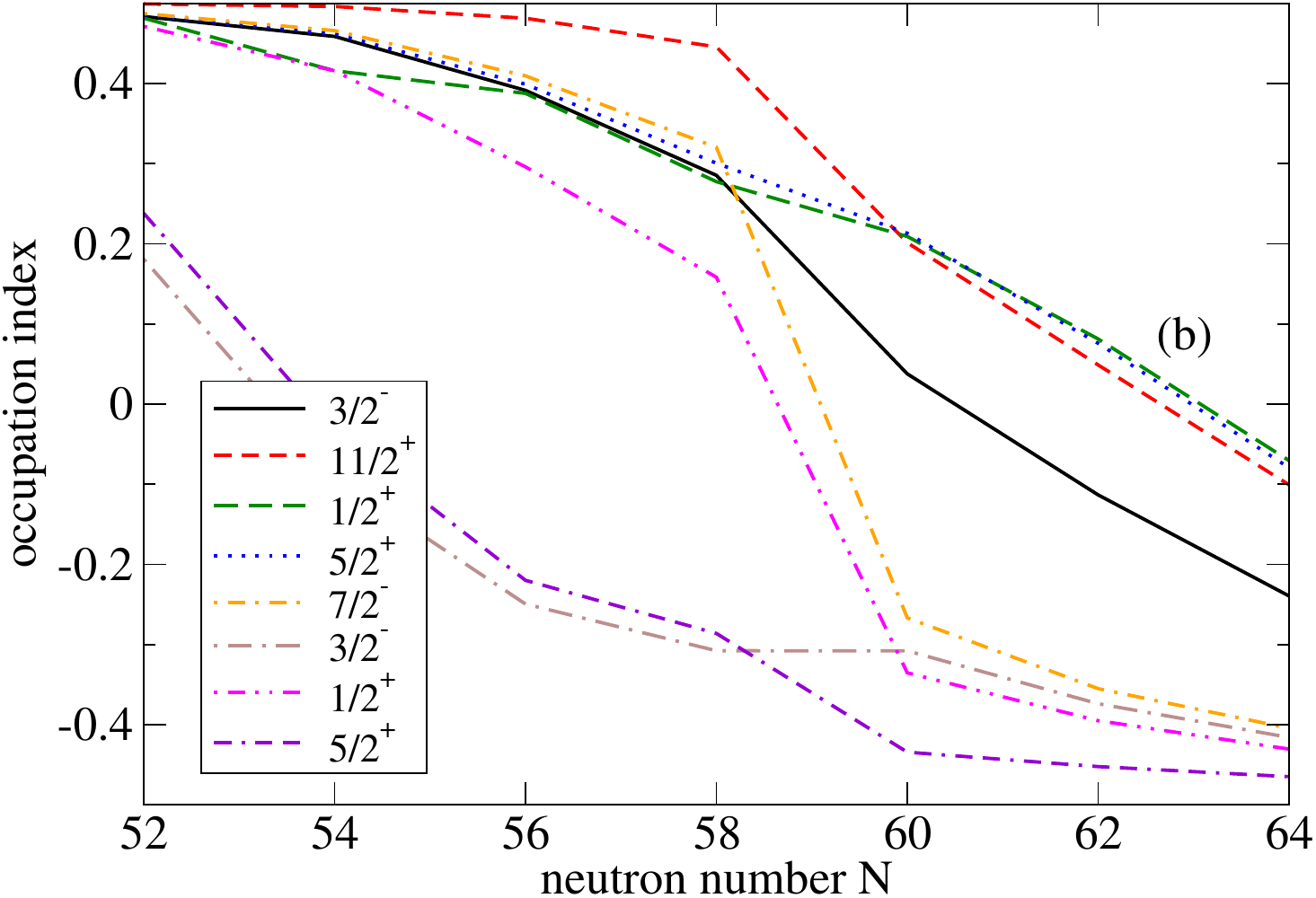}
\end{minipage}
\caption{Occupation index $i$ for the proton (a) and neutron (b) states closest to the Fermi level in the zirconium chain (SV-bas parametrization).}
\label{Zr-oindex}
\end{figure}

\begin{figure}[!htb]
\begin{minipage}[t]{.45\textwidth}
\includegraphics[width=0.95\textwidth]{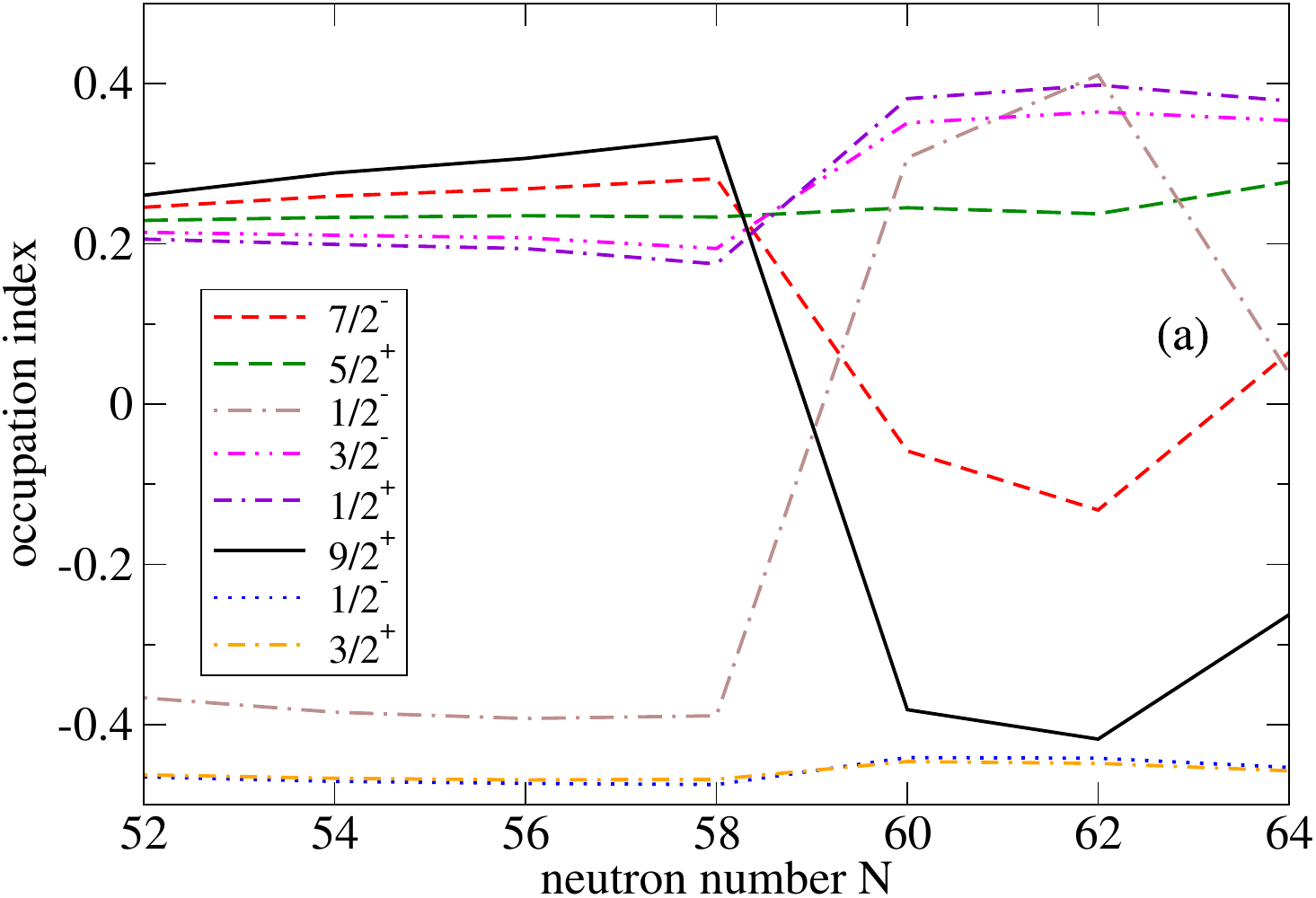}
\end{minipage}
\begin{minipage}[t]{.45\textwidth}
\includegraphics[width=0.95\textwidth]{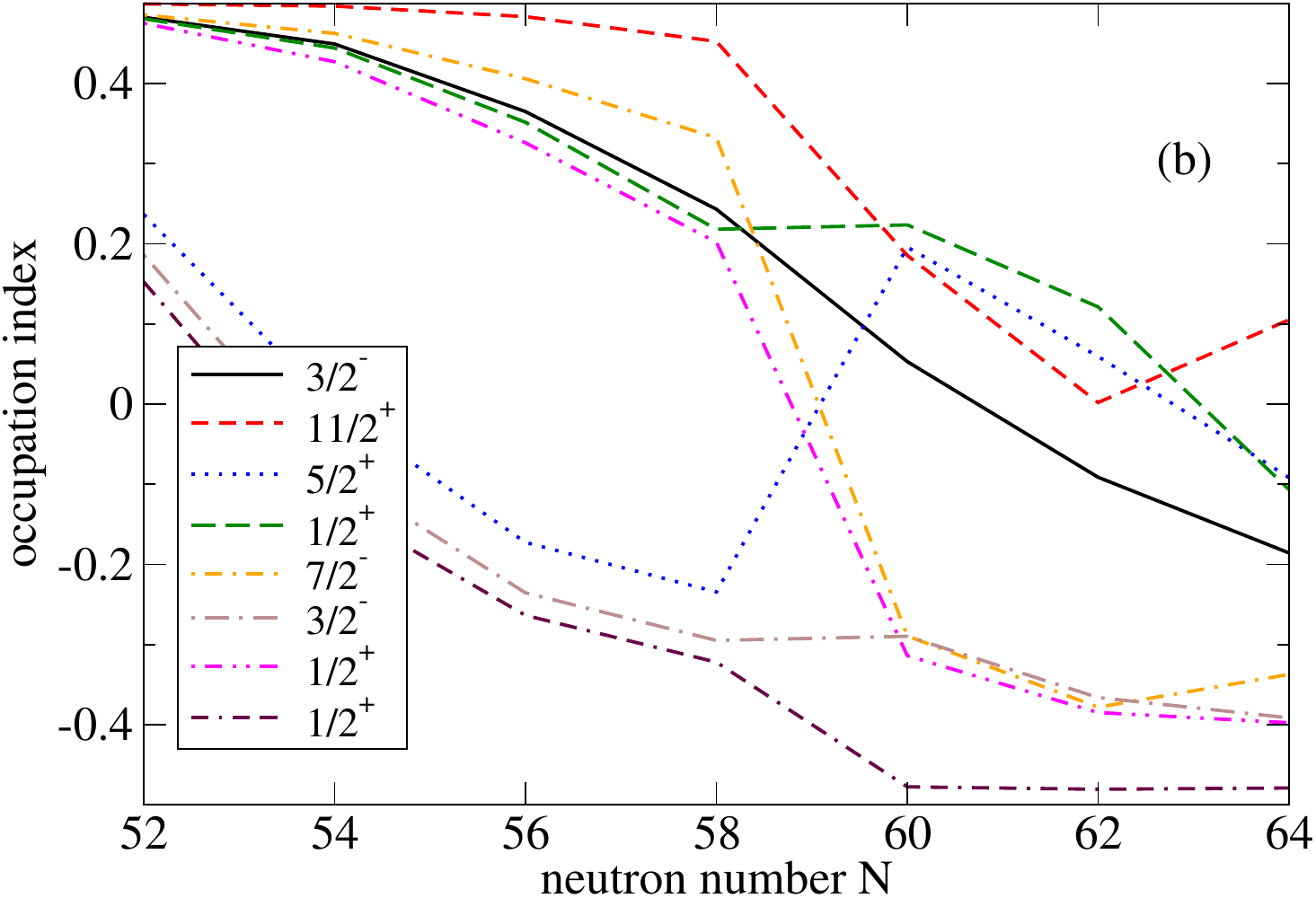}
\end{minipage}
\caption{Occupation index $i$ for the proton (a) and neutron (b) states closest to the Fermi level in the molybdenum chain (SV-bas parametrization).}
\label{Mo-oindex}
\end{figure}

\begin{figure}[!htb]
\begin{minipage}[t]{.45\textwidth}
\includegraphics[width=0.95\textwidth]{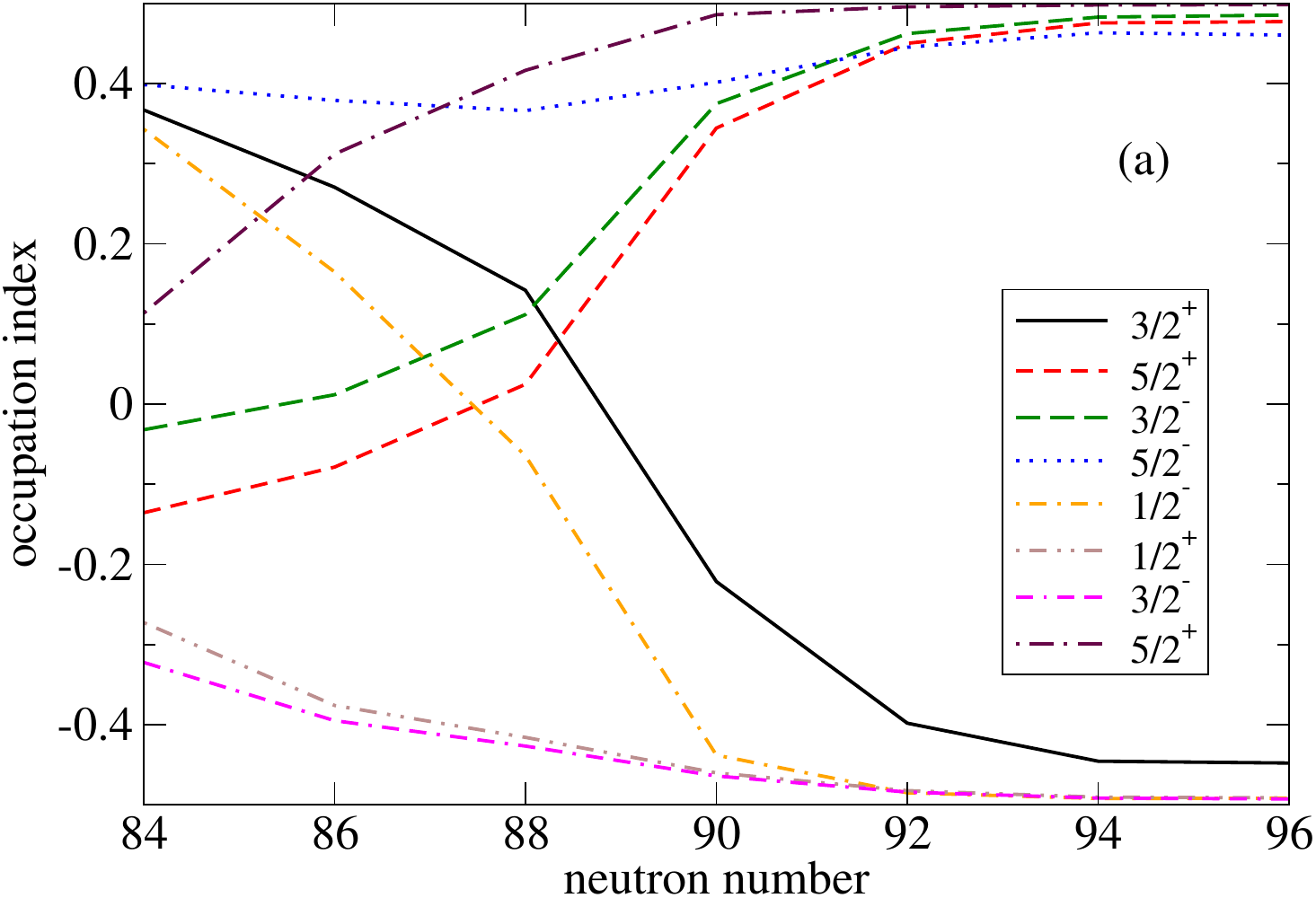}
\end{minipage}
\begin{minipage}[t]{.45\textwidth}
\includegraphics[width=0.95\textwidth]{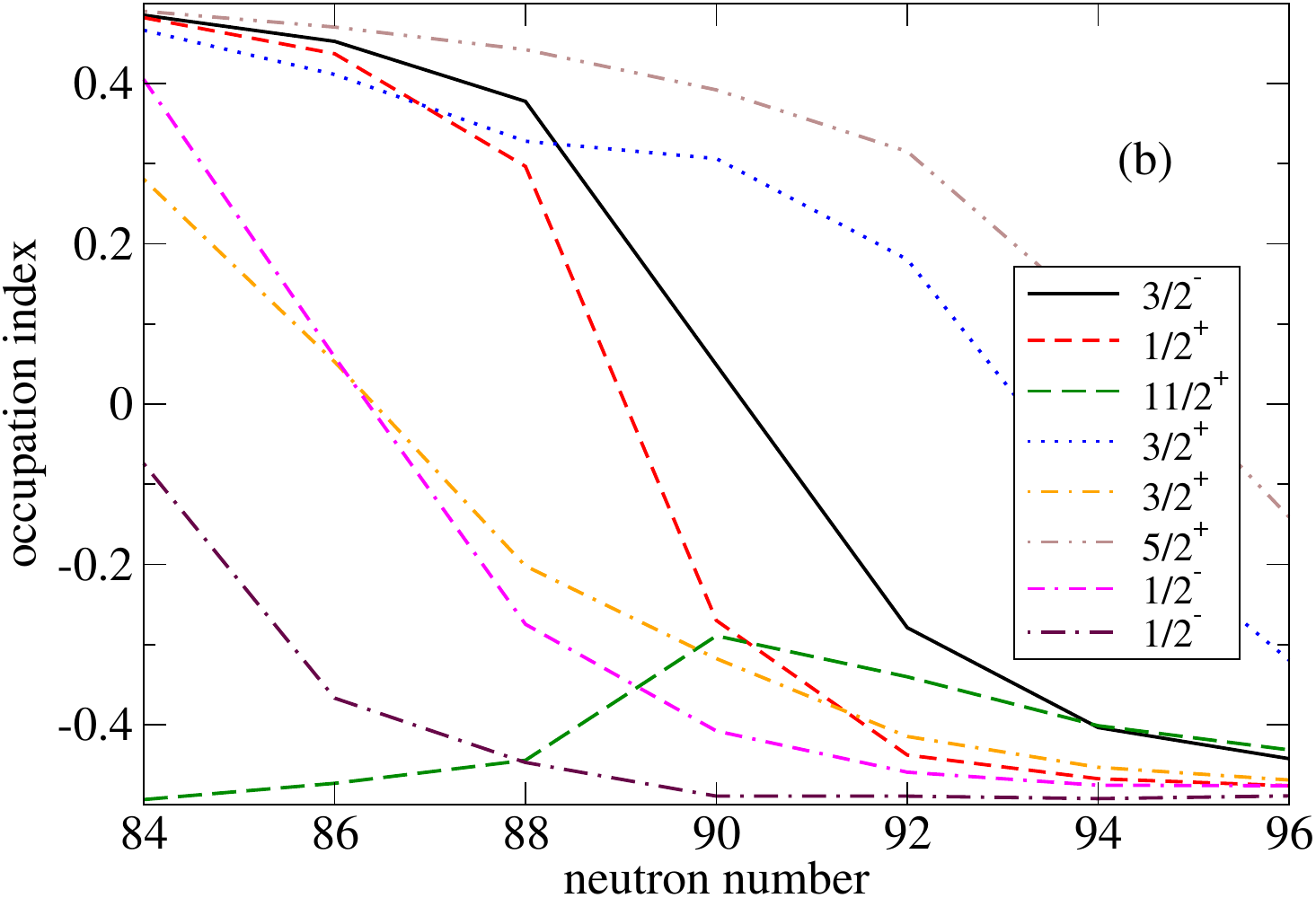}
\end{minipage}
\caption{Occupation index $i$ as a function of the neutron number for the proton (a) and neutron (b) states closest to the Fermi level in the neodymium chain (SV-bas parametrization).}
\label{Nd-oindex}
\end{figure}

If we look closely at the single-quasiparticle (qp) states close to the Fermi level obtained in our SHF+BCS calculations, we can identify microscopic signatures of the CPS on the
single-particle level. Here, we restrict ourselves to the SV-bas parametrization only, because for other parametrizations similar results were obtained.
The states resulting from the SHF+BCS calculations are characterized by the projection of the total angular momentum $j_z$ on the symmetry axis and their parity  $\pi$.
In Figs.~\ref{Se-oindex}--\ref{Nd-oindex} the occupation index $i = 0.5 - v^2$ for the qp states closest to the Fermi levels for the Se, Zr, Mo and Nd chains is shown 
(the index $i$ approaches 0
when the qp state aproaches the Fermi level and is positive for particle states and negative for hole states; in addition it aproaches 0.5 for particle and -0.5 for hole
 states if the qp states are distant from the Fermi level or the pairing interation is weak). 
 
One can observe a typical proton-qp level crossing around the phase transition
points: for the Se chain $1/2^-$ gets occupied while $5/2^-$ unoccupied, for the Zr and Mo chains the intruder $9/2^+$ gets occupied while $1/2^-$ unoccupied, for
the Mo chain $7/2^-$ gets occupied in addition, and for the Nd chain $3/2^+$ and the intruder $1/2^-$ get occupied while $5/2^+$ and the intruder $3/2^-$ unoccupied and the
proton pairing gets weaker.

For the Se chain and $N=40$ the neutron pairing gets weaker, at $N=38$ neutron-qp states $1/2^-$ and intruder $3/2^+$ cross the Fermi level and the intruder $9/2^+$ lies
closer to the Fermi level and later between $N=42$ and 44, $5/2^-$, $3/2^-$ and the intruder $5/2^+$ cross the Fermi level. For the Zr and Mo chains, $1/2^+$ and intruders
$3/2^-$ and $7/2^-$ cross the Fermi level and become occupied, while for Mo only $5/2^+$ becomes unoccupied. In the case of the Nd chain, $3/2^-$ and the intruder $1/2^+$ 
get occupied around $N=90$.

To summarize, we have found three possible microscopic signatures of criticality (QPT): (i) flatness of PECs, (ii) 
position index for neutron intruder states approaching a maximum value, and (iii) occupation index of selected proton qp-levels along the
isotopic chain changes its sign (i.e. some particle states become hole states and v.v. at the QPT point).

\begin{figure}[!htb]
\begin{minipage}[t]{.45\textwidth}
\includegraphics[width=0.95\textwidth]{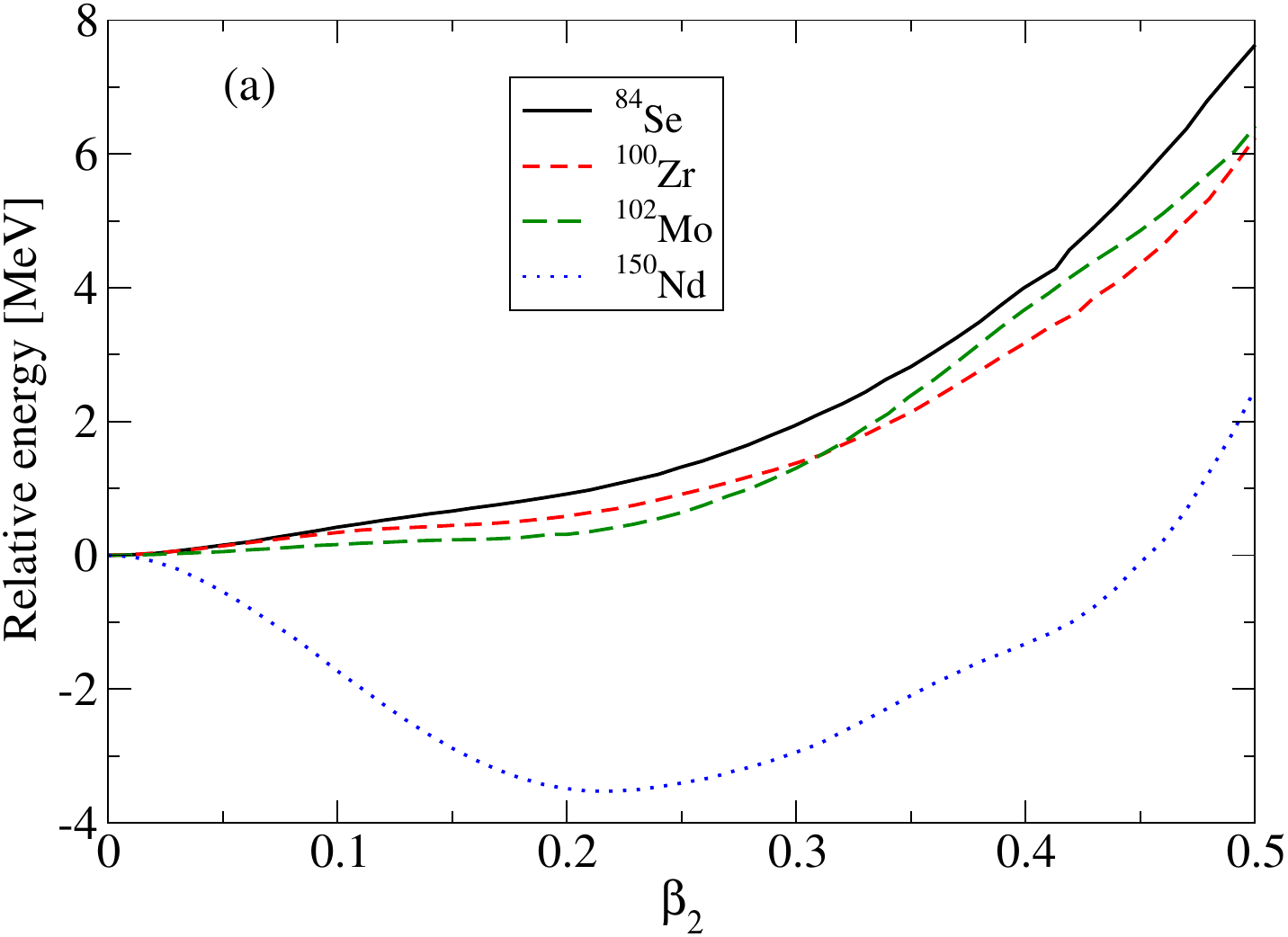}
\end{minipage}
\begin{minipage}[t]{.45\textwidth}
\includegraphics[width=0.95\textwidth]{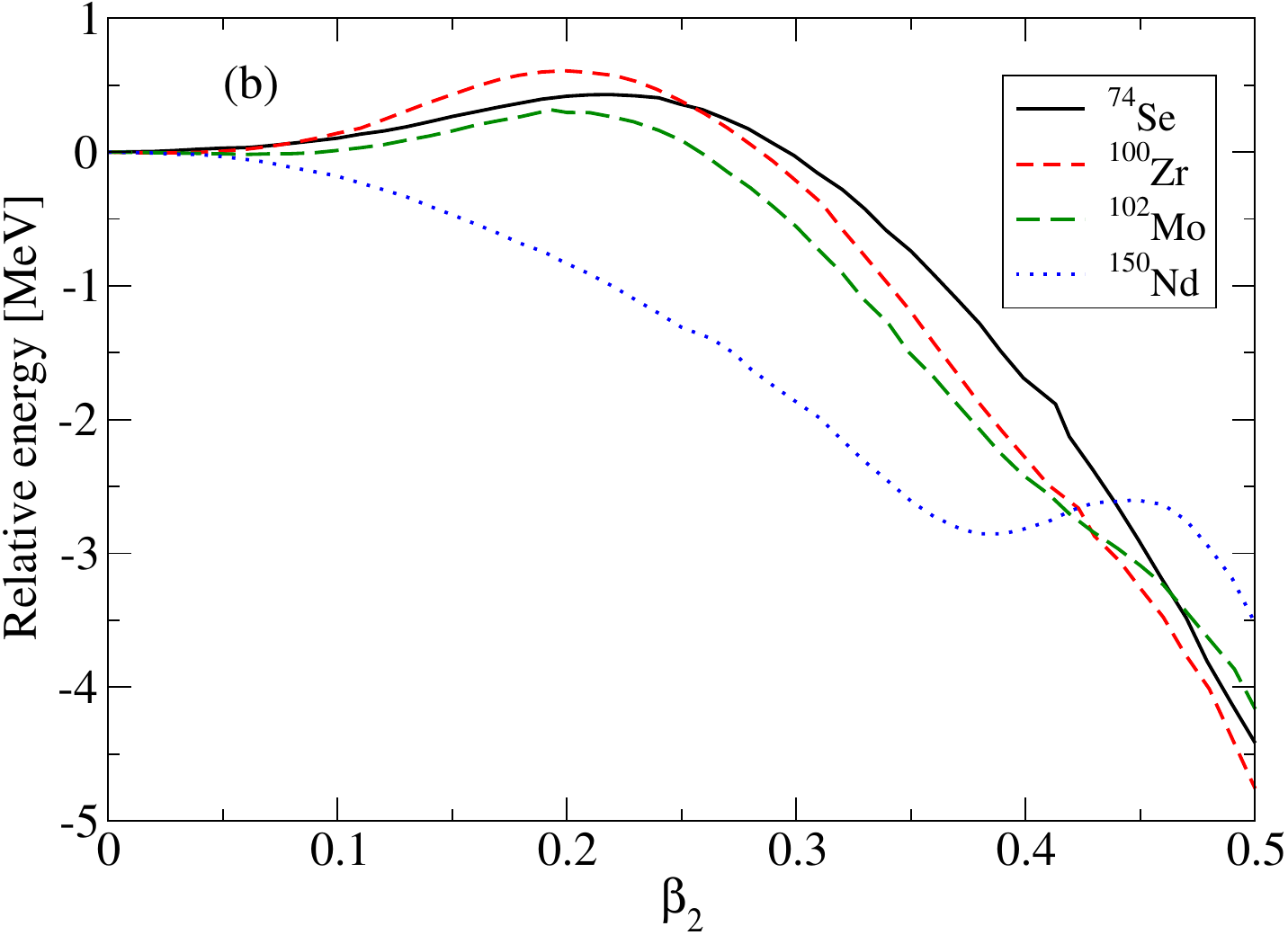}
\end{minipage}
\caption{Symmetric (a) and antisymmetric (b) parts of the potential derived from the microscopic PECs for the SV-bas parametrization and the investigated
$N=40$, 60 and 90 nuclei.}
\label{FigVsym-asym}
\end{figure}

\begin{figure}[!htb]
\begin{minipage}[t]{.45\textwidth}
\includegraphics[width=0.95\textwidth]{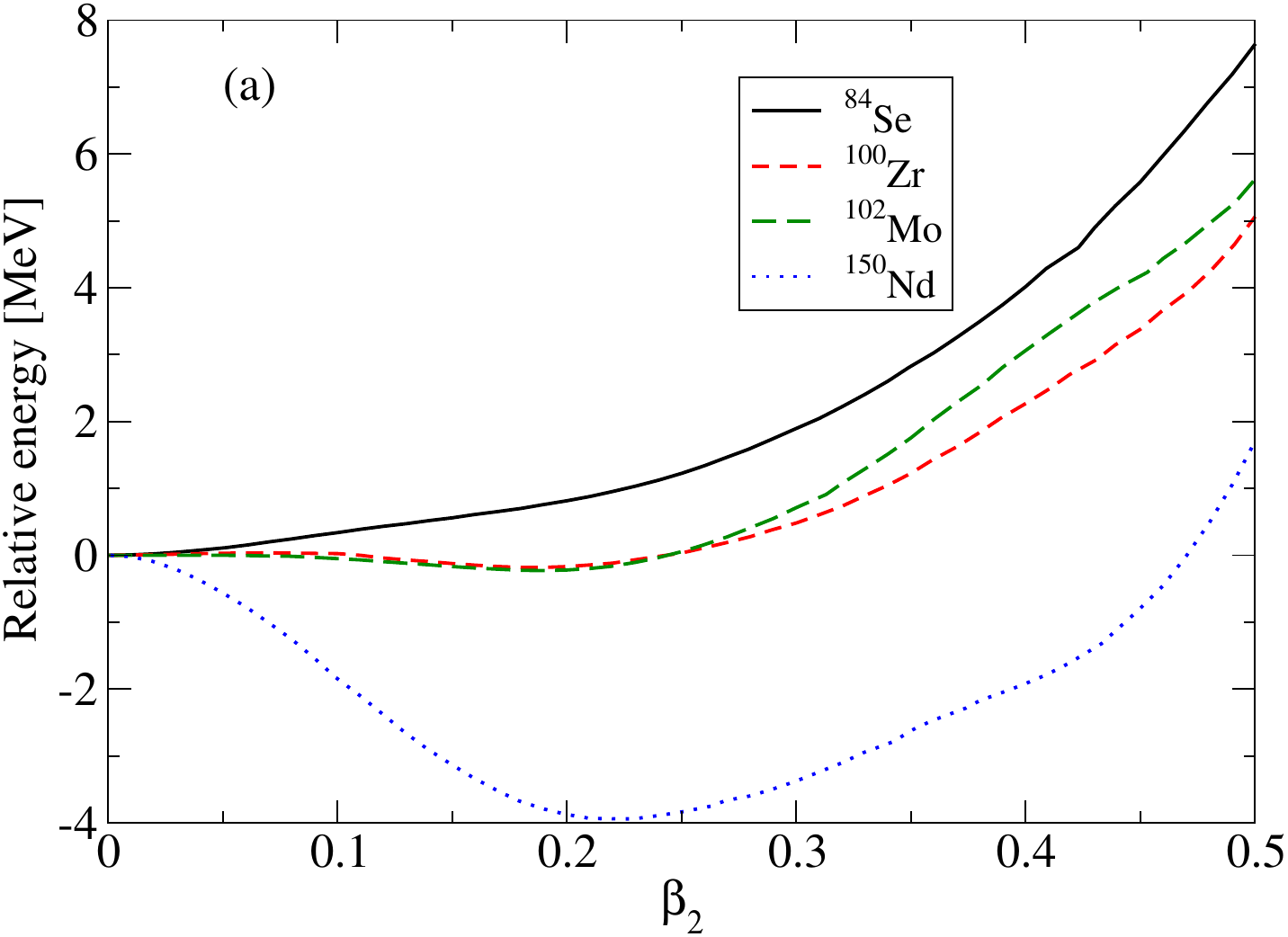}
\end{minipage}
\begin{minipage}[t]{.45\textwidth}
\includegraphics[width=0.95\textwidth]{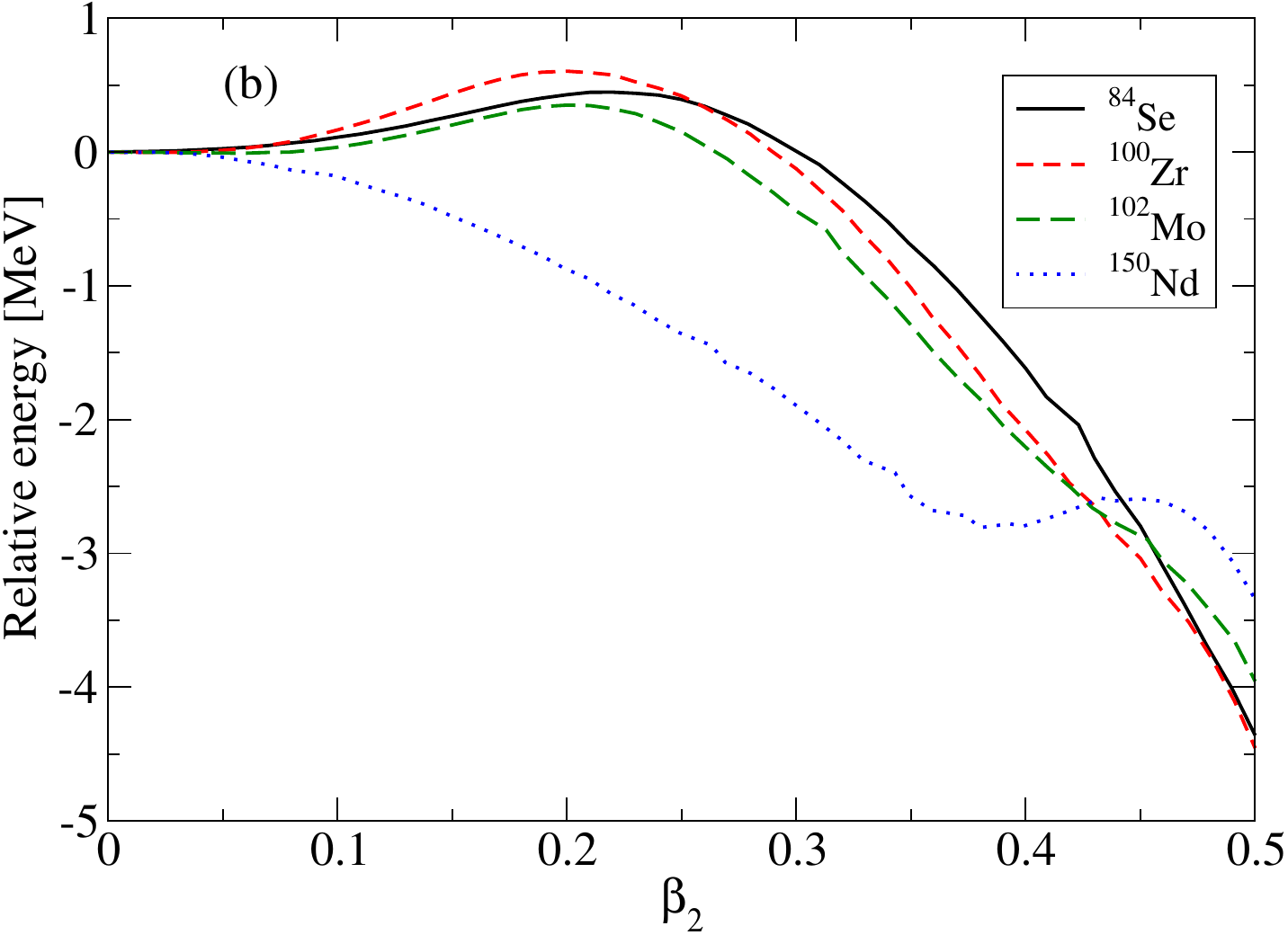}
\end{minipage}
\caption{Symmetric (a) and antisymmetric (b) parts of the potential derived from the microscopic PECs for the SV-tls parametrization and the investigated
$N=40$, 60 and 90 nuclei.}
\label{FigVsym-asym-tls}
\end{figure}

To bridge from the microscopic calculations within the SHF+BCS model towards macroscopic calculations within the Algebraic Collective Model (ACM), we can use the procedure described in Ref.~\cite{Klupfel2008}.
In macroscopic calculations we need a potential $V_{\mathrm{ACM}}$ as a function of two deformation parameters, $\beta$ and $\gamma$. It can be constructed from the microscopic
PEC as a function of $\beta_2$, $V(\beta_2)$ and expanded up to the first order term in $\cos{3\gamma}$:
\begin{equation}
V(\beta, \gamma) =\frac{1}{2} [V(\beta_2) + V(-\beta_2)] + \frac{1}{2} [V(\beta_2) - V(-\beta_2)] \cos{3\gamma}
\label{Vsym-asym}
\end{equation}
The potential obtained from Eq.~\ref{Vsym-asym} is composed of two parts, a $\gamma$-independent symmetric part and a $\gamma$-dependent antisymmetric part. The 
results for the investigated $N=40$, 60 and 90 nuclei are shown in Figs.~\ref{FigVsym-asym} and \ref{FigVsym-asym-tls} for two parametrizations, SV-bas and 
SV-tls. Qualitatively, we get a $\gamma$-independent well for the symmetric part of the potential and a decreasing antisymmetric $\gamma$-dependent potential for low and medium quadrupole deformations.

\section{Algebraic collective model}
\label{ACM}

The Algebraic Collective Model (ACM) \cite{Rowe2004,Rowe2005,Rowe2009}, introduced as a computationally tractable version of the collective model of Bohr and Mottelson (BMM) \cite{Bohr1998a,Bohr1998b} restricted to quadrupole rotational and vibrational degrees of freedom, is characterized by a well defined algebraic structure.
Unlike the conventional U(5)$\supset$SO(5)$\supset$SO(3) dynamical subgroup chain used, for example, in the Frankfurt program \cite{Hess1981,Gneuss1971}, as well as in the IBM \cite{Arima1976,Iachello1987}, the ACM
makes use of the subgroup chain \cite{Szpikowski1980,Rowe1998,Rowe2005b,DeBaerdemacker2007,DeBaerdemacker2009}
\begin{equation}
 {\rm SU}(1,1) \times {\rm SO}(5) \supset {\rm U}(1) \times {\rm SO}(3) \supset {\rm SO}(2) 
\end{equation}
to define basis wave functions as products of $\beta$ wave functions and SO(5) spherical harmonics.

Several advantages result from this choice of the dynamical subgroup chain:

(i) With the now available SO(5) Clebsch-Gordan coefficients \cite{Caprio2009,Welsh2008}, and explicit expressions for SO(5) reduced matrix elements,
matrix elements of BMM operators can be calculated analytically \cite{Rowe2005}.

(ii) By appropriate choices of SU(1,1) modified oscillator representations, the $\beta$ basis wave  functions range \cite{Rowe2005} from those of the U(5)$\supset$SO(5) harmonic vibrational model to those of the rigid $\beta$ wave function of the SO(5)-invariant model of Wilets and Jean \cite{Wilets1956}.

(iii) With these SU(1,1) representations, collective model calculations converge an order of magnitude more rapidly for deformed nuclei than in U(5)$\supset$SO(5) bases
\cite{Rowe2009}.

Thus, the ACM  combines the advantages of  the BMM and the IBM and makes collective model calculations a simple routine procedure
\cite{Caprio2009,Welsh2016}. A pedagogical treatment of the geometrical and algebraic foundations of the ACM can be found in the book by Rowe and Wood \cite{Rowe2010}.

Recall, that both E(5) \cite{Iachello2000} and X(5) \cite{Iachello2001} critical point symmetries correspond to special solutions of the Bohr Hamiltonian \cite{Bohr1952}, in which an infinite square well potential in the quadrupole ($\beta$) degree of freedom is assumed. The question of whether the assumption of a flat potential in both models is justified has been raised and investigated using various methods. Limiting ourselves to the regions of interest, we mention the following cases.

a) The need for a potential with a non-flat bottom in $^{152}$Sm within the geometric collective model has been early realized \cite{Zhang1999}. A sextic potential possessing two minima has been used in the Bohr Hamiltonian for the description of some Nd \cite{Budaca2018}, Mo \cite{Budaca2019b}, and Se \cite{Budaca2019a} isotopes. A potential with two minima (spherical and deformed) has been recently employed in the Bohr Hamiltonian for the description of the Zr isotopes \cite{Mardyban2020,Mardyban2022}. The existence of a bump in the PES of good experimental X(5) cases might be related to the confined $\beta$-soft (CBS) rotor model \cite{Pietralla2004}, employing an infinite square well potential displaced from zero, as well as to the relevance of the Davidson potential $\beta^{2}+\frac{\beta_{0}^{4}}{\beta^{2}}$ \cite{Davidson1932,Bonatsos2007}, where $\beta_{0}$ is the minimum of the potential.

b) Potential energy surfaces (PES) for the $N=90$ isotones, calculated in the relativistic mean field (RMF) framework  \cite{Fossion2006,Niksic2007}, also exhibit a shape with two minima. An alternative approach \cite{Nomura2008,Nomura2010}, allowing for efficient calculation of spectra and transition rates, is based on the use of an IBM Hamiltonian, the parameters of which are determined by fitting the IBM PES to the PES obtained from RMF calculations. The results of these studies do not predict flat potential energy surfaces (PES) for the $N=90$ isotones \cite{Fossion2006,Niksic2007,Nomura2010,Nomura2019}, which are the best experimental manifestations of X(5), while a similar picture occurs also in the $N=60$ \cite{Nomura2016} and $N=40$ \cite{Nomura2017} isotones. 

c) Shell model calculations in the regions of interest have been rather limited, because of obvious computational hurdles. However, recent calculations \cite{Kaneko2015} for the Se isotopes have pointed out the importance of the tensor force \cite{Otsuka2020,Otsuka2022} in shaping up the nuclear properties. While the central force has a minimum at zero, the tensor force possesses a minimum at non-zero distance from the center (see Fig. 1 of \cite{Kaneko2015}), its shape roughly resembling a Kratzer potential \cite{Kratzer1920,Bonatsos2013}. Monte-Carlo shell model calculations have recently reached  the Zr isotopes, studying the QPT appearing in them at $N=60$ \cite{Togashi2016,Lay2023}. 

From the above the importance of having a flexible potential within the ACM to address these questions becomes evident.  

In our implementation of the ACM, observables of interest are formed by taking sums of products of several generating operators, that are SO(3) invariant, such as 
$\hat{\beta^{2}}$, $\hat{\beta^{-2}}$, $\beta\frac{\mathrm{d}}{\mathrm{d}\beta}$, $\Delta^{2}$.
The Hamiltonians that can be analysed are polynomials in these operators with coefficients that may be arbitrary real numbers or functions of the quantum numbers on which they operate.

For the purposes of our study, the starting point will be the Hamiltonian in the form
\begin{eqnarray}
\hat H & = & x_{1}{\nabla^2}+x_{2}+x_{3}\beta^{2}+x_{4}\beta^{4}+\frac{x_{5}}{\beta^{2}}+x_{6}\beta \cos{3\gamma}+  
x_{7}\beta^3 \cos{3\gamma}
+ x_{8}\beta^5 \cos{3\gamma} + \frac{x_{9}}{\beta}\cos{3\gamma} + x_{10}\cos^2{3\gamma} \nonumber \\
& +  & x_{11}\beta^{2}\cos^{2}{3\gamma} 
+x_{12}\beta^{4}\cos^{2}{3\gamma} +\frac{x_{13}}{\beta^{2}}\cos^{2}{3\gamma}+\frac{x_{14}}{\hbar^{2}}[\hat \pi \otimes \hat q\otimes \hat \pi]_0, 
\label{HACM}
\end{eqnarray}
where
\begin{equation}
 \nabla^2 = \frac{1}{\beta^4} \frac{\partial}{\partial\beta} \beta^4 \frac{\partial}{\partial\beta} + \frac{1}{\beta^2} \hat\Lambda
\end{equation}
is the  Laplacian on the 5-dimensional collective model space  and $\hat\Lambda$ is the $SO(5)$ angular momentum operator
\cite{Rowe2010}.
Such a Hamiltonian, expressed in terms of the quadrupole deformation parameters $\beta$ and $\gamma$ serves as a useful starting point for a description of a wide range of nuclear collective spectra. As the above Hamiltonian is a rational function of the basic observables $\hat q $ and $\hat \pi$, its matrix elements are obtained efficiently via available analytic expressions.



\section{Numerical results obtained within the ACM}
\label{Num_ACM}

\begin{figure}[!htb]
\includegraphics[width=0.45\textwidth]{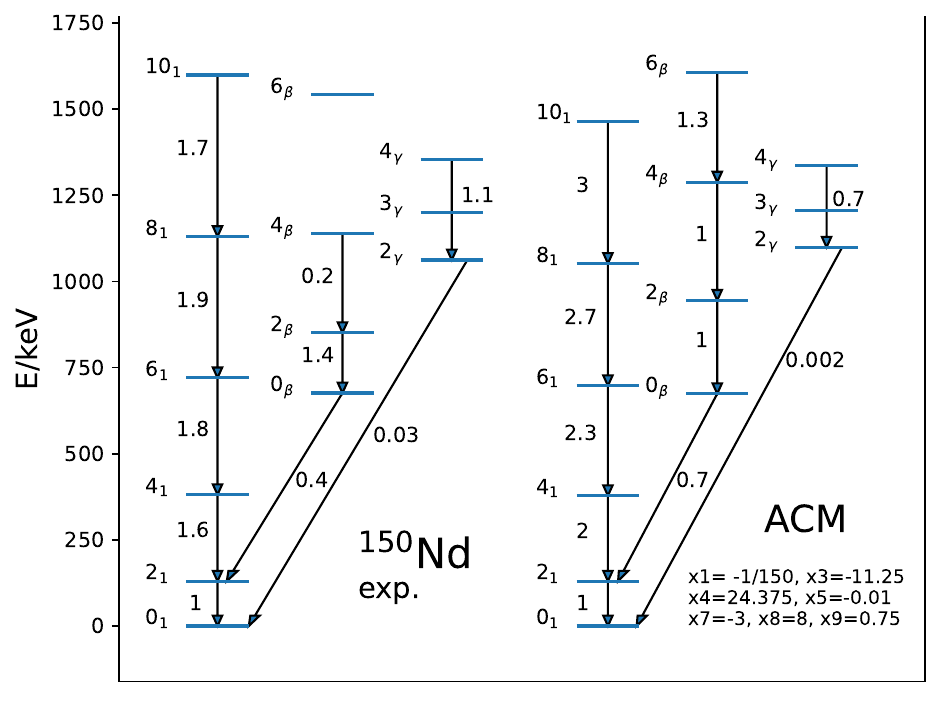}
\caption{Comparison of experimental (left) and ACM (right) spectra for $^{150}$Nd. The energy difference of the $2^+_1$ and $0^+_1$ states is normalized 
to the experimental value in $^{150}$Nd. The $B(E2)$ transition rates are expressed in units of the $B(E2,2_{1}^{+}\rightarrow 0_{1}^{+}) = 1$ transition.
The parameters used in the ACM calculations are displayed in the right corner.} 
\label{150Nd_ACM}
\end{figure}

\begin{figure}[!htb]
\includegraphics[width=0.45\textwidth]{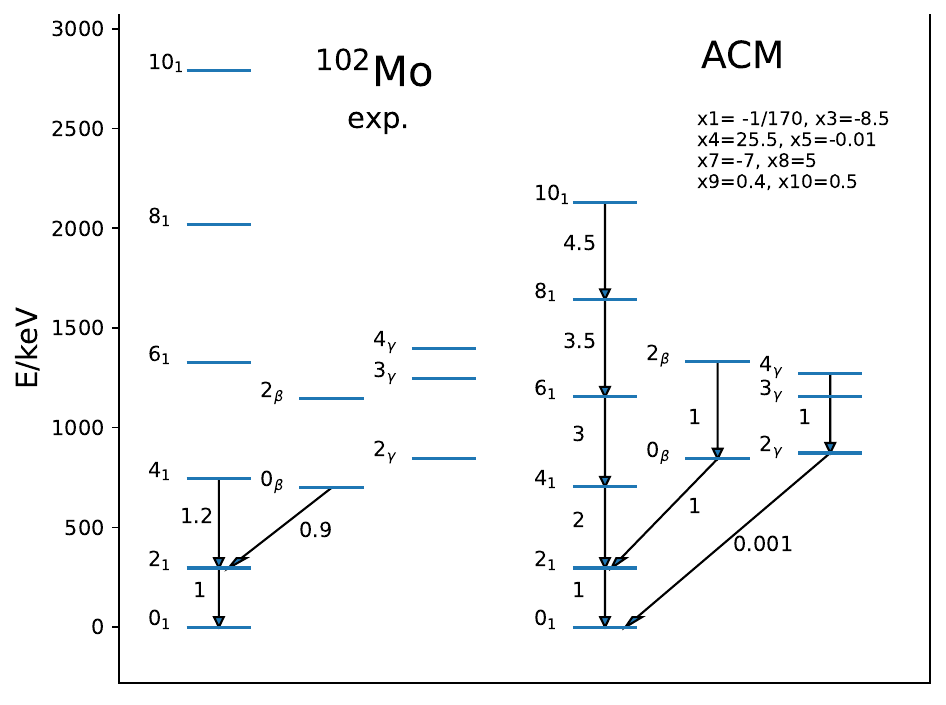}
\caption{Comparison of experimental (left) and ACM (right) spectra for $^{102}$Mo. The energy difference of the $2^+_1$ and $0^+_1$ states is normalized 
to the experimental value in $^{102}$Mo. The $B(E2)$ transition rates are expressed in units of the $B(E2,2_{1}^{+}\rightarrow 0_{1}^{+}) = 1$ transition.
The parameters used in the ACM calculations are displayed in the right corner.} 
\label{102Mo_ACM}
\end{figure}

\begin{figure}[!htb]
\includegraphics[width=0.45\textwidth]{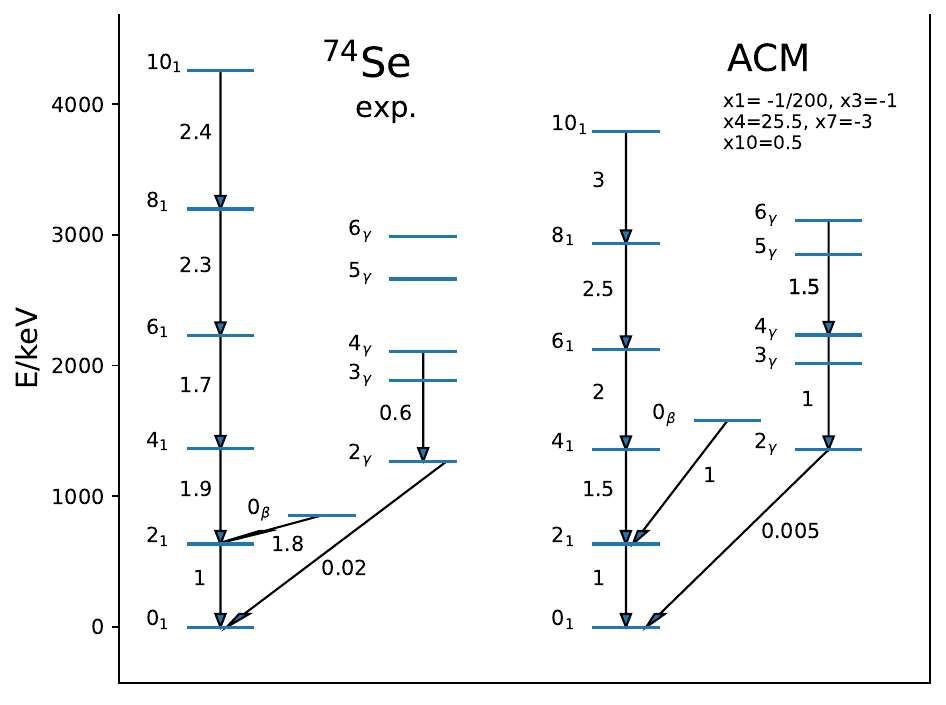}
\caption{Comparison of experimental (left) and ACM (right) spectra for $^{74}$Se. The energy difference of the $2^+_1$ and $0^+_1$ states is normalized 
to the experimental value in $^{74}$Se. The $B(E2)$ transition rates are expressed in units of the $B(E2,2_{1}^{+}\rightarrow 0_{1}^{+}) = 1$ transition.
The parameters used in the ACM calculations are displayed in the right corner.} 
\label{74Se_ACM}
\end{figure}

In the numerical application of the ACM to the studied nuclei we fitted the parameters of the Hamiltonian of Eq. (\ref{HACM})
to the experimentally known low-lying members of the ground, beta and gamma bands in $^{150}$Nd, $^{102}$Mo and $^{74}$Se. 
Both theoretical and experimental spectra are shown in Figs.~\ref{150Nd_ACM}--\ref{74Se_ACM}.
As for the $^{100}$Zr, the used ACM model does not give satisfactory results.
In an attempt to describe the very low energy of the $\beta$ bandhead ($R_{\beta/2_1} = 1.56$),
the structure of the ground and excited bands is severely disturbed.
We intend to address this issue in our future studies.

When fitting the experimental spectra, we take into account the fact that the first excited $K=0$ state band is not necessarily the $\beta$ band \cite{Garrett2001}. There are still many questions about the true nature of this state \cite{SharpeySchafer2008,SharpeySchafer2010,SharpeySchafer2011,SharpeySchafer2019}. Thus, we focus the comparison to the ground state band and to the first excited $2^{+}$ states, which are undoubtedly the $\gamma$ bandheads. In doing so, we try to achieve the best possible description of the energy of the first excited $K=0$ state, whose low energy is a signature of the shape phase coexistence, but it is challenging, especially in case of $^{74}$Se. As a result of the attempt to describe the low-lying $K=0$ states we observe centrifugal stretching in the ground band, all the more prominent when the energy of the first excited $K=0$ state decreases.

Even if the three nuclei are candidates for critical point nuclei of a first-order QPT, we have already emphasized that the energy conditions in them are very different. For example, the ratio
$R_{42}$ decreases from 2.93 in $^{150}$Nd to 2.51 in $^{102}$Mo and 2.15 in $^{74}$Se.
This is very strongly reflected by the value of the parameter $x_3$ which increases from $^{150}$Nd to $^{74}$Se. 

For $^{150}$Nd the agreement of the experimental and theoretical spectrum is very good. Along with the energies, the moments of inertia of the bands are also relatively well described. The $\beta$ bandhead is significantly lower than the $\gamma$ bandhead, but relatively high compared to the other two studied nuclei, the ratio $R_{\beta 2}=E(0_{2})/E(2_{1})$ being 5.2. Obtaining a low-lying $\beta$ band is generally a big problem mainly for lighter nuclei but in this case the energies of both $\beta$ and $\gamma$ bandheads are well fitted by the model. We also see that the ratio 
$B(E2, 0^{+}_{2} \rightarrow  2^{+}_{1})/B(E2, 2^{+}_{1} \rightarrow  0^{+}_{1})$ is 0.7 to be compared to the experimental value of 0.37. A big value of this ratio (above 0.3) is believed to be a signature of the QPT. 

In case of $^{102}$Mo the ratio $R_{\beta 2}=E(0_{2})/E(2_{1})$ is about 2.4. In an attempt to fit such a low-lying state  a pronounced centrifugal stretching in the ground state band is hard to avoid. The ratio 
$B(E2, 0^{+}_{2} \rightarrow  2^{+}_{1})/B(E2, 2^{+}_{1} \rightarrow  0^{+}_{1})$=1 is in an excellent agreement with the experimental value of 0.95.

$^{74}$Se is the lightest studied isotope with $R_{42}$ having a near-vibrational value of 2.15. Thus to get an overall good agreement with the experimetal spectrum a higher value of the parameter $x_3=-1$ is needed. Some centrifugal stretching is again observed in the ground band, but the energies and moments of inertia of the $\gamma$ band are very good.
The energy of the $\beta$ bandhead is less satisfactory, being about twice as high as its experimental counterpart. Consequently, the theoretical ratio $B(E2, 0^{+}_{2} \rightarrow  2^{+}_{1})/B(E2, 2^{+}_{1} \rightarrow  0^{+}_{1})$=1 is much smaller than the experimental value of 7.

\begin{figure}[!htb]
\begin{minipage}[t]{.45\textwidth}
\includegraphics[width=0.95\textwidth]{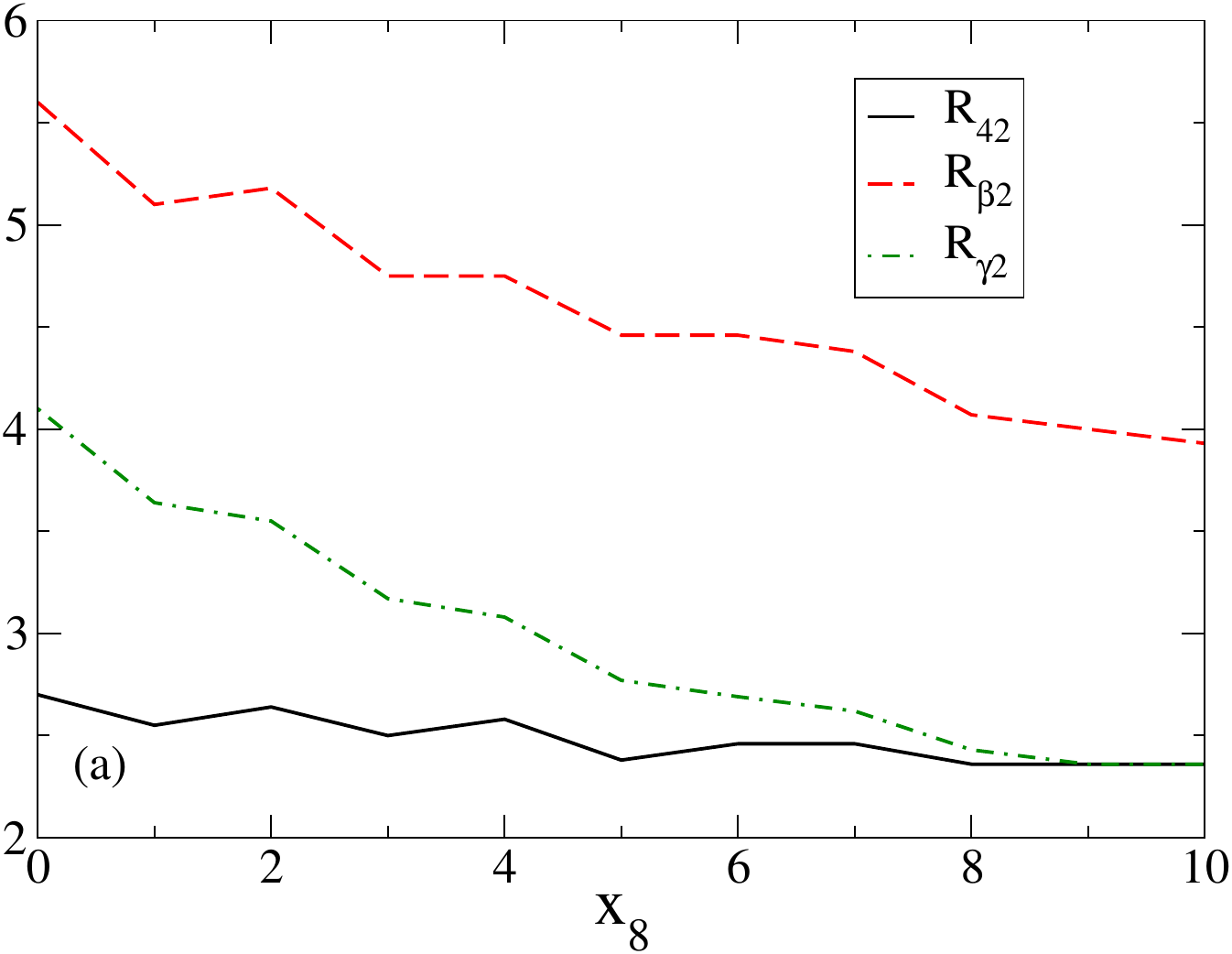}
\end{minipage}
\begin{minipage}[t]{.45\textwidth}
\includegraphics[width=0.95\textwidth]{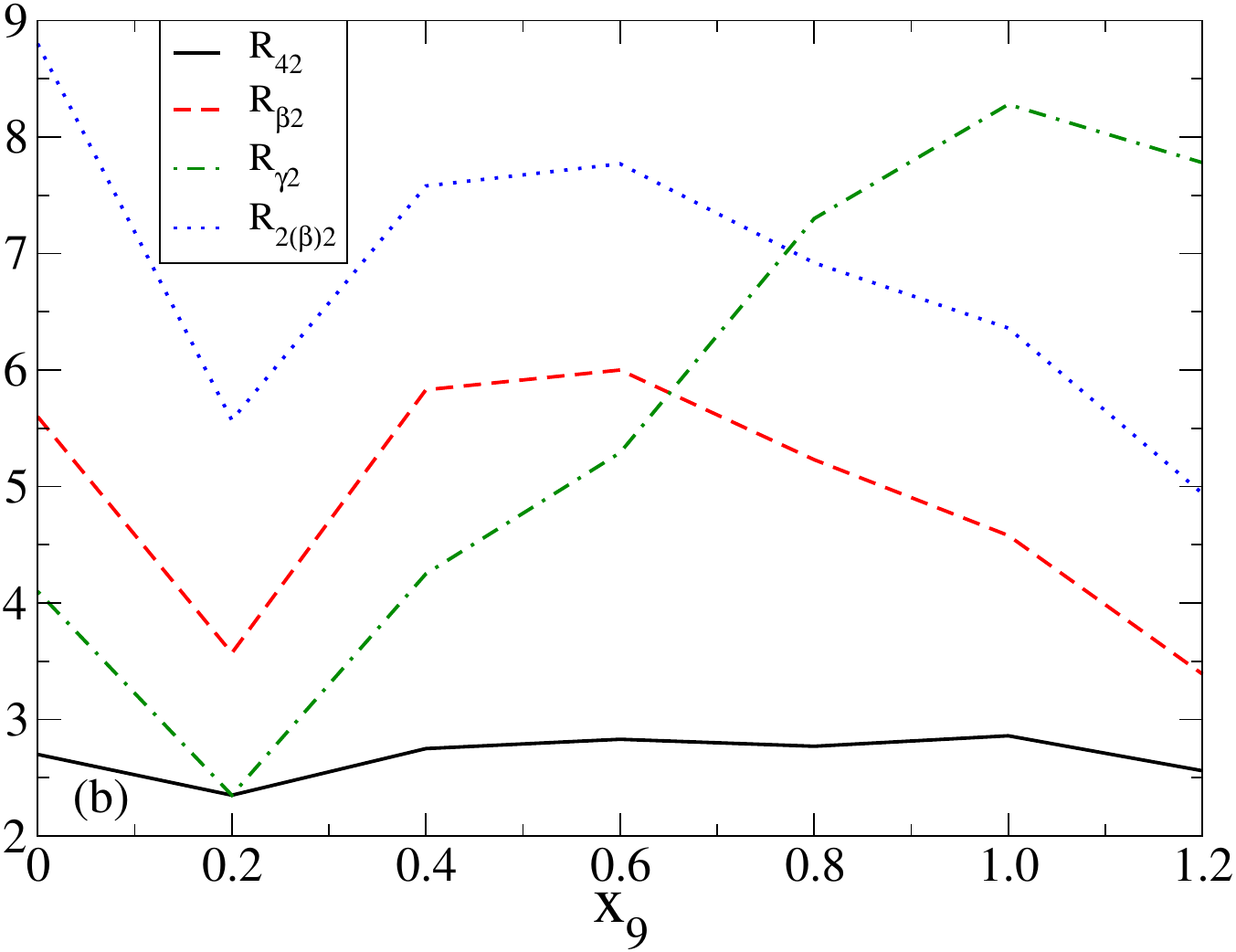}
\end{minipage}
\caption{Ratios $R_{42}$, $R_{\beta 2}$, and $R_{\gamma 2}$ as functions of the ACM parameter $x_8$ (a) and ratios $R_{42}$, $R_{\beta 2}$, and $R_{\gamma 2}$ as functions 
of the ACM parameter $x_9$ (b).}
\label{x8x9_ACM}
\end{figure}

Let us comment briefly on some of the parameters used in the calculation. 
All in all, the parameters $x_{5}$ and $x_{8}$ allow to reduce the ratios $R_{42}$, $R_{\beta 2}$, $R_{\gamma 2}=E(2_{\gamma})/E(2_{1})$  and $R_{2(\beta)2}=E(2_{\beta})/E(2_{1})$, the effect of the parameter $x_{8}$ being more important (see Fig.~\ref{x8x9_ACM}).
We may also notice that while in $^{150}$Nd the $2^{+}$ member of the $\gamma$ band is the third excited $2^{+}$ state, 
the situation changes in $^{102}$Mo where the $2^{+}$ member of the $\gamma$ band is the second excited $2^{+}$ state. 
(The second or third excited $2^{+}$ state is identified as a $\gamma$ bandhead based on the basis of a stronger $B(E2, 3^{+}_{1} \rightarrow  2^{+}_{2,3})$ transition).
The relative position of the $2_{\gamma}$ and $2_{\beta}$ states is strongly sensitive to the parameter $x_{9}$, as is illustrated in Fig.~\ref{x8x9_ACM}. It can be seen that from a certain value of this parameter, the relative position of the $2_{\gamma}$ and $2_{\beta}$ states changes. Therefore, the largest value of $x_{9}$ for $^{150}$Nd (among the three studied cases) is obvious.
It should be noted that the term proportional to parameter $x_{10}$ in Hamiltonian of Eq.~(\ref{HACM}) can induce triaxiality.
This term is not needed for $^{150}$Nd, where microscopic calculations result in an effective $\gamma$ value of $0^{\circ}$
(see Table~\ref{Sky3D}). In the case of $^{102}$Mo and $^{74}$Se, a small parameter value of $x_{10} = 0.5$ leads to an overall improvement
of the fit of the experimental spectra. This seems to be consistent with microscopic calculations, where, especially
in the case of $^{74}$Se, the effective value of the parameter 
$\gamma = 22^{\circ}$.

\begin{figure}[!htb]
\includegraphics[width=\textwidth]{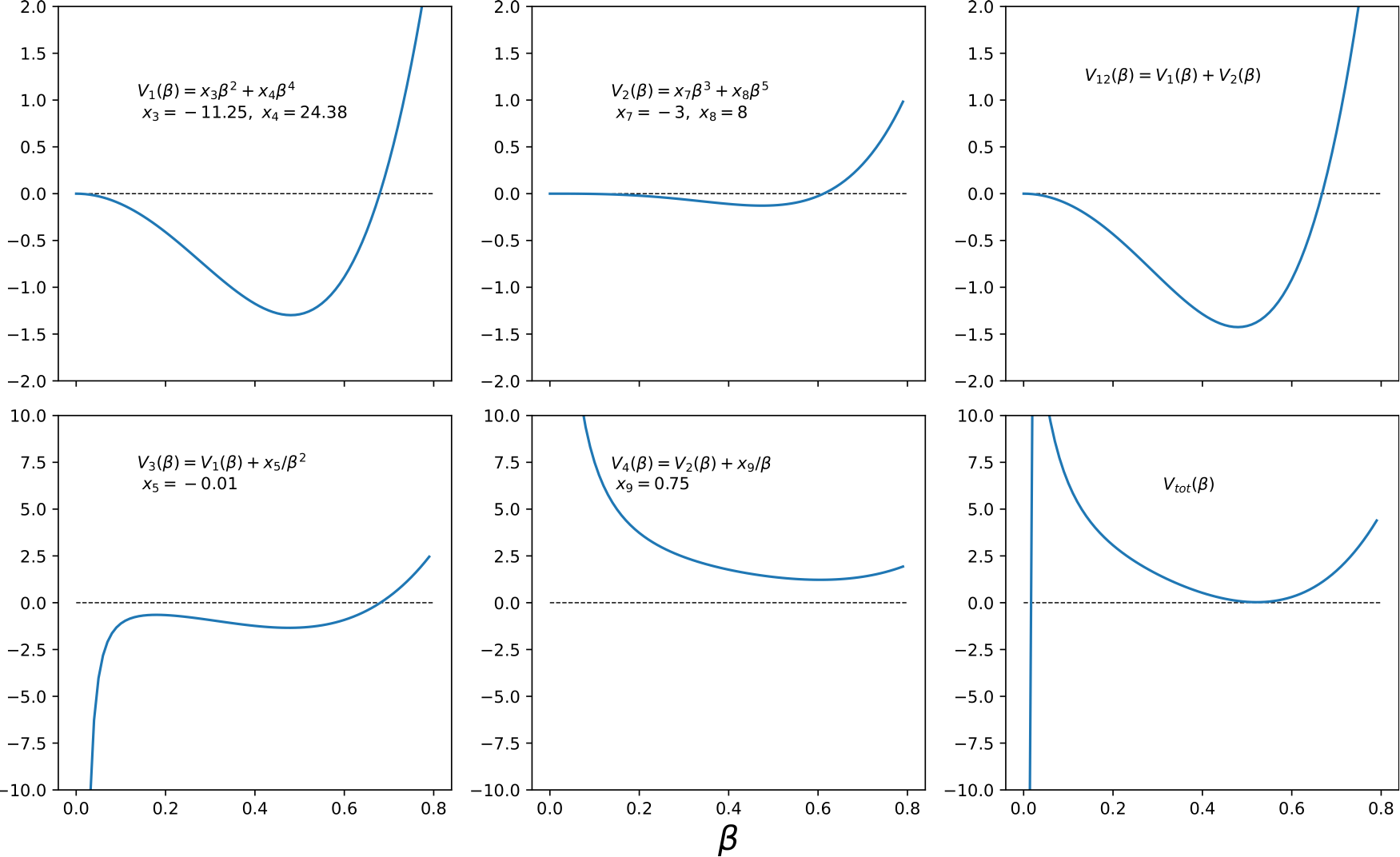}
\caption{Symmetric (left) and antisymmetric (middle) parts of the ACM potential and the total ACM potential (right) obtained from the fit to the experimental spectra of $^{150}$Nd. Potential terms diverging for $\beta=0$ are excluded in the upper panel and included in the lower panel.}
\label{150NdACMpot}
\end{figure}

\begin{figure}[!htb]
\includegraphics[width=\textwidth]{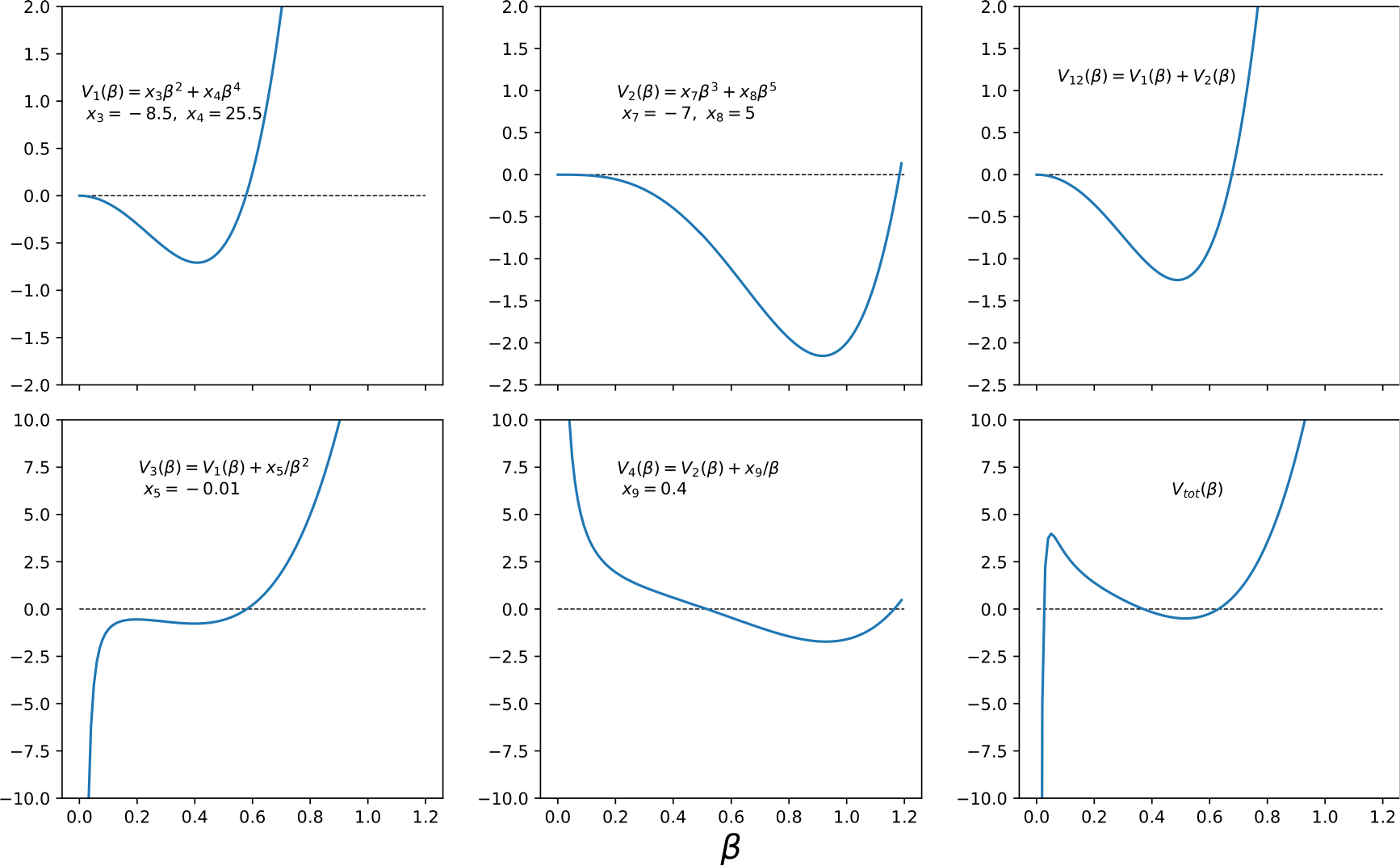}
\caption{Symmetric (left) and antisymmetric (middle) parts of the ACM potential and the total ACM potential (right) obtained from the fit to the experimental spectra of $^{102}$Mo. Potential terms diverging for $\beta=0$ are excluded in the upper panel and included in the lower panel.}
\label{102MoACMpot}
\end{figure}

\begin{figure}[!htb]
\includegraphics[width=\textwidth]{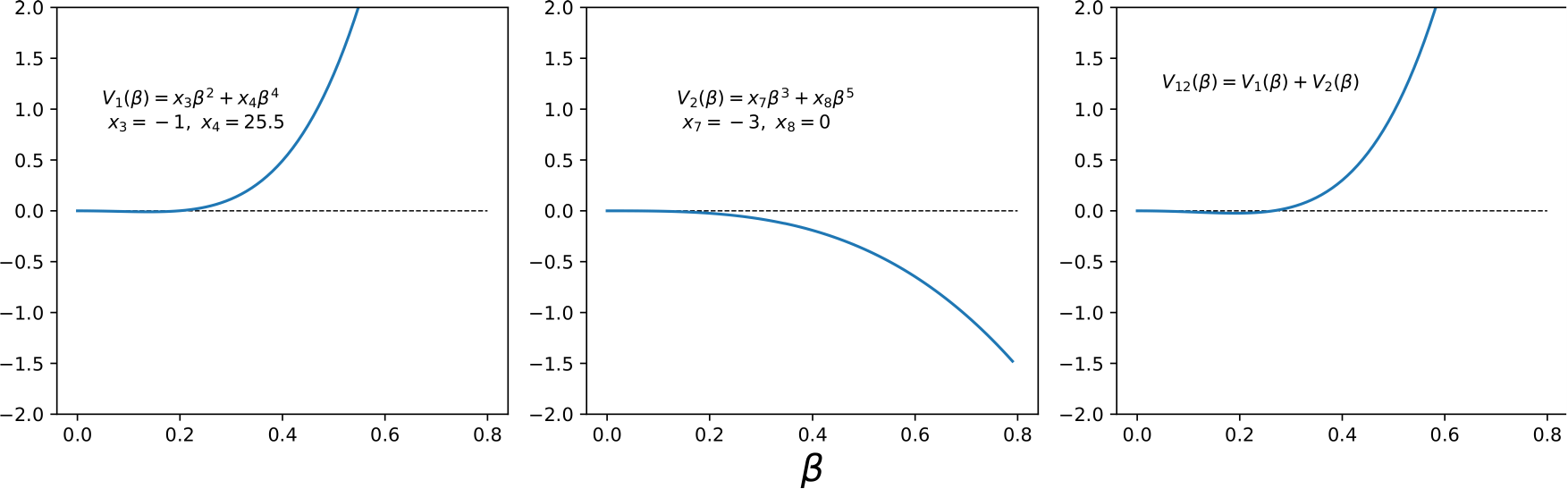}
\caption{Symmetric (left) and antisymmetric (middle) parts of the ACM potential and the total ACM potential (right) obtained from the fit to the experimental spectra of $^{74}$Se.}
\label{74SeACMpot}
\end{figure}

In Figs.~\ref{150NdACMpot}-\ref{74SeACMpot} potentials used in the ACM fits for $^{150}$Nd, $^{102}$Mo and $^{74}$Se are plotted. Their similarity to the microscopic symmetric and antisymmetric potentials (Figs.~\ref{FigVsym-asym}-\ref{FigVsym-asym-tls}) for low and medium quadrupole deformation values is clearly visible, especially for $^{74}$Se and $^{102}$Mo, where the symmetric parts ($V_1(\beta)$) exhibit a minimum at low or medium deformations while the antisymmetric parts ($V_2(\beta)$) are descreasing. The potentials presented for $^{102}$Mo and $^{74}$Se are shown without the term $x_{10}\cos^2{3\gamma}$ in Eq.~(\ref{HACM}) for simplicity, and thus under the assumption of zero triaxiality.
The cases of $^{150}$Nd and $^{102}$Mo are qualitatively very similar. 
The upper left panel (the symmetric potential part not depending on $\cos{3\gamma}$ and containing positive even powers of $\beta$) looks like a Davidson potential, widely used in the context of QPTs \cite{Bonatsos2007}, except that it does not go to infinity at the origin. The upper middle panel (the antisymmetric potential part that is $\cos{3\gamma}$ dependent and contains positive odd powers of $\beta$) again looks like a Davidson potential. 
Their sum (upper right panel) indicates that the result is dominated by the even powers.
In the lower left and middle panels we see that the negative even powers of $\beta$ lead to minus infinity at the origin, while the negative odd powers lead to plus infinity at the origin. 
The final result (lower right panel) indicates that the negative even powers dominate at the origin, but the negative odd powers dominate a little further to the right. It is clear that it is shaped up by the Davidson potential of the positive even powers, the minus infinity potential contributed by the negative even powers near the origin, and the plus infinity potential contributed by the  negative odd powers a little further to the right.
The final result resembles very much the confined beta-soft (CBS) model of Pietralla {\it et al.}, which uses an infinite square well potential displaced from the origin. This has been used for both deformed nuclei \cite{Pietralla2004}  and gamma-soft nuclei \cite{Bonatsos2006b}. In other words, the main contribution of the negative odd powers is to produce the displacement from the origin. 

In the case of $^{74}$Se it seems that the negative powers are not needed, since no displacement from the origin seems to be required. The total potential is dominated by the positive even powers of $\beta$. A Davidson potential is created by the positive even powers, but it is very shallow and looks flat. The same holds for the total potential. Thus in the case of $^{74}$Se the total potential looks very similar to the square well potential starting from the origin, which is used in the $E(5)$ and $X(5)$ CPS. The only difference is that the right wall of the potential does not rise abruptly to infinity, but rises more smoothly, resembling the sloped wall potential used by Caprio 
\cite{Caprio2004} in the $E(5)$ and $X(5)$ frameworks. The main effect of the sloped wall is that it allows the beta band to go lower in energy. It also allows for $R_{4/2}$
ratios lower than the 2.91 value predicted by $X(5)$. For example, in Fig.~6 of Ref.~\cite{Caprio2004} one can see that the spectra and $B(E2)$s for $^{150}$Nd are well reproduced by a sloped well which has $R_{4/2}=2.667$ (see also Fig.~5 of Ref.~\cite{Caprio2004} for sloped wall results producing $R_{4/2}=2.643$). However, $^{74}$Se has the experimental value of $R_{4/2}=2.148$, thus this sloped wall alone cannot describe it.  

\section{Conclusions}
\label{Concl}

In the present investigation three regions of the nuclear chart ($N=40$, 60, 90), in which a QPT from spherical to deformed nuclei is observed, are studied on equal footing in three different ways, a) by consideration of the empirical systematics of spectra and $B(E2)$ transition rates, b) through microscopic mean field calculations using the Skyrme-Hartree-Fock + BCS approach, and c) in the framework of the Bohr Hamiltonian, using the Algebraic Collective Model. Several similarities and differences among these three regions are pointed out.

The empirical systematics of spectra and $B(E2)$ transition rates indicate the uniform occurrence of a first-order QPT in all three regions. In contrast, they also show that the numerical values of the various indicators of collectivity differ in the three regions, showing values corresponding to high deformation at $N=90$, intermediate deformation at $N=60$, and low deformation at $N=40$.  From this point of view, the $N=90$ QPT appears to be close to the parameter-independent predictions given by the X(5) critical point symmetry, while the $N=40$ region appears to be closer to the X(3) CPS. 
 
The potential energy curves obtained through microscopic Skyrme-Hartree-Fock + BCS calculations for several different parametrizations provide a flat-bottomed potential at non-zero prolate deformation in the $N=90$ region, while in the $N=60$ region some degree of flatness  is obtained on the oblate side. In the $N=40$ region signs of flatness appear around zero deformation.

In the framework of the Algebraic Collective Model, the free parameters in front of the various terms in the Bohr Hamiltonian are fitted to the data. The resulting potential energy curves in the $N=90$ and $N=60$ regions resemble a Davidson potential possessing a deformed minimum, being at the same time displaced from the origin, in analogy to the confined $\beta$-soft (CBS) model, applicable in deformed and transitional nuclei. In contrast, the PEC obtained in the $N=40$ region resembles a flat potential starting from the origin, as expected within the E(5) and X(5) CPSs, albeit with a sloped right wall instead of a vertical one. The radically different picture obtained for $N=40$ can be attributed to the fact that valence protons and neutrons in this case occupy the same major shell.

Several open questions call for further investigations. 
It has been shown that the symmetric and antisymmetric parts of the microscopic PECs and the ACM potential behave in a similar way at least
for low and middle $\beta_2$ values. To establish a closer link between
these two approaches one has to take into account also vibrational and rotational zero-point corrections in the PEC calculations and include higher order terms in $\beta_2$ and $\cos{3\gamma}$ to obtain a more realistic ACM potential. The rigid kinetic term in the ACM Hamiltonian should be also replaced by a deformation dependent ($\beta_2$ and $\cos{3\gamma}$) term derived from the microscopic mass tensor that 
might solve the problem of rotational band stretching in $^{100}$Zr. 

The preference for prolate shapes in the $N=90$ region and for oblate shapes in the $N=60$ region should be further investigated. Since spectra alone do not provide differentiation between prolate and oblate shapes, relevant investigations should be focused on the determination of the sign of the relevant quadrupole moments, which is opposite for oblate nuclei in relation to prolate nuclei.
In light and medium-mass nuclei, in which the valence protons and neutrons occupy the same major shell, the influence of the SU(4) Wigner supermultiplet symmetry should be taken into account \cite{Kota2017}.

\acknowledgments
This work was supported by the project SP2023/25 financed by the Czech Ministry of Education, Youth 
and Sports and by the GA\v{C}R grant 22-14497S. 
This paper is dedicated to the memory of Adam Pr\'{a}\v{s}ek, an excellent student of V\v{S}B-Technical University of Ostrava who worked on the ACM calculations
and passed away in Grenoble Alpes on October 1, 2023.

\end{document}